\documentclass[10pt,journal,compsoc]{IEEEtran}

\ifCLASSOPTIONcompsoc
  \usepackage[nocompress]{cite}
\else
  \usepackage{cite}
\fi

\ifCLASSINFOpdf
  \usepackage[pdftex]{graphicx}
\else
\fi

\usepackage{comment}
\usepackage{tabu}                      
\usepackage{booktabs}  
\usepackage[font=footnotesize]{caption}
\usepackage{subcaption}
\usepackage{dblfloatfix}

\usepackage{xprintlen}

 \usepackage{array,multirow, makecell}
 \usepackage{float}
\newcommand\Tstrut{\rule{0pt}{2.6ex}}       
\newcommand\Bstrut{\rule[-1.4ex]{0pt}{0pt}} 
\newcommand{\TBstrut}{\Tstrut\Bstrut} 

\usepackage{amssymb}
\usepackage{etoolbox}
\robustify{\fbox}

\usepackage{hyperref}
\usepackage[dvipsnames]{xcolor}
\usepackage{tikz}
\usepackage{marvosym}
\usepackage{soul}
\usepackage{tcolorbox}
\usepackage{xifthen}
\usepackage{xspace}
\newboolean{showchecklists}
\setboolean{showchecklists}{true}

\newboolean{showcards}
\setboolean{showcards}{true}

\newboolean{showlinks}
\setboolean{showlinks}{true}

\newboolean{showdels} 
\setboolean{showdels}{false}

\newboolean{highlightac} 
\setboolean{highlightac}{true}

\newboolean{delsyesacno}
\setboolean{delsyesacno}{true}
\ifthenelse{\boolean{delsyesacno}}{
    \setboolean{highlightac}{false}
    \setboolean{showdels}{true}
}{
}

\newboolean{finalpresentation}
\setboolean{finalpresentation}{true}
\ifthenelse{\boolean{finalpresentation}}{
    \setboolean{highlightac}{false}
    \setboolean{showdels}{false}
}{
}

\definecolor{notiongray}{gray}{0.11}

\DeclareRobustCommand{\TODO}[4][orange]{
    \HollowBox~\textcolor{#1}{
        \textbf{#2} 
        \ifthenelse{\isempty{#3}}{\ignorespaces}{: #3 } 
        \ifthenelse{\isempty{#4}}{\ignorespaces}{( #4 )}
    } 
}

\DeclareRobustCommand{\hideTODO}[4][orange]{}

\DeclareRobustCommand{\TODOdone}[4][Gray]{
    \Checkedbox~\textcolor{Gray}{
        \st{\textbf{#2}} 
        \ifthenelse{\isempty{#4}}{}{(#4)}
    }
}

\DeclareRobustCommand{\hideTODOdone}[4][Gray]{}

\DeclareRobustCommand{\TODOcancelled}[4][Gray]{
    \CrossedBox~\textcolor{Gray}{\st{#2} {#4}}
}

\DeclareRobustCommand{\hideTODOcancelled}[4][Gray]{}

\DeclareRobustCommand{\paper}[4][Purple]{\textcolor{#1}{
        \textbf{#2 \cite{#2}}
        \ifthenelse{\isempty{#3}}{\ignorespaces}{: #3 } 
        \ifthenelse{\isempty{#4}}{\ignorespaces}{( #4 )}
    }
}

\ifthenelse{\boolean{showlinks}}{
    \DeclareRobustCommand{\LINK}[3][RoyalBlue]{\href{#3}{\textcolor{#1}{$\rightarrow$~{#2}}}}

}{
    \DeclareRobustCommand{\LINK}[3][RoyalBlue]{}

}

\definecolor{ct20_blue_d}{HTML}{4c78a8}
\definecolor{ct20_blue_l}{HTML}{9ecae9}
\definecolor{ct20_orange_d}{HTML}{f58518}
\definecolor{ct20_orange_l}{HTML}{ffbf79}
\definecolor{ct20_green_d}{HTML}{54a24b}
\definecolor{ct20_green_l}{HTML}{88d27a}
\definecolor{ct20_yellow_d}{HTML}{b79a20}
\definecolor{ct20_yellow_l}{HTML}{f2cf5b}
\definecolor{ct20_teal_d}{HTML}{439894}
\definecolor{ct20_teal_l}{HTML}{83bcb6}
\definecolor{ct20_red_d}{HTML}{e45756}
\definecolor{ct20_red_l}{HTML}{ff9d98}
\definecolor{ct20_pink_d}{HTML}{d67195}
\definecolor{ct20_pink_l}{HTML}{fcbfd2}
\definecolor{ct20_purple_d}{HTML}{b279a2}
\definecolor{ct20_purple_l}{HTML}{d6a5c9}

\colorlet{cDeleted}{ct20_red_d}
\colorlet{cNewcont}{ct20_blue_d}
\colorlet{cChanged}{ct20_purple_d}

\ifthenelse{\boolean{highlightac}}{
    \DeclareRobustCommand{\newcont}[1]{\textcolor{cNewcont}{#1} \ignorespaces}
    \newenvironment{Newcont}[1][]{\color{cNewcont}}{}{}
    \DeclareRobustCommand{\changed}[2]{\textcolor{cChanged}{#2} \ignorespaces}
    \newenvironment{Changed}[1][]{\color{cChanged}}{}{}

}{
    \DeclareRobustCommand{\newcont}[1]{\textcolor{black}{#1} \ignorespaces}
    \DeclareRobustCommand{\changed}[2]{\textcolor{black}{#2} \ignorespaces}
    \newenvironment{Newcont}[1][]{}{}{}
    \newenvironment{Changed}[1][]{}{}{}
    
}

\ifthenelse{\boolean{showdels}}{
    \DeclareRobustCommand{\deleted}[1]{\textcolor{cDeleted}{\st{#1}} \ignorespaces}
    \newenvironment{Deleted}[1][]{\color{cDeleted}}{}{}
    \DeclareRobustCommand{\changed}[2]{\textcolor{cDeleted}{#1} \ignorespaces}
}{
    \DeclareRobustCommand{\deleted}[1]{\ignorespaces}
    \excludecomment{Deleted}
}

\ifthenelse{\boolean{showcards}}{
    \DeclareRobustCommand{\cardref}[2][Notecard]{
    \hyperref[card:#2]{($\downarrow$~{#2})}}
    \DeclareRobustCommand{\secref}[2][Section-link]{
    \hyperref[sec:#2]{($\uparrow$~{#2})}}
    \DeclareRobustCommand{\figref}[2][Figure-link]{
    \hyperref[sec:#2]{($\nearrow$~{#2})}}
}{
    \DeclareRobustCommand{\cardref}[2][Notecard]{}
    \DeclareRobustCommand{\secref}[2][Notecard]{}
    \DeclareRobustCommand{\figref}[2][Notecard]{}
}
\DeclareRobustCommand{\note}[3][Cerulean]{
    \textcolor{#1}{
        #2
        \ifthenelse{\isempty{#3}}{\ignorespaces}{(#3)}
    }
}
\DeclareRobustCommand{\hidenote}[3][Cerulean]{\ignorespaces}

\DeclareRobustCommand{\tempsec}[2]{\csname #1\endcsname{\textcolor{
ForestGreen}{#2}}}

\DeclareRobustCommand{\point}[2][Brown]{\textcolor{#1}{$\bullet$ #2}\newline}
\DeclareRobustCommand{\arrow}[2][Brown]{\textcolor{#1}{$\rightarrow$
#2}\newline}

\DeclareRobustCommand{\assoc}[2][hide]{
    \ifthenelse{\equal{#1}{show}}
    {\cardref{#2}}
    {}
}

\newcommand{\comm}[1]{\textcolor{ForestGreen}{(#1)}}
\newcommand{\hidecomm}[1]{\ignorespaces}

\DeclareRobustCommand{\graybox}[1]{\colorbox{gray}{\textcolor{white}{#1}}}

\DeclareRobustCommand{\lbox}[1]{\fbox{\sffamily\footnotesize #1}}

\ifthenelse{\boolean{showchecklists}}{
    \newenvironment{checklist}[2][Checklist]
    {
        \newcommand{\todo}[4][{#2}]{\par \TODO[##1]{##2}{##3}{##4}}
        \newcommand{\tododone}[4][{#2}]{\par \TODOdone[##1]{##2}{##3}{##4}}
        \newcommand{\todocancelled}[4][{#2}]{\par \TODOcancelled[##1]{##2}{##3}{##4}}
        
        \newcommand{\hidetodo}[4][{#2}]{\hideTODO[##1]{##2}{##3}{##4}}
        \newcommand{\hidetododone}[4][{#2}]{\hideTODOdone[##1]{##2}{##3}{##4}}
        \newcommand{\hidetodocancelled}[4][{#2}]{\hideTODOcancelled[##1]{##2}{##3}{##4}}
        \setCardColor{#2}
        \begin{tcolorbox}
        \begin{small}
        \textbf{#1}\par
    }
    {
        \end{small} 
        \end{tcolorbox}
    }
}
{
    
}

\DeclareRobustCommand{\card}[5][Black]{
    \setCardColor{#1}
    \begin{small}
        \ifthenelse{\isempty{#4}}%
        {
            \begin{tcolorbox}[title={{#2} \tiny(card:#5) },lower separated=true] 
            {#3}
            \end{tcolorbox}
        }
        {
            \begin{tcolorbox}[sidebyside,title={\begin{center}{\tiny (card:{#5})}\end{center} \par {#2} },lower separated=true] 
            {#3}
            \tcblower
            {#4}
            \end{tcolorbox}
        }
        \label{card:#5}
    \end{small}
}

\DeclareRobustCommand{\setCardColor}[1]{
    \tcbset{colback=#1!5!white,colframe=#1!75!black,fonttitle=\bfseries}
}

\def \sec{section}
\def \secref#1{\S~\ref{sec:#1}}

\def \gof{Goodness-of-Fit}

\def \conI{Con-1}
\def \conII{Con-2}
\def \conIII{Con-3}
\def \conIV{Con-4}
\def \conV{Con-5}
\def \keyI{Key-I}
\def \keyII{Key-II}
\def \reqI{Req-1}
\def \reqII{Req-2}
\def \reqIII{Req-3}
\def \reqIV{Req-4}

\def \dpI{\dgI{}}
\def \dpII{\dgII{}}
\def \dpIII{\dgIII{}}
\def \dpIV{\dgIV{}}

\def \dgI{DG-1}
\def \dgII{DG-2}
\def \dgIII{DG-3}
\def \dgIV{DG-4}

\def \gdlI{Gdl-1}
\def \gdlII{Gdl-2}
\def \gdlIII{Gdl-3}
\def \gdlIV{Gdl-4}

\def \visrec{VisRec}

\def \hist{\emph{Hist}}
\def \sp{\emph{SP}}
\def \dcp{\emph{wDCP}}
\def \wdcp{\emph{wDCP}}
\def \splom{\emph{SPLOM}}
\def \rsplom{\emph{rSPLOM}}

\def \pc{\emph{PC}}

\def \psc{\emph{PSc}}

\def \coIline{\emph{1D-Line}}
\def \coIbox{\emph{1D-Box}}
\def \coIhist{\emph{1D-Hist}}
\def \coIIgrid{\emph{2D-Grid}}
\def \coIIgridminus{\emph{2D-Grid\textsuperscript{ (-)}}}
\def \coIIsup{\emph{2D-Sup}}
\def \coIIjux{\emph{2D-Jux}}

\def \histfull{\emph{Histogram}}
\def \spfull{\emph{Scatterplot}}
\def \dcpfull{\emph{Density Contourplot}}
\def \splomfull{\emph{Scatterplot Matrix}}
\def \rsplomfull{\emph{reduced SPLOM}}
\def \pcfull{\emph{Parallel Coordinates}}

\def \pscfull{\emph{Point Scales}}

\def \coIlinefull{\emph{1D - Linegraph}}
\def \coIboxfull{\emph{1D - Boxplot}}
\def \coIhistfull{\emph{1D - Cumulative Histogram}}
\def \coIIgridfull{\emph{2D - Gridview}}
\def \coIIsupfull{\emph{2D - Superpositioned views}}
\def \coIIjuxfull{\emph{2D - Juxtapositioned views}}

\def \VPSA{VPSA}

\definecolor{copt}{RGB}{238,34,12}

\DeclareRobustCommand{\topt}{\raisebox{-0.3em}{\includegraphics[height=1.2em]{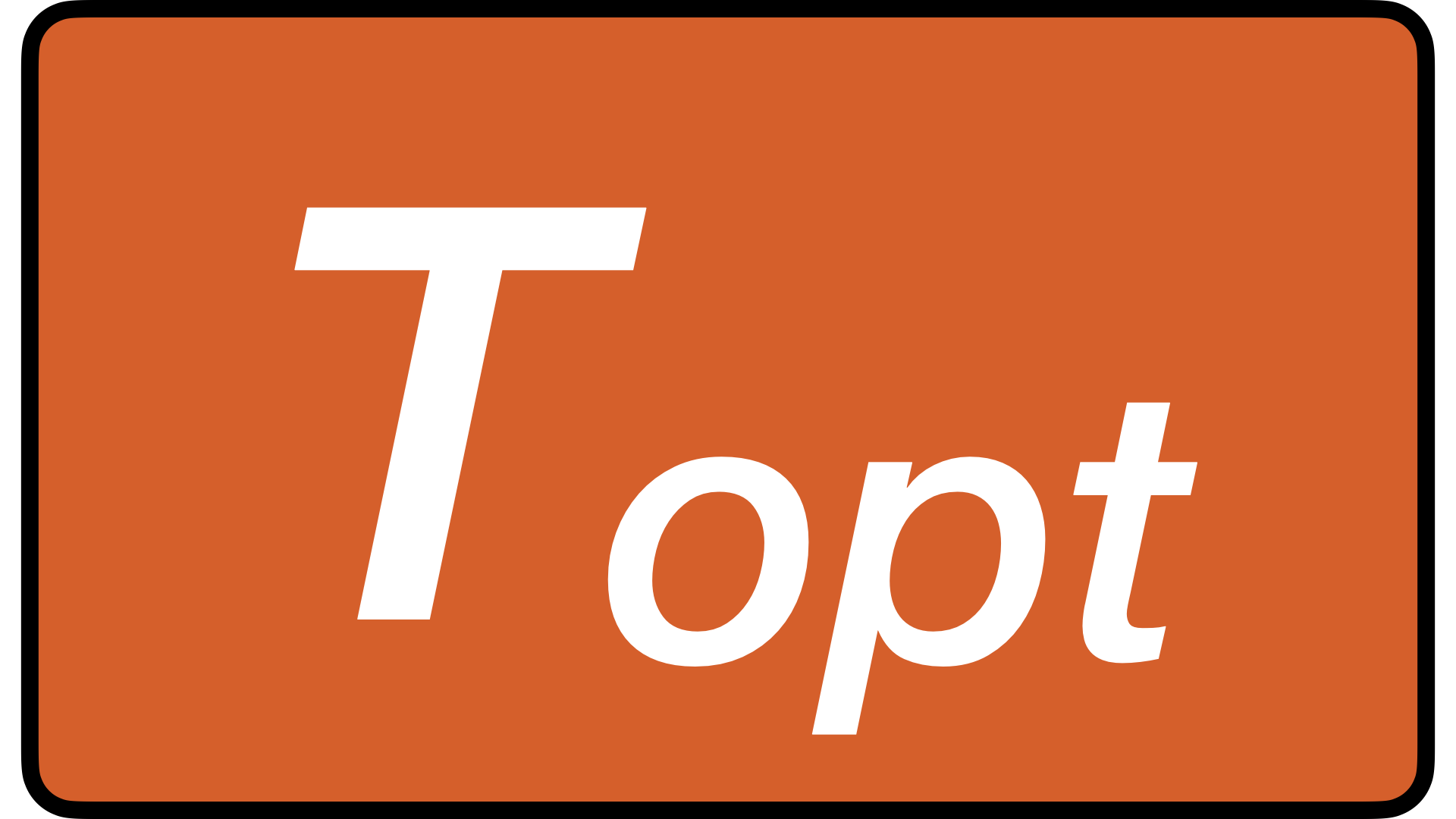}}}
\DeclareRobustCommand{\tfit}{\raisebox{-0.3em}{\includegraphics[height=1.2em]{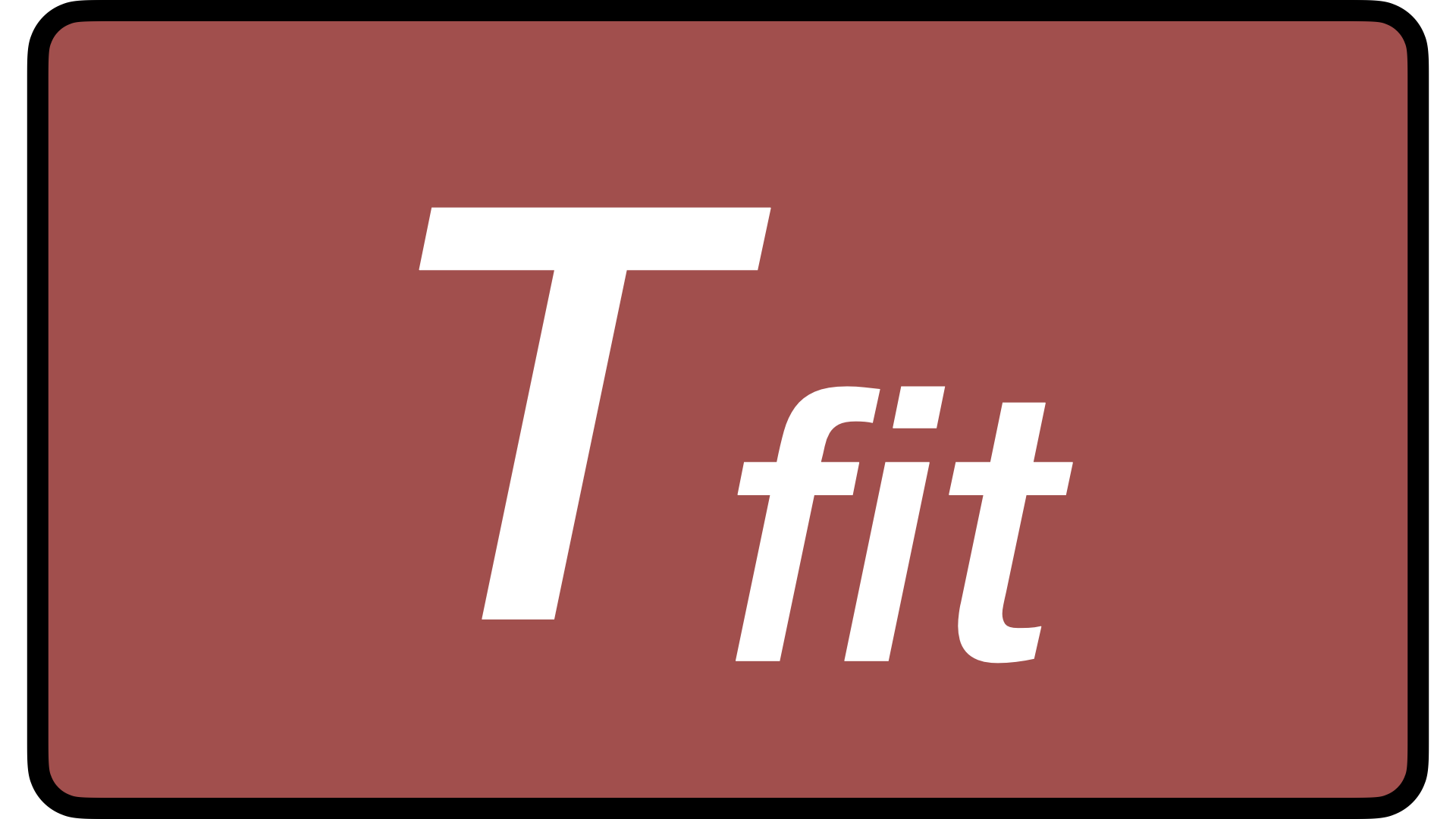}}}
\DeclareRobustCommand{\tunc}{\raisebox{-0.3em}{\includegraphics[height=1.2em]{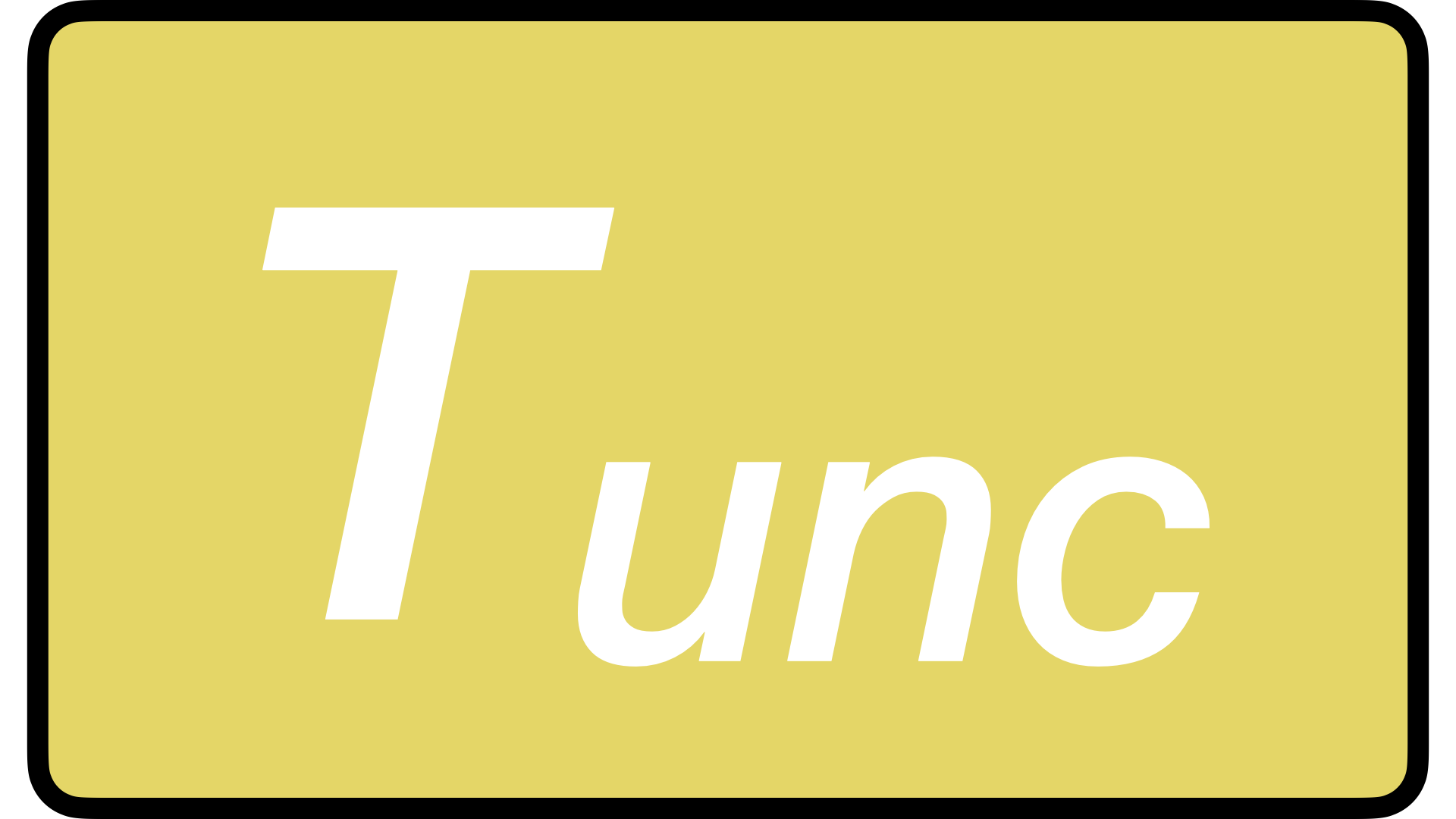}}}
\DeclareRobustCommand{\tout}{\raisebox{-0.3em}{\includegraphics[height=1.2em]{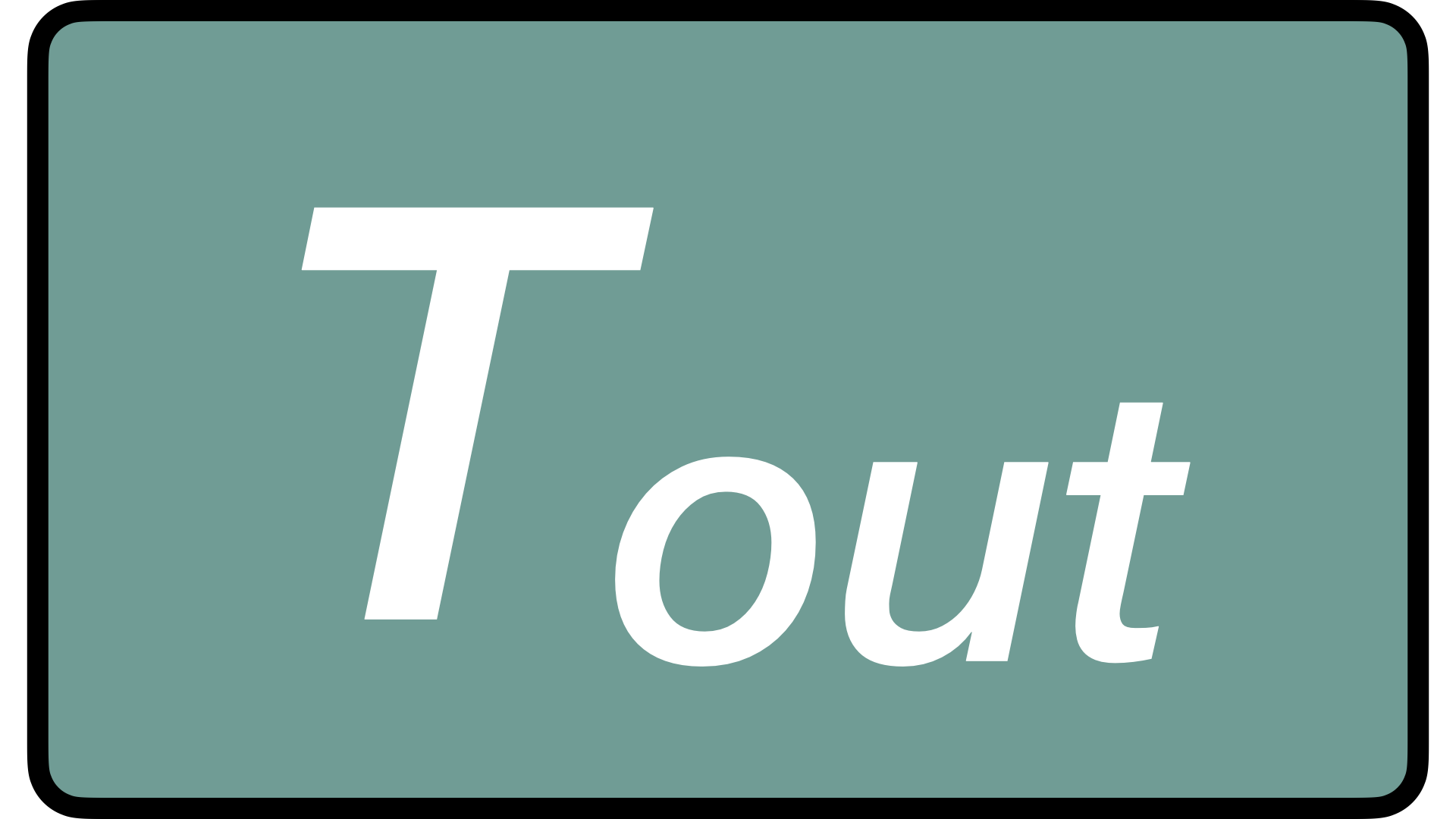}}}
\DeclareRobustCommand{\tsens}{\raisebox{-0.3em}{\includegraphics[height=1.2em]{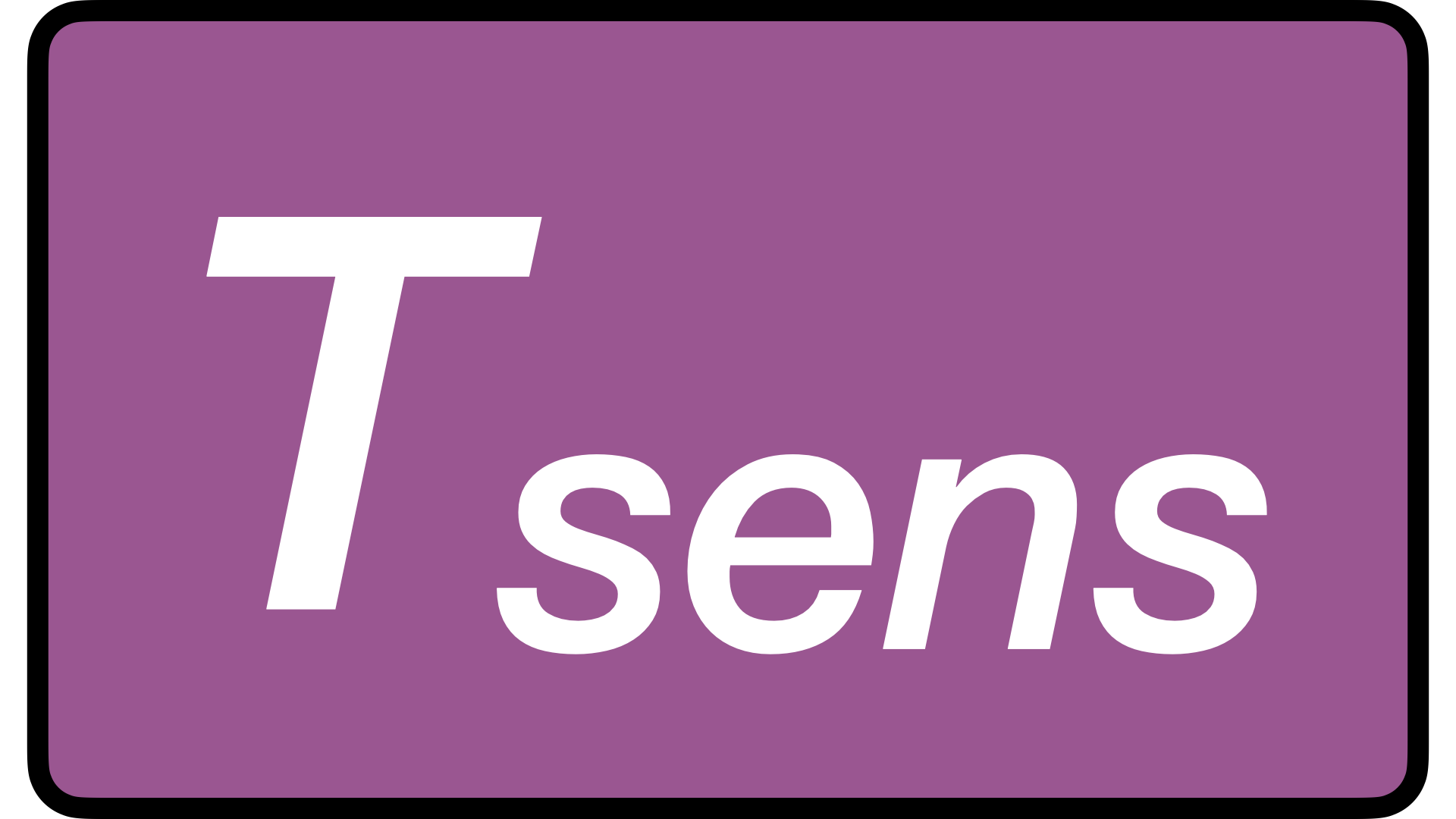}}}
\DeclareRobustCommand{\tpart}{\raisebox{-0.3em}{\includegraphics[height=1.2em]{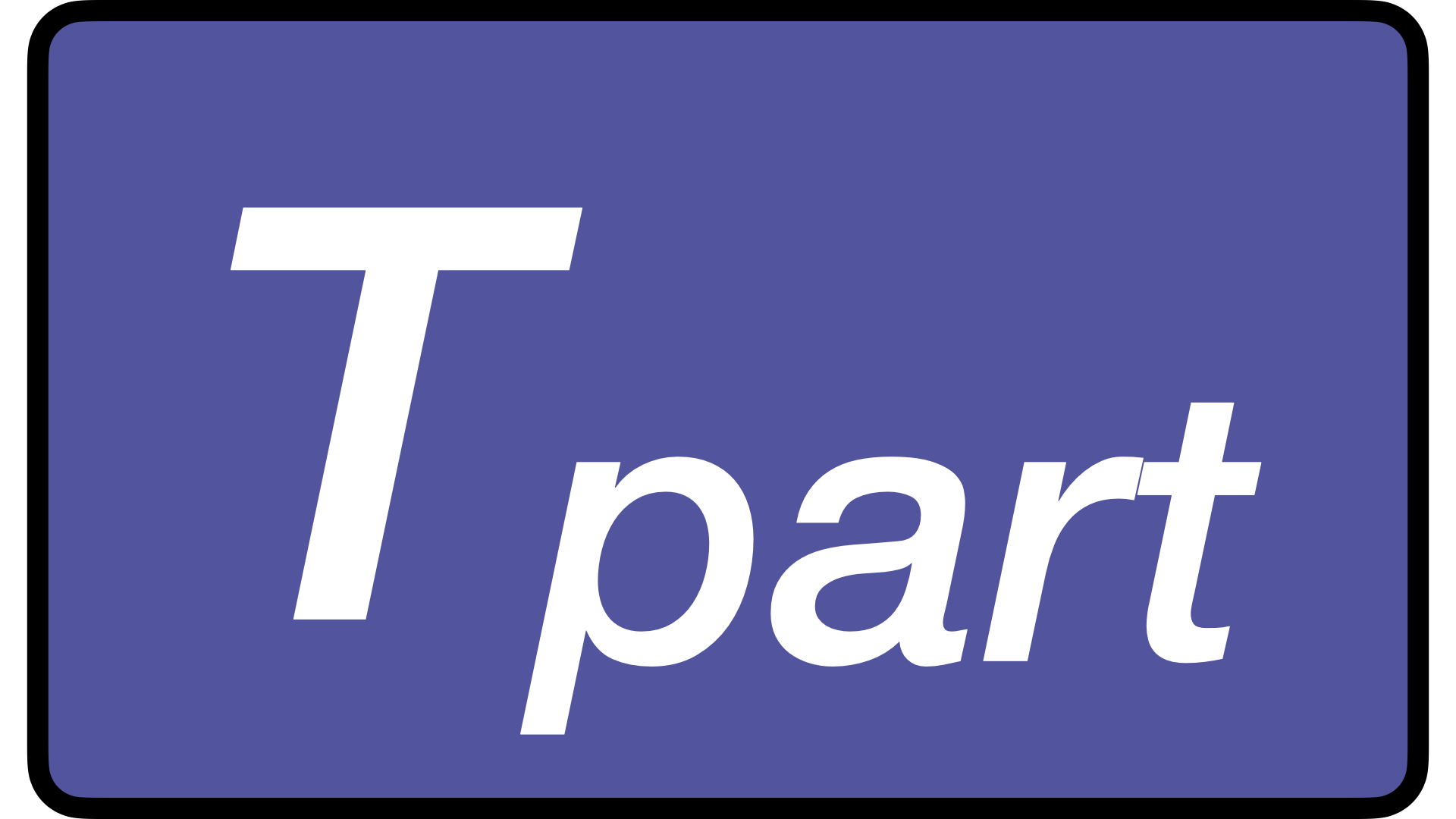}}}

\DeclareRobustCommand{\ldsp}{\raisebox{-0.31em}{\includegraphics[height=1.2em]{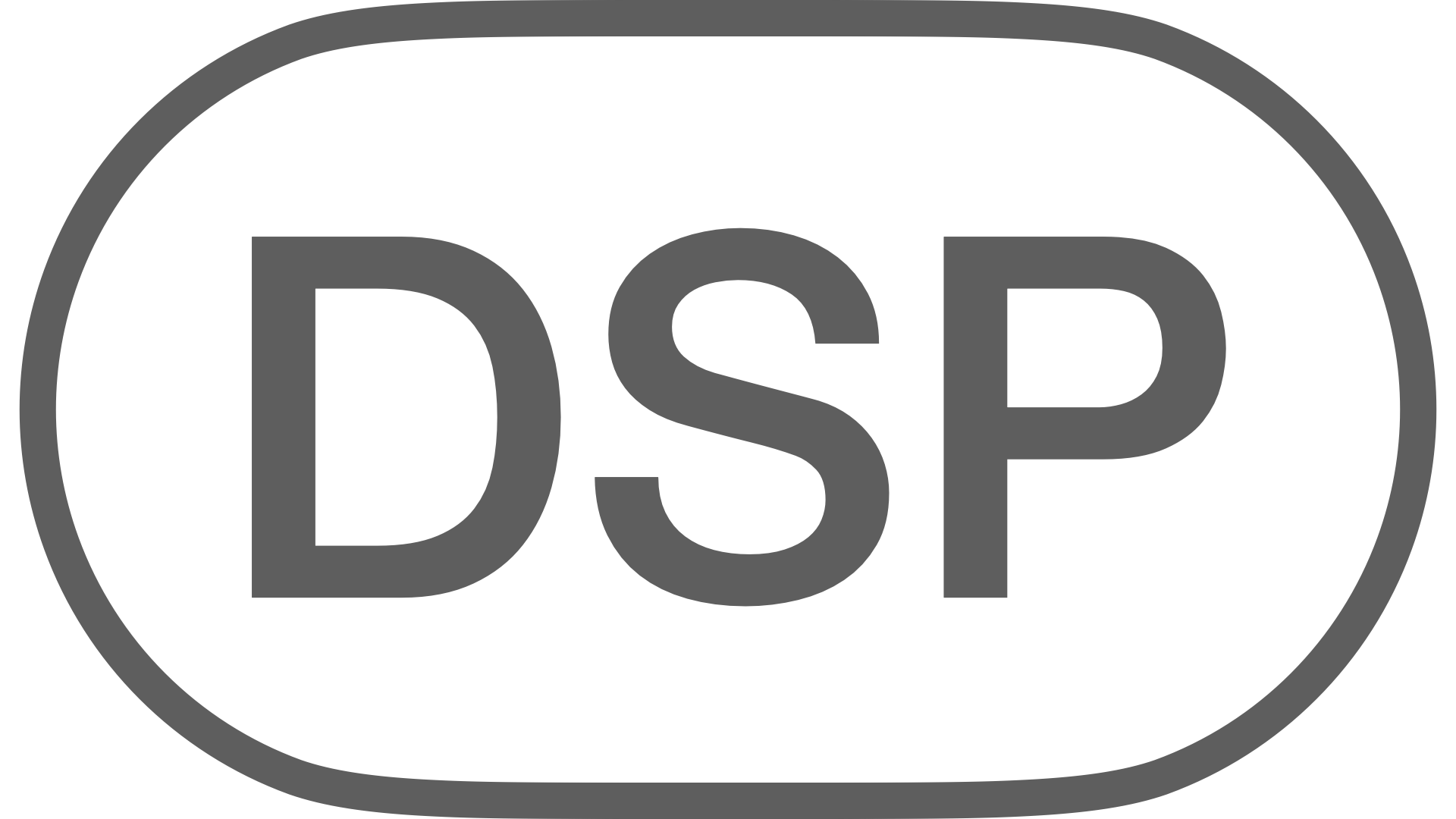}}}

\DeclareRobustCommand{\ldp}{\raisebox{-0.31em}{\includegraphics[height=1.2em, trim={0 0 11cm 0},clip]{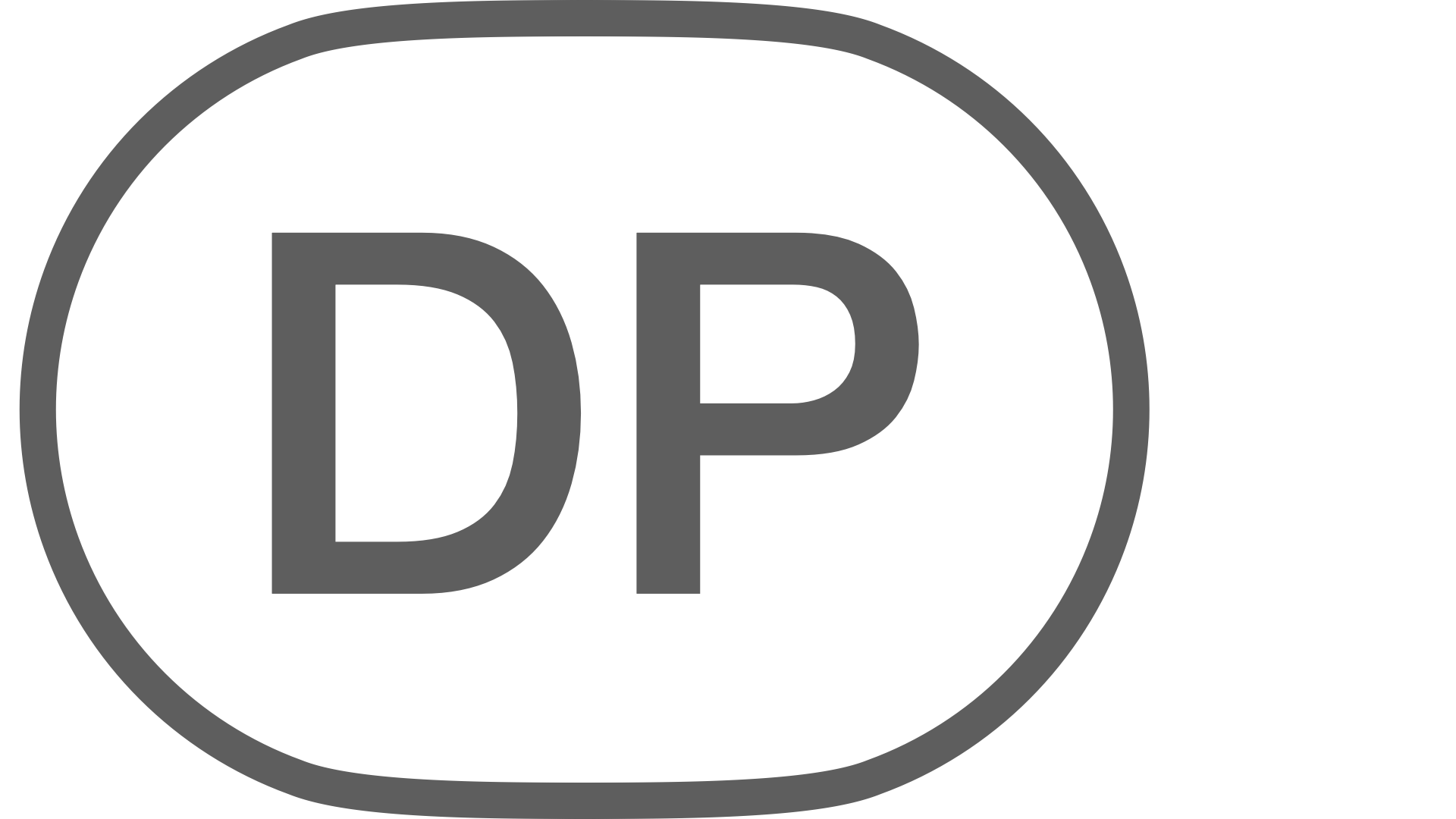}}}

\DeclareRobustCommand{\lexI}{\raisebox{-0.31em}{\includegraphics[height=1.2em]{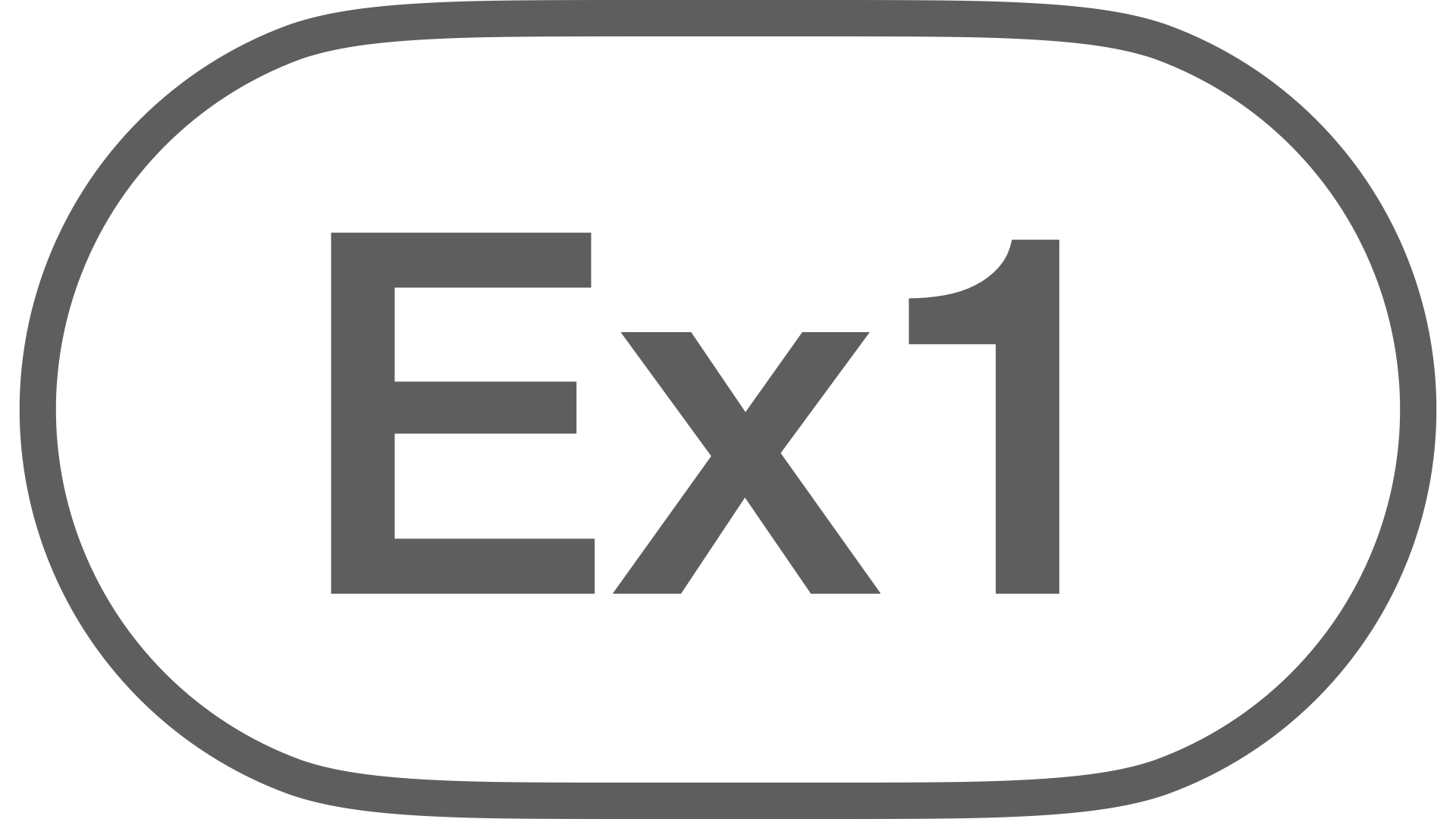}}}

\DeclareRobustCommand{\lexIII}{\raisebox{-0.31em}{\includegraphics[height=1.2em]{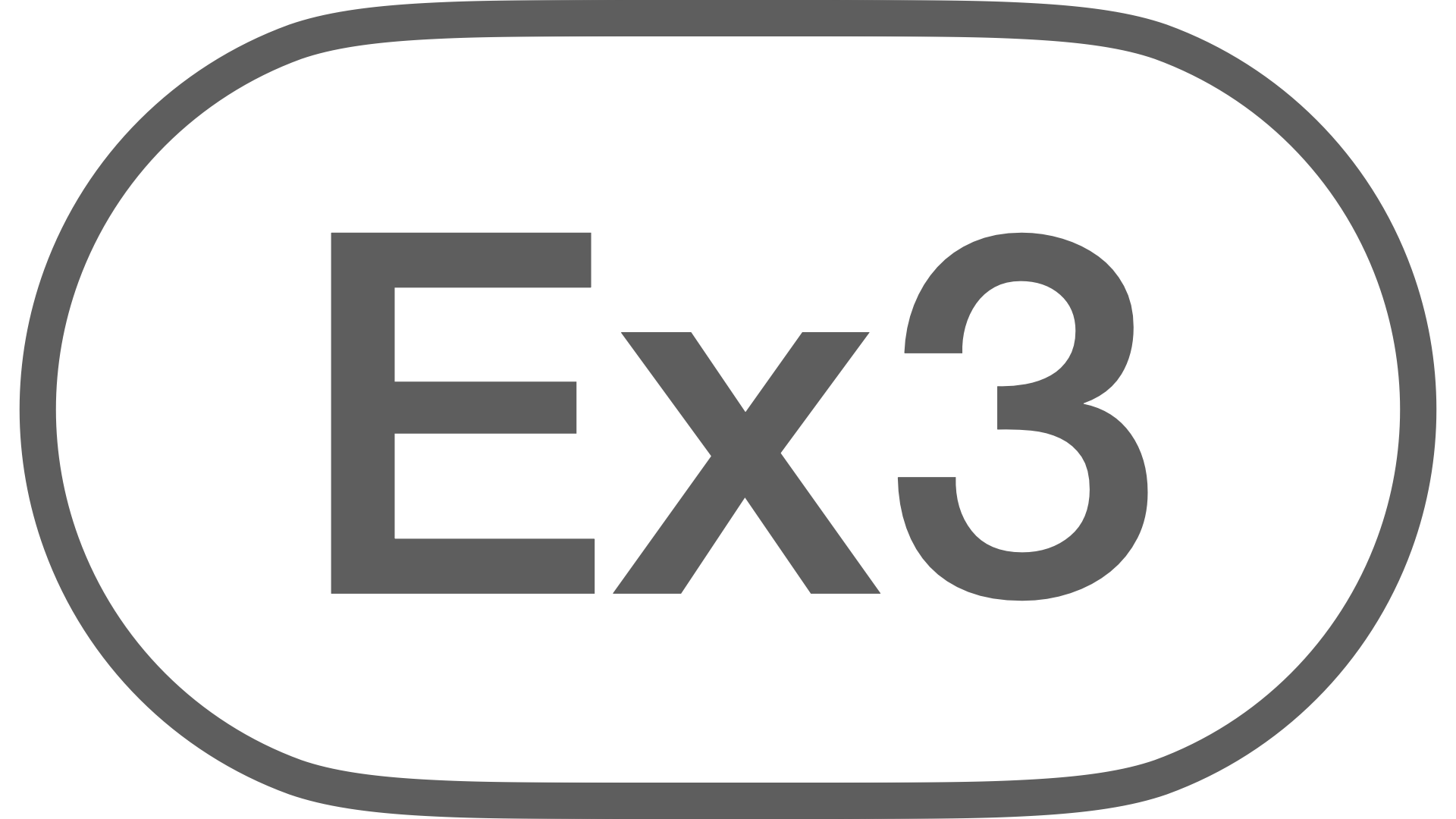}}}

\DeclareRobustCommand{\lgp}{\raisebox{-0.31em}{\includegraphics[height=1.2em, trim={0 0 11cm 0},clip]{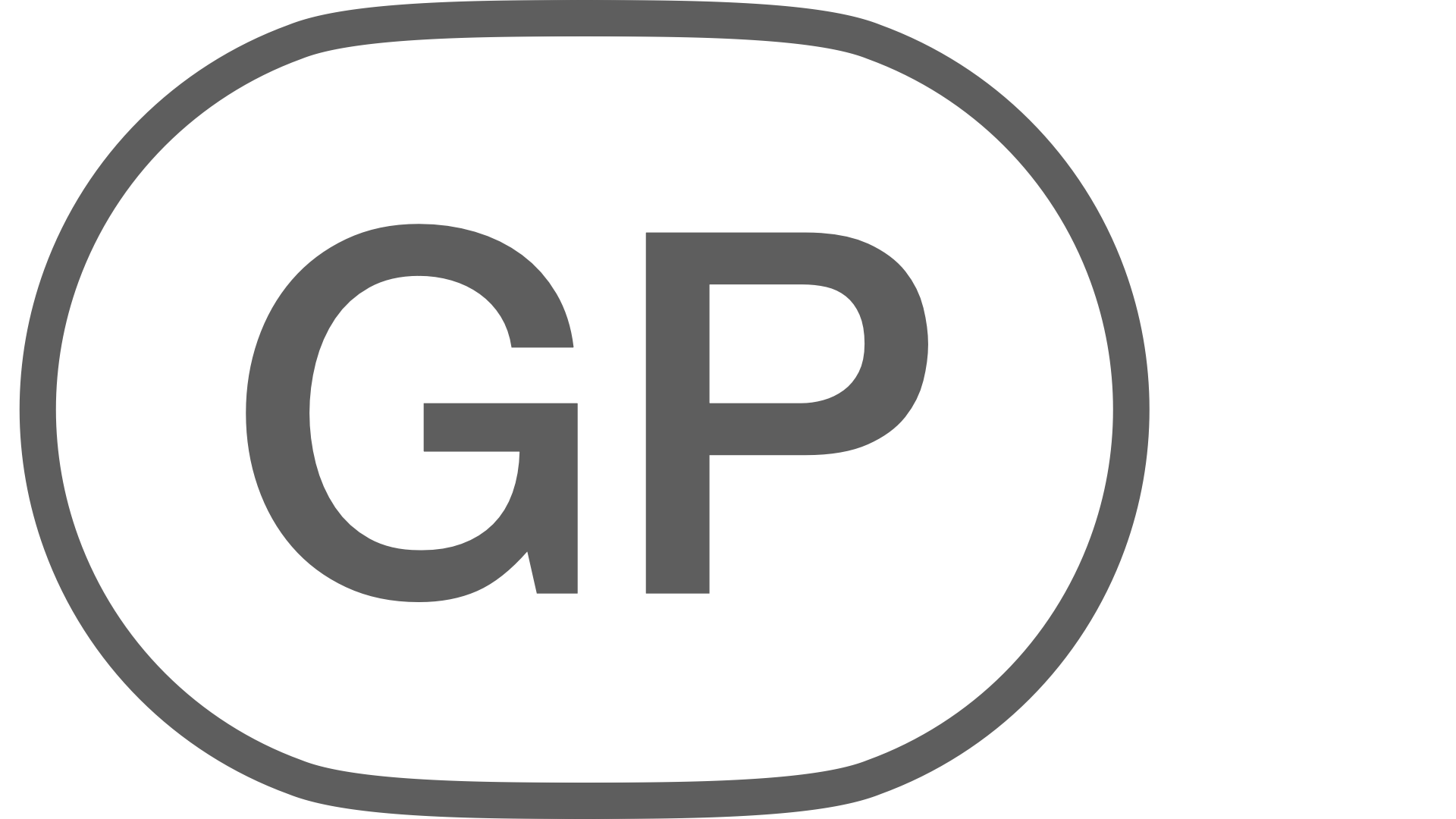}}}

\DeclareRobustCommand{\lova}{\raisebox{-0.31em}{\includegraphics[height=1.2em]{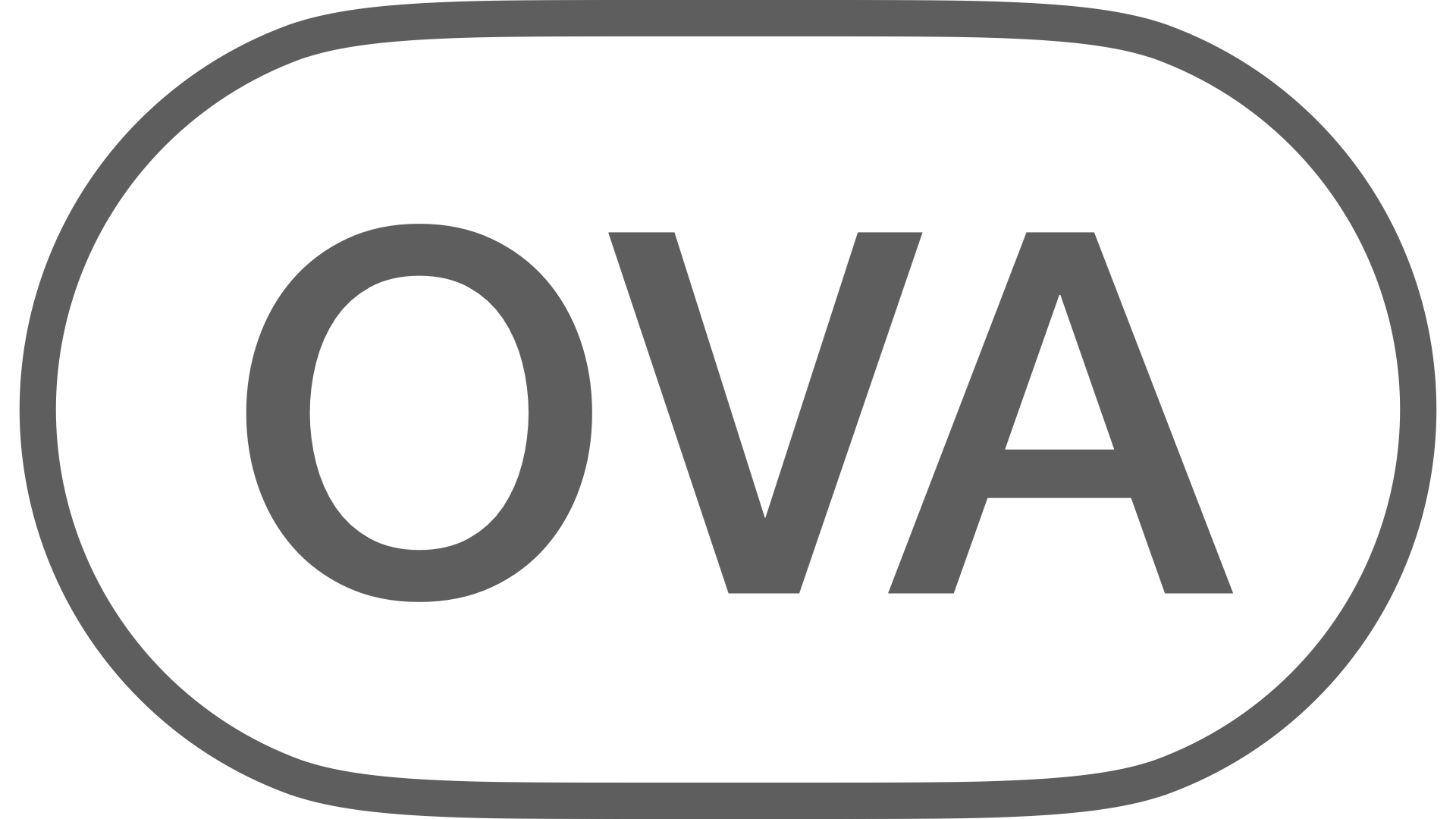}}}

\DeclareRobustCommand{\lsp}{\raisebox{-0.31em}{\includegraphics[height=1.2em, trim={0 0 11cm 0},clip]{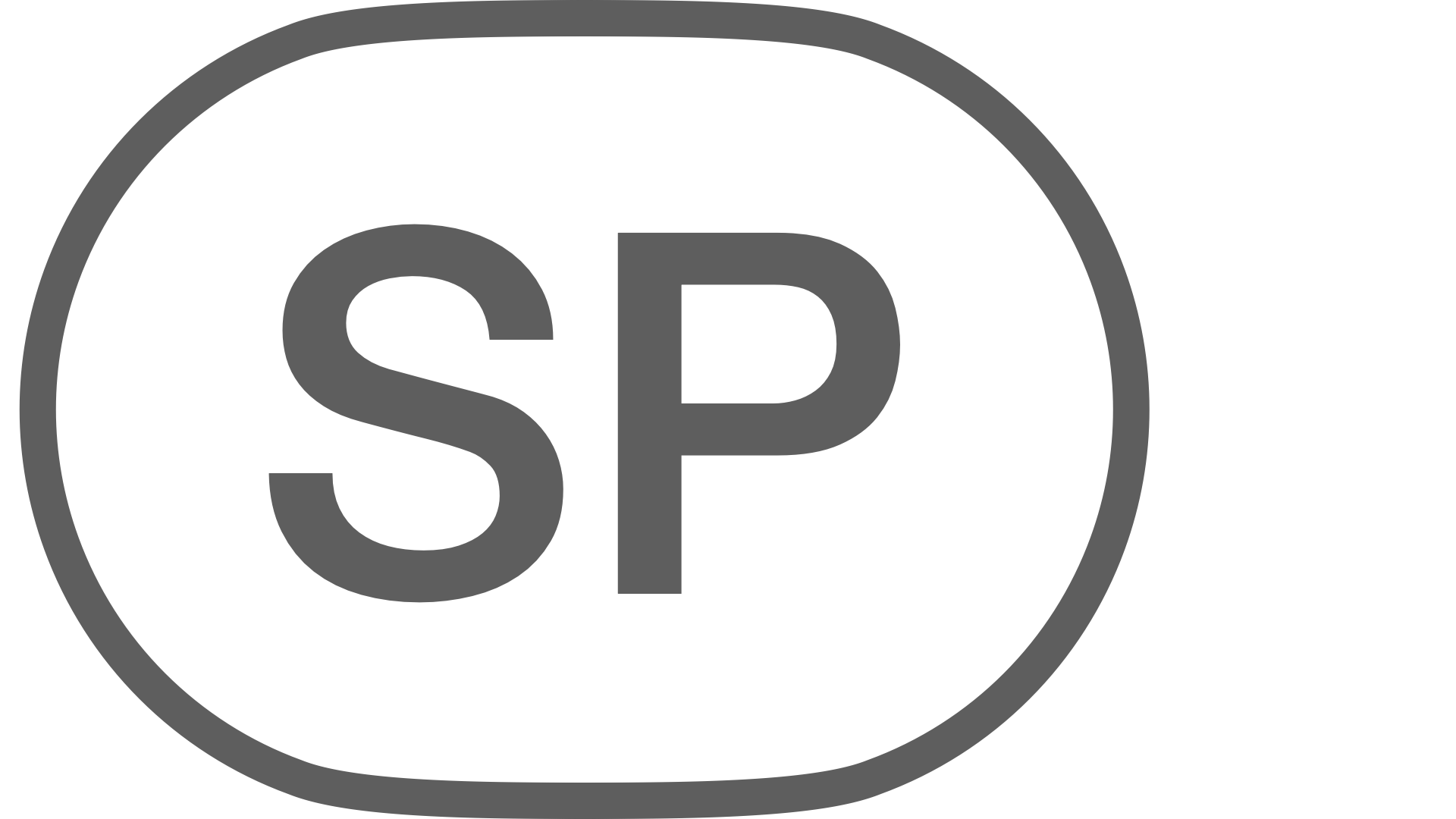}}}

\DeclareRobustCommand{\ltb}{\raisebox{-0.31em}{\includegraphics[height=1.2em, trim={0 0 11cm 0},clip]{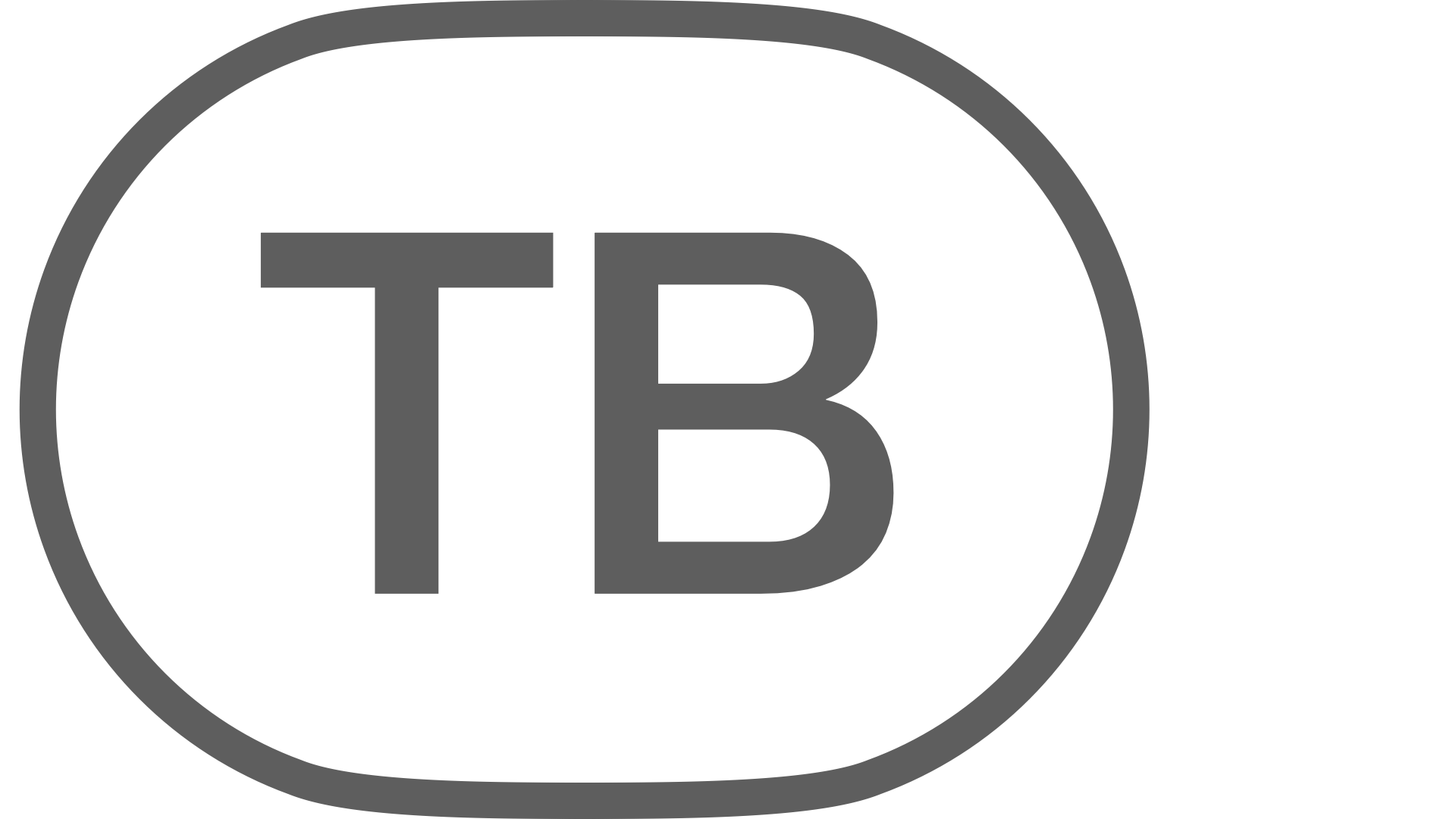}}}

\DeclareRobustCommand{\lvdb}{\raisebox{-0.31em}{\includegraphics[height=1.2em]{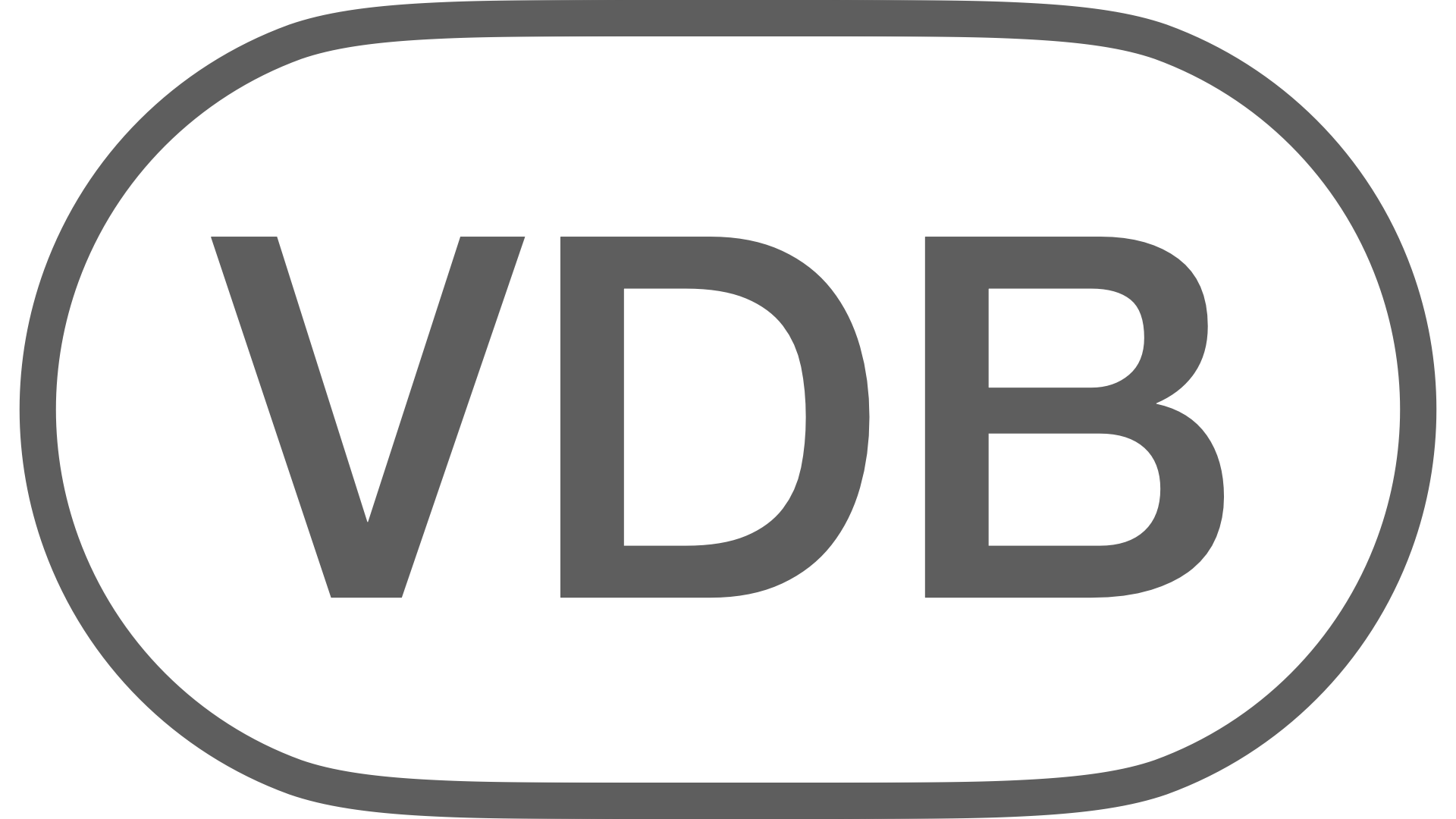}}}

\DeclareRobustCommand{\lvri}{\raisebox{-0.31em}{\includegraphics[height=1.2em]{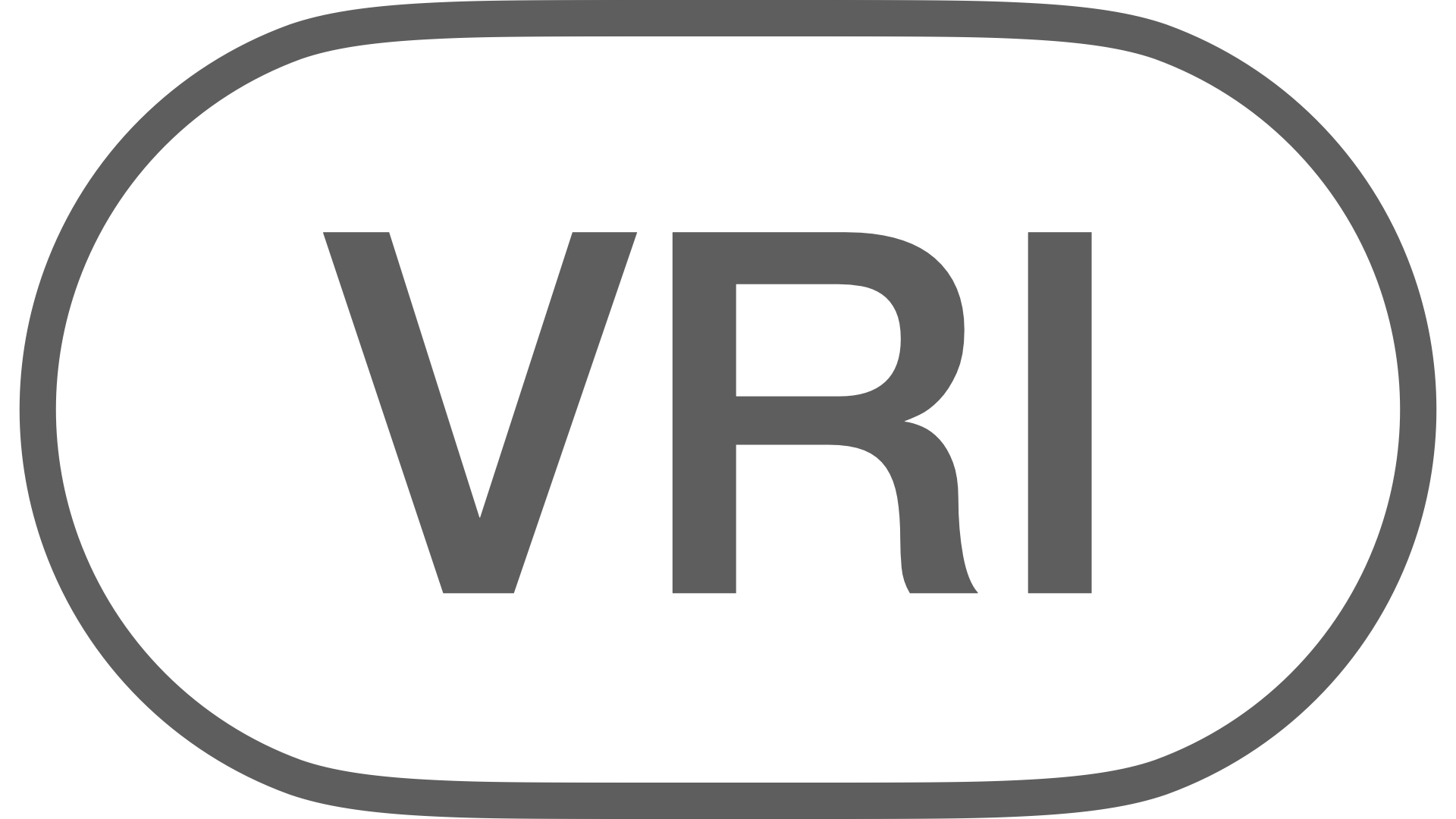}}}

\DeclareRobustCommand{\vrtstoch}{\raisebox{-0.2em}{\includegraphics[height=1em, trim={0 5.6cm 0 0},clip]{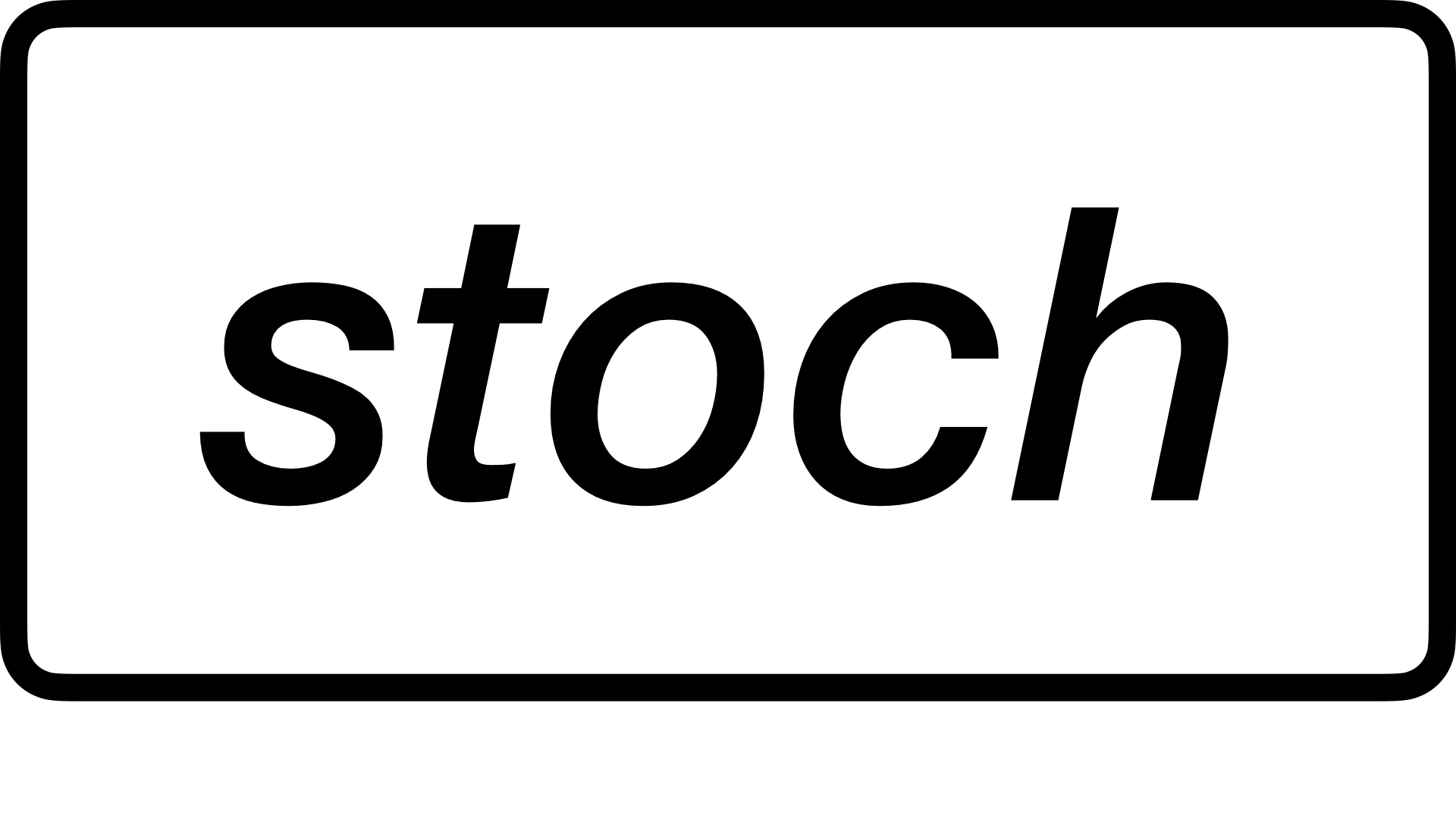}}}

\def \incp{\emph{In\textsubscript{CP}}}
\def \inep{\emph{In\textsubscript{EP}}}
\def \outdir{\emph{Out\textsubscript{dir}}}
\def \outder{\emph{Out\textsubscript{der}}}
\def \outdirder{\emph{Out\textsubscript{dir,der}}}

\definecolor{keynotegreen}{HTML}{60D937}
\definecolor{keynoteyellow}{HTML}{FEAF00}
\definecolor{keynoteblue}{HTML}{00A2FF}

\colorlet{lightkeynotegreen}{keynotegreen!10}
\colorlet{lightkeynoteyellow}{keynoteyellow!10}
\colorlet{lightkeynoteblue}{keynoteblue!10}

\definecolor{inputs}{HTML}{74AF7C}
\definecolor{directout}{HTML}{E8B454}
\definecolor{derivedout}{HTML}{E59F94}
\definecolor{co}{HTML}{3E8CF6}
\definecolor{coI}{HTML}{489FF8}
\definecolor{coII}{HTML}{4472B0}
\usepackage{circledsteps}

\definecolor{cc20_green}{HTML}{2ca02c}
\definecolor{cc20_purple}{HTML}{9467bd}
\definecolor{cc20_red}{HTML}{d62728}

\definecolor{nicered}{HTML}{FF725C}
\colorlet{smalln}{nicered}

\definecolor{niceblue}{HTML}{6cc5b0}
\colorlet{details}{niceblue}
\definecolor{niceblueII}{HTML}{4269d0}
\colorlet{details}{niceblueII}

\definecolor{nicegreen}{HTML}{3ca951}
\colorlet{designspace}{nicegreen}

\definecolor{niceyellow}{HTML}{efb118}
\colorlet{tableau}{niceyellow}

\newcommand{\colored}[2][black]{\textcolor{black}{#2}} 

\begin{document}
%
\title{RSVP for VPSA : A Meta Design Study on \\Rapid Suggestive Visualization Prototyping\\ for Visual Parameter Space Analysis}
%
%
%
%

\author{Manfred~Klaffenboeck,
        Michael~Gleicher,
        Johannes~Sorger,
        Michael~Wimmer,
        and~Torsten~Möller
\IEEEcompsocitemizethanks{\IEEEcompsocthanksitem M. Klaffenboeck and M. Wimmer are with the Institute of Visual Computing and Human Centered Technology at TU Wien, Austria; Email: \{lastname\}@cg.tuwien.ac.at.\protect\\
\IEEEcompsocthanksitem J. Sorger was with the Complexity Science Hub, Vienna, Austria.\protect\\
\IEEEcompsocthanksitem M. Gleicher is with the 
Department of Computer Sciences at the University of Wisconsin - Madison. 
Email: gleicher@cs.wisc.edu\protect\\
\IEEEcompsocthanksitem T. Möller is with the Faculty of Computer Science at the University of Vienna and the Research Network Data Science at Uni Wien. Email: torsten.moeller@univie.ac.at}
\thanks{Manuscript received April 30, 2023; revised February 1, 2024.}}

%
%

\markboth{Journal of \LaTeX\ Class Files,~Vol.~14, No.~8, April~2024}%
{Shell \MakeLowercase{\textit{et al.}}: Bare Demo of IEEEtran.cls for Computer Society Journals}
%




\IEEEtitleabstractindextext{%
  \includegraphics[width=\linewidth, trim={0 530pt 0 0},clip]{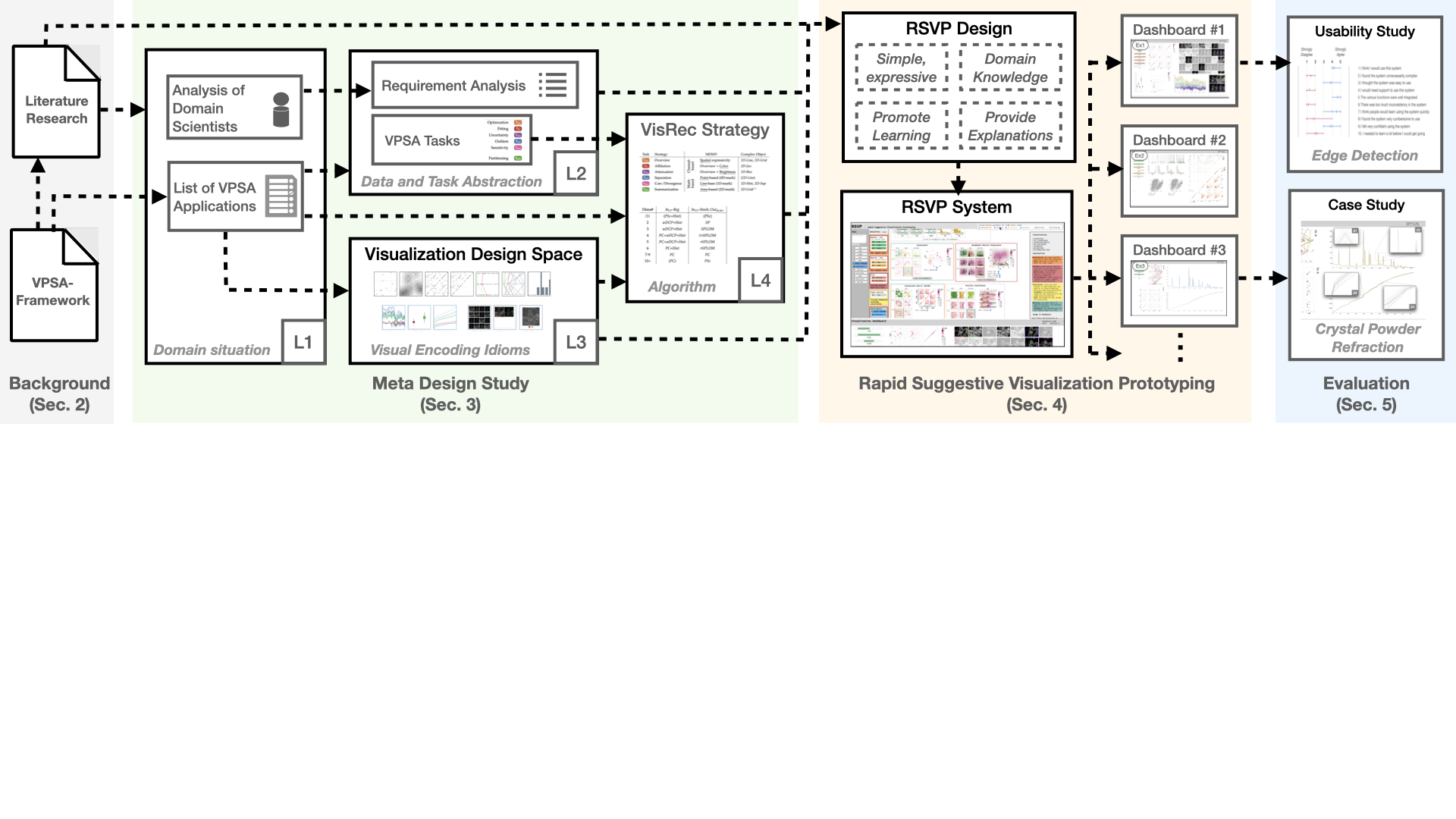}
  \captionof{figure}{Overview of our overall process to design a mixed-initiative system that provides domain scientists with the opportunity to perform VPSA without the need for a visualization expert. We conducted a \textit{Meta Design Study} by surveying existing VPSA literature and extracting relevant information.
This procedure allowed us to analyze \textit{requirements}, extract a \textit{visualization design space}, and devise a \textit{task-oriented VisRec strategy}. 
We implemented our findings in RSVP, the \textit{Rapid Suggestive Visualization Prototyping} system. 
It enables domain scientists to rapidly create and experiment with visualization dashboards tailored to their specific models and data.
We externally \textit{evaluated} RSVP's efficiency through a usability study and a real-world case study.\\}
  \addtocounter{figure}{-1}
  \label{fig:flow-teaser}
\begin{abstract}
Visual Parameter Space Analysis (VPSA) enables domain scientists to explore input-output relationships of computational models. 
Existing VPSA applications often feature multi-view visualizations designed by visualization experts for a specific scenario, making it hard for domain scientists to adapt them to their problems without professional help. 
We present RSVP, the Rapid Suggestive Visualization Prototyping system encoding VPSA knowledge to enable domain scientists to prototype custom visualization dashboards tailored to their specific needs. 
The system implements a task-oriented, multi-view visualization recommendation \deleted{(VisRec) }{}strategy over a visualization design space optimized for VPSA to guide users in meeting their analytical demands.
We derived the VPSA knowledge implemented in the system by conducting an extensive meta design study over the body of work on VPSA. 
We show how this process \changed{could}{can} be used to perform a data and task abstraction, extract a common visualization design space, and derive a task-oriented VisRec strategy.
User studies indicate that the system is user-friendly and can uncover novel insights. 
\end{abstract}

}

\maketitle

\IEEEdisplaynontitleabstractindextext

%
\IEEEpeerreviewmaketitle

\IEEEraisesectionheading{\section{Introduction}\label{sec:introduction}}
\label{sec:introduction}

The use of input-output-oriented models is ubiquitous in modern research, spanning problem domains from climate research\cite{Potter2009,Wang2017} to complex engineering tasks\cite{Matkovic2009,Cibulski2020}, from parametric design problems\cite{Bruckner2010,Matejka2018} to machine learning and deep learning\cite{Muehlbacher2017, Hamid2019}. 
To comprehensively understand a model's behavior, it is crucial to go beyond mere trial and error testing of parameter configurations. 
A more holistic approach performed by some domain experts is to systematically sample the model, relating those sampled input parameters to the respective results and analyzing them together visually.
This process is known as \emph{Visual Parameter Space Analysis (\VPSA)}~\cite{Sedlmair2014}.

VPSA provides powerful means for scientists to investigate the behavior of complex models under different conditions and reveal patterns and relationships inherent in the model that might otherwise stay hidden.
It can aid domain scientists in identifying optimal parameter settings, exploring the underlying model uncertainty in different regions, and finding model boundaries for input parameterizations based on their corresponding outputs.

Despite promising evidence of its usefulness\cite{Sedlmair2014}, VPSA has not been adopted by the broader research community as a general problem-solving strategy. 
A reason for this might be the lack of a visualization tool that allows domain experts to apply this type of visual analysis to their input-output models.
VPSA applications designed to solve high-level problems often comprise multiple visualizations capable of performing multi-dimensional analysis.
Designing such a multi-view visualization with its various tradeoffs often exceeds the visualization knowledge of domain scientists.
In some cases, domain scientists and visualization experts collaborate in design studies to develop customized solutions for specific problems.
However, this process requires time, resources, and mutual interest, which may exclude users seeking to utilize VPSA for less significant problems\cite{Sedlmair2012}.

We introduce a mixed initiative system \cite{Horvitz1999} called RSVP (\textit{Rapid Suggestive Visualization Prototyping)} which encodes VPSA knowledge and guides users in designing visualization dashboards for their data and needs. 
To gather and implement the necessary knowledge, we developed a meta-design strategy to extract tacit knowledge from VPSA papers and transform it into explicit knowledge\cite{Federico2017}.
We refer to this approach as a \textit{meta design study}.
The study yielded a comprehensive requirement analysis, a VPSA-specific visualization design space, and a visualization recommendation (VisRec) strategy for creating multi-view dashboards to solve typical VPSA tasks.
RSVP implements these findings, making an otherwise complex and time-consuming task actionable and comprehensible for non-visualization experts.
The system aims to be focused, transparent, and to promote learning to \textit{minimize the cost} of using it and \textit{build trust} in it -- two factors we identified as latent needs of domain scientists when adopting a new visualization system.

\label{sec:contributions}
%
%
\textbf{\textit{Outline \& Contributions:}} Our overall approach is illustrated in \autoref{fig:flow-teaser}.
\begin{Deleted}
After introducing important concepts and related work (\autoref{sec:020_background}), we present the results of a \textbf{Meta Design Study} which we devised and conducted over the body of knowledge on VPSA, complementing and extending previous work in the field~\cite{Sedlmair2014}. 
The artifacts from this process include a \textit{comprehensive VPSA requirement analysis}, an \textit{expressive VPSA visualization design space}, and a \textit{practical VisRec strategy} over this design space (\autoref{sec:meta-design-study}).
%
Subsequently, we introduce the \textbf{Rapid Suggestive Visualization Prototyping (RSVP)} system, which incorporates the design knowledge from the meta-design study into a user-friendly drag-and-drop interface for creating multi-view dashboards. The system additionally supports the latent needs of domain experts (\autoref{sec:rsvp-system}).
Finally, we present a \textbf{multi-faceted evaluation} of our approach, which includes a \textit{qualitative result inspection} (QRI, (\autoref{sec:qri})), a \textit{usability study}, and a \textit{real-world case study} ((\autoref{sec:evaluation})) to demonstrate the validity and efficacy of our approach.
\end{Deleted}
\begin{Changed}
We introduce important concepts and related work (\autoref{sec:020_background}), followed by our main contributions:
\begin{itemize}
    \item We devised and conducted a \textbf{\textit{Meta Design Study}} over the body of knowledge on VPSA, complementing and extending previous work in the field~\cite{Sedlmair2014}. 
The artifacts from this process include a \textbf{comprehensive VPSA requirement analysis}, an \textbf{expressive VPSA visualization design space}, and a \textbf{practical VisRec strategy} 
(\autoref{sec:meta-design-study}).
    \item The \textbf{\textit{Rapid Suggestive Visualization Prototyping (RSVP)}} system, which incorporates the design knowledge from the meta-design study into a user-friendly drag-and-drop interface for creating multi-view dashboards
    (\autoref{sec:rsvp-system}).
    \item A \textbf{\textit{multi-faceted evaluation}}, which includes a \textbf{qualitative result inspection} (QRI, (\autoref{sec:qri})), a \textbf{usability study}, and \textbf{two real-world case studies} ((\autoref{sec:evaluation})) to demonstrate the validity and efficacy of our approach.
\end{itemize}

%

%

We discuss findings, design aspects, limitations, and future research opportunities in \autoref{sec:discussion} and present our conclusion in \autoref{sec:conclusion}.
\end{Changed}

\section{Background}
\label{sec:background}
\label{sec:020_background}
First, we \deleted{will} introduce VPSA using a practical example. 
Next, we \deleted{will} define the term \textit{meta design study} and discuss \textit{domain experts} as a general user group. 
Finally, we will review related work on visualization recommendation, exploration, and construction tools.

\subsection{VPSA - Introduction}
\label{sec:vpsa-intro}

\newcont{VPSA is a general approach to perform input-output-oriented, multi-dimenional analysis. }
The overall VPSA workflow is presented in \autoref{fig:vpsa-intro}.
\changed{We introduce essential \VPSA{} concepts and terminology using a well-known edge detection algorithm as a practical \newcont{running} example.}{We use a well-known edge detection algorithm\cite{Canny1986} as a \textbf{running example} to introduce important VPSA concepts and terminology.}
\newcont{We opted to use a generic image to avoid potential domain complexities. However, image segmentation, similar to the one presented, is often necessary in biological\cite{Pretorius2011} and medical applications\cite{Torsney-Weir2011}, as well as in material sciences\cite{Froehler2016}.}

\addtocounter{figure}{1}
\begin{figure*}[ht]
    \centering
     \begin{subfigure}[m]{0.36\textwidth}
         \centering
         \includegraphics[width=\textwidth]{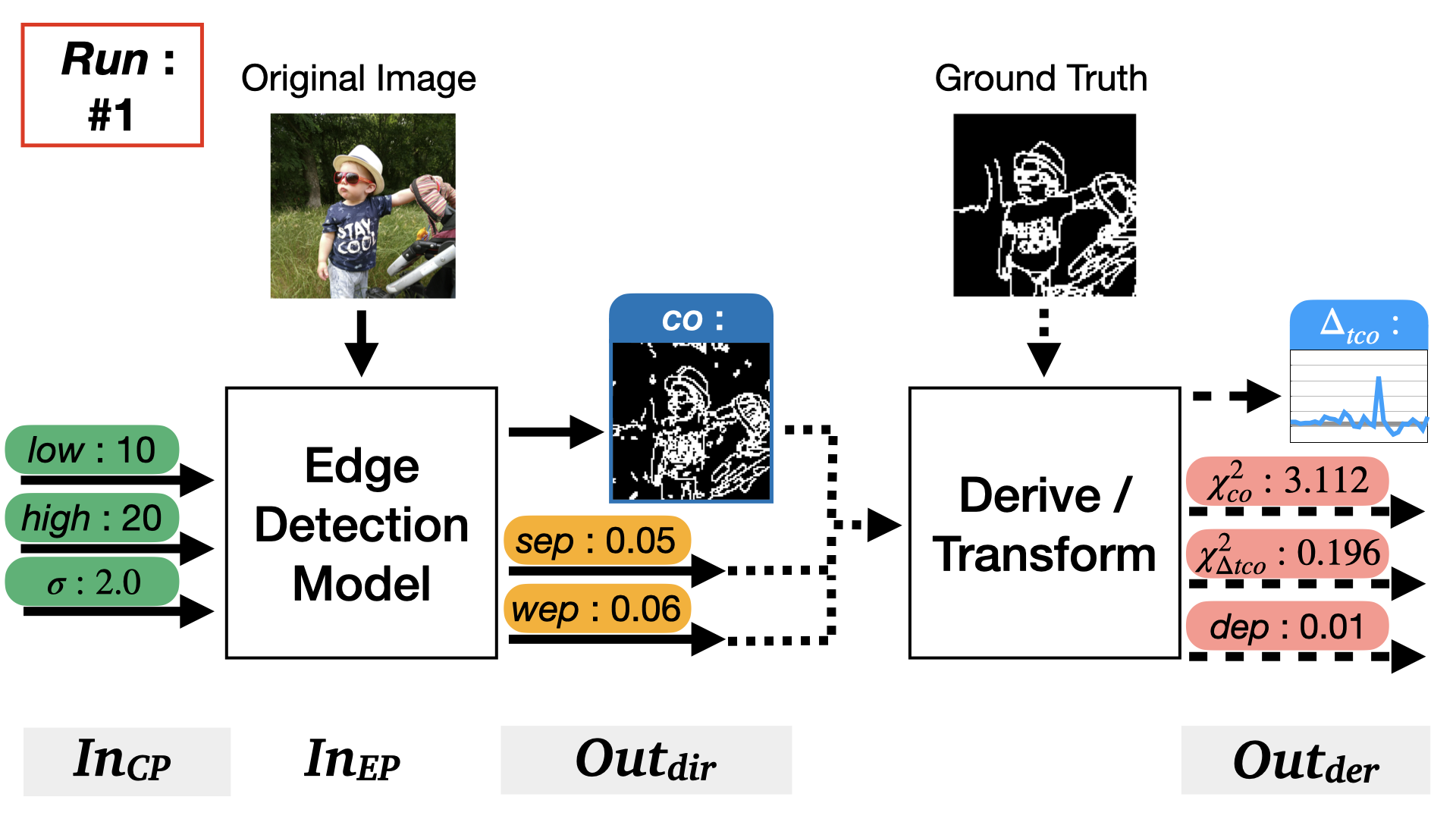}
         \caption{Practical Dataflow Example}
         \label{fig:vpsa-dataflow}
     \end{subfigure}
     \begin{subfigure}[m]{0.04\textwidth}
         \centering
         \includegraphics[width=\textwidth]{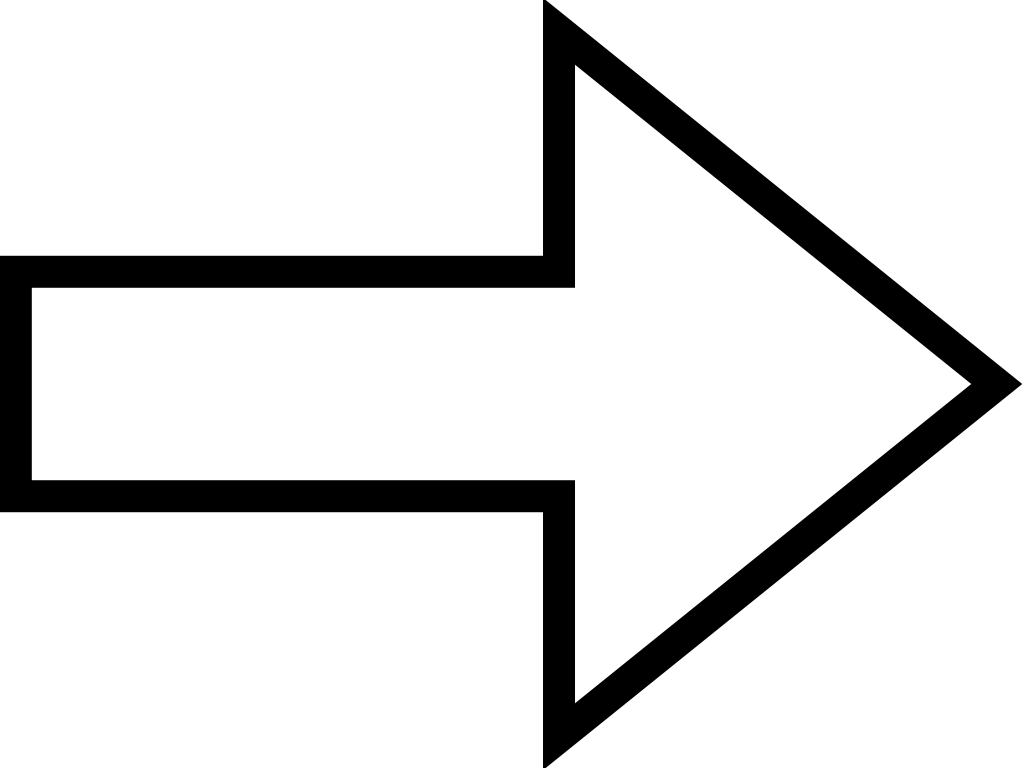}
         \caption*{\tiny{sample}}
     \end{subfigure}
     \begin{subfigure}[m]{0.27\textwidth}
         \centering
         \includegraphics[width=\textwidth]{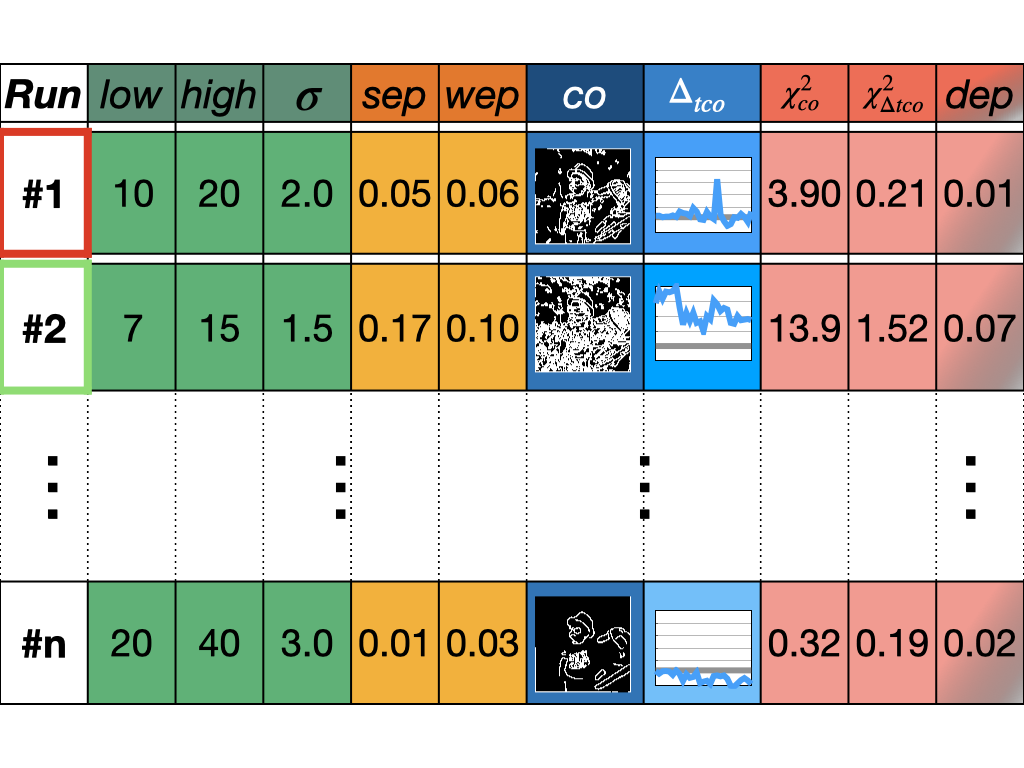}
         \caption{Sampled data}
         \label{fig:sampled-data}
     \end{subfigure}
     \begin{subfigure}[m]{0.04\textwidth}
         \centering
         \includegraphics[width=\textwidth]{figures/VPSA-intro/VPSA-Intro-Arrows.002.png}
         \caption*{\tiny{analyze}}
     \end{subfigure}
     \begin{subfigure}[m]{0.27\textwidth}
         \centering
         \includegraphics[width=\textwidth]{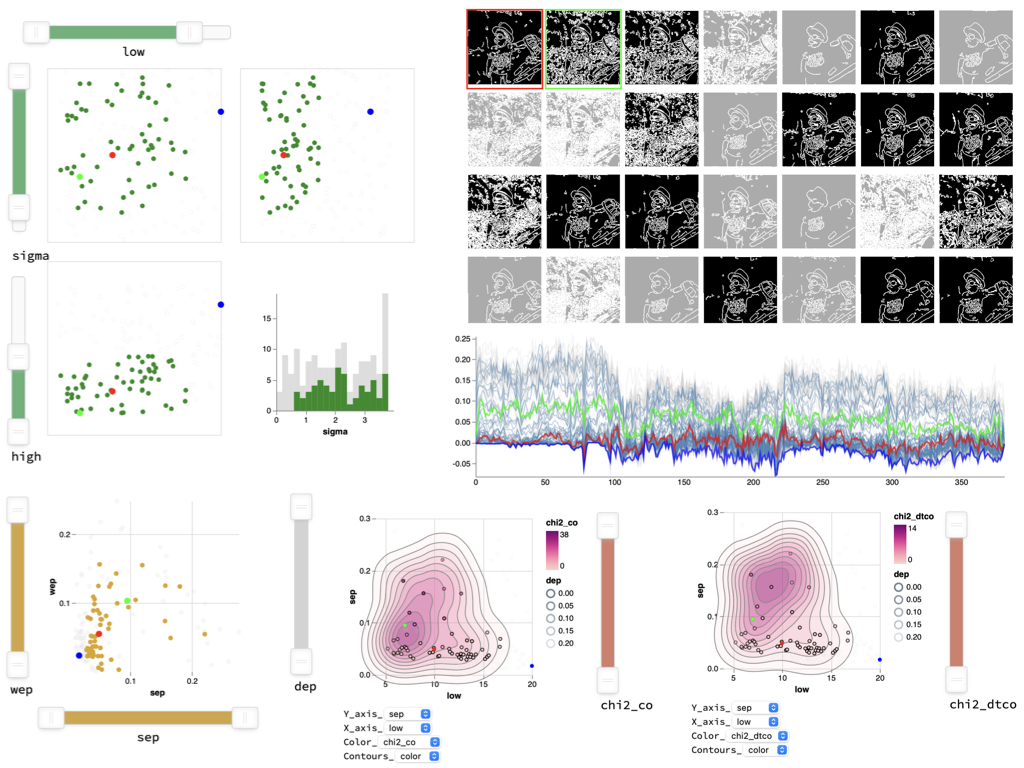}
         \caption{Interactive VPSA Session}
         \label{fig:vpsa-session}
     \end{subfigure}
    \caption{
    Overview of the VPSA workflow.
     (a) depicts the dataflow model for VPSA, using an edge detection algorithm as a practical example. 
The model gets run multiple times using some kind of sampling strategy. 
The 
\Circled[fill color=inputs, outer color=white]{ inputs }, 
\Circled[fill color=directout, outer color=white]{ direct outputs }, 
\Circled[fill color=derivedout, outer color=white]{ derived outputs },
and resulting 
\Circled[fill color=co, outer color=white]{ complex objects } 
(shades of blue)
from this process are gathered and stored in a data structure, such as a data table 
(b),
which serves as the data source for the visual analysis process (c).
\newcont{The colors used for the visualizations in (c) are designed to match the colors in (a) and (b) in order to show the main data variable they are trying to encode.}    }
       \label{fig:vpsa-intro}
\end{figure*}

\changed{\emph{Running Example:}}{\textbf{\textit{Running Example:}}}
\label{sec:running-example}
The edge detection model \changed{featured in \autoref{fig:vpsa-dataflow}}{(see \autoref{fig:vpsa-dataflow})} takes the unsegmented image and three numerical parameters (\textit{low}, \textit{high}, $\sigma$) for internal thresholds as inputs.
It outputs a binary contour outline (\textit{co}) and two statistical parameters (\textit{sep} and \textit{wep}).
For analytical purposes, the contour outline gets transformed into a 1D projection and compared to the same projection of a hand-drawn \textit{"Ground Truth"} image, resulting in the difference between these two transformed contour outlines ($\Delta_{tco}$).
The \gof{} between the contour outline and the ground truth ($\chi^2_{co}$) and between their respective 1D projections ($\chi^{2}_{\Delta_{tco}}$) is measured using Pearson’s Chi-square test.
\begin{Deleted}
   \st{It calculates the sum of differences between (O)bserved and (E)xpected outcomes:}

\begin{equation}
\label{eq:chi2}
\chi^2=\sum_{i=1}^{N}\frac{(O_i-E_i)^2}{E_i}
\end{equation}

\st{A low $\chi^2$ value means a high correlation between the compared objects.}
\end{Deleted}

\changed{\emph{Concepts and Terminology:}}{\textbf{\textit{Concepts and Terminology:}}}
\autoref{fig:vpsa-dataflow} is organized \changed{as}{in the form of} the \textbf{data flow model} for VPSA \cite{Sedlmair2014}.
The three input parameters are control parameters (\incp{}), which the user can directly manipulate. The "Original Image" is an environmental parameter (\inep{}), often outside the user's direct control, meaning if a specific image needs to be segmented, it cannot be freely exchanged for another.
The segmented image (\emph{co}) and the statistical parameters (\emph{sep} and \emph{wep}) are direct outputs (\outdir{}). These may need further processing for analytical efficiency, resulting in derived outputs (\outder{}). This example illustrates how VPSA handles various data types, classified broadly into \textit{multi-dimensional/multi-variate (MDMV)} and \textit{complex objects}. In our example, all numerical parameters are MDMV, while the segmented image (\emph{co}) and its transformed 1D-representation ($\Delta_{tco}$) are examples of complex objects
A complex object is a semantic unit that can not be described with a single quantitative/ordinal/categorical variable without losing information.
%
Control parameters affect the segmentation process and require adjustment for the desired model output. A non-VPSA approach achieves this through trial-and-error, where the user adjusts parameters after inspecting unsatisfactory results. VPSA replaces this tedious (manual sampling) process by using (systematic) \textbf{sampling} to vary control parameters.
\textit{Regular} or \textit{stochastic} sampling is often used for this purpose, and the corresponding inputs and outputs are stored in a single data table (\autoref{fig:sampled-data}).
The columns of the data table reprise the \emph{dimensions} to be analyzed, and each row represents the inputs and outputs of a single \emph{run}. 
Data organized in such a way can then be analyzed visually (like in the example dashboard in \autoref{fig:vpsa-session}).
Users adopting VPSA typically aim to perform a certain set of \textit{analysis tasks}, such as finding optimal parameterizations, identifying potential outliers, or partitioning the data.
 \autoref{tab:vpsa-tasks} provides an overview of typical VPSA tasks, as outlined by Sedlmair et al.\cite{Sedlmair2014}.
Please note that the colors used to classify the individual tasks are different from the colors used for inputs and outputs in \autoref{fig:vpsa-intro}.
To perform these tasks, VPSA users need to navigate a multi-dimensional problem space. 
The most common \textbf{navigation strategy} is the \textit{global-to-local approach}, where users start with an overview of the data and then explore details. 
Less commonly used strategies include the \textit{local-to-global} approach, where users start with a specific sample point and explore alternatives, and the \textit{steering} approach, where users adjust parameters during simulation runtime.

 \renewcommand{\arraystretch}{1.5}
\begin{table}[!b]
    \caption{List of typical VPSA-Tasks according to the conceptual framework by Sedlmair at al. \cite{Sedlmair2014}}
    \centering
    \begin{tabular}{r l p{0.6\columnwidth}}
         \multicolumn{2}{c}{ \textbf{Task}} & \textbf{Description} \\
         \hline
         Optimzation& \topt{} &Find the best parameter setting \\
         Fitting & \tfit{} & Find where actual model data occurs \\
         Uncertainty & \tunc{} & Determine the reliability of the output\\
         Outliers & \tout{} & Find odd or special outputs\\
         Sensitivity &\tsens{} & Identify input regions with high or low impact on the output\\
         Partitioning &\tpart{} & Identify different types of model behavior
    \end{tabular}
    \label{tab:vpsa-tasks}
\end{table}
\renewcommand{\arraystretch}{1.0}

\subsection{Meta Design Study}
\label{sec:meta-design-study-definition}


\begin{Deleted}
   \st{According to Munzner, \textit{design studies} are a problem-driven form of research where visualization experts analyze \emph{"the problems of some real-world user[s] and attempt to design a solution that helps them more effectively"}}\cite{Munzner2014}.
\st{Classical design studies are user-centric, focusing on connecting with real-world users at an early stage}\cite{Sedlmair2012}. 

\st{In the more recently proposed \emph{data-driven design study}}\cite{Oppermann2020}, 
\st{the initial focus is on obtaining and abstracting data.
The collaboration with domain scientists begins later but is still necessary for the development process. 
Instead of collaborating with domain scientists directly, our approach synthesizes information from written reports of other design studies to develop a general solution for various problem domains and a wider group of users. 
Since this strategy shares many similarities with \textit{meta analysis}}\cite{LipseyMW2001}\st{, we propose to call it \textbf{meta design study}. }
\end{Deleted}
\begin{Changed}
Traditional design studies \cite{Sedlmair2012} and data-driven design studies~\cite{Oppermann2020} require the early collaboration with domain scientists to gather domain knowledge.
In contrast, we apply a design approach that synthesizes this kind of information from written reports of existing design studies.
Since this strategy shares many similarities with \textit{meta analysis}\cite{LipseyMW2001}, we call this a \textbf{meta design study}. 
A meta design study is suitable when the goal is to develop a more general tool that serves the needs of a wider group of users within a certain problem domain.
\end{Changed}
A meta design study offers a principled approach to performing requirements analysis and extracting a visualization design space from analyzed applications and reports. 
Additionally, it provides a mechanism for deriving a data and task-oriented visualization recommendation strategy over the extracted design space.
Practical details will be further explained in \autoref{sec:meta-design-study}.

\changed{\st{We want to highlight that, similar to other design studies, our approach follows}}{Meta design studies follow} Munzner's nested model for visualization design\cite{Munzner2009} in a top-down manner\cite{Munzner2014}:
Gathering and analyzing the papers is learning about the \emph{domain situation} \lbox{L1}.
The requirement extraction process includes \textit{data and task abstraction} \lbox{L2}.
Extracting the visualization design space from the literature is \deleted{practically} acquiring knowledge about \textit{visual encoding idioms} \lbox{L3},
and deriving a visualization recommendation strategy over the visualization design space operates on an \textit{algorithmic level} \lbox{L4}.



\label{sec:domain-experts}
\changed{\st{A downside of a meta design study is that it may not capture the latent needs of \textbf{domain scientists}. Thus, a separate analysis of this target group is required.}}{Since there is no immediate interaction with \textbf{domain scientists}, we extract general knowledge about this user group from existing literature in a similar fashion.} 
Several sources \changed{\st{mention}}{state} that domain experts often have little time and prefer visualizations capable of providing immediate data insights\cite{Sedlmair2011, Russell2016}.
\changed{\st{On the other hand, somewhat contrary to the previous statement, domain scientists want to explore their various options}\cite{Boukhelifa2019, Ma'ayan2020}.}{Somewhat contrary to this statement, domain scientists also seem to want to explore their various options \cite{Boukhelifa2019, Ma'ayan2020}.}
It is beneficial for domain scientists to use familiar data when learning about unknown visualizations\cite{ Slingsby2012, YeY2020}.
Unlike casual users, domain scientists analyze visualizations in a more structured way\cite{MayrEva2016}, and they are skilled in transferring their domain knowledge from familiar to unfamiliar visualizations\cite{Gegenfurtner2013}.
Furthermore, showing transitions from simple to more complex visualizations fosters an understanding of the latter\cite{Ruchikachorn2015}. 
However, domain scientists are a heterogeneous user group with different backgrounds, interests, and levels of visualization literacy, which is why they should be treated similarly to visualization novices\cite{Wong2019, Grammel2012}.
From a high-level perspective, these guidelines try to \textbf{\emph{minimize the cost}}\cite{van-Wijk2005} of using a visualization tool.
%

Furthermore, \deleted{in our experience,} domain experts do not like being told what the (supposedly) best solution is without being able to explore alternative options \cite{Ma'ayan2020, Cibulski2020}. 
\deleted{Other authors conducting experiments or design studies involving domain experts report similar insights and experiences.}
We interpret this behavior as a \changed{\st{necessary exercise}}{necessity} for those experts to build a high enough \textbf{\emph{level of trust}} \cite{Dasgupta2017} in a proposed solution. 
We deem this a crucial factor when designing a general-purpose tool intended to be used by domain scientists that utilizes a recommendation strategy.
\subsection{Related Work}
\label{sec:related-work}

In this section, we present related work in terms of visualization recommendation, exploration, and construction.\\\\
\label{sec:vis-rec}
\textbf{Visualization Recommendation (VisRec)} is a versatile and extensively researched field \cite{Zhu2020}. Here, we provide background information on various aspects of VisRec that are important for our further discussion.

\changed{\emph{Data-oriented \visrec{}:}}{\textbf{Data-oriented \visrec{}:}} APT\cite{Mackinlay1986} was amongst the first tools to provide visualization recommendations. 
It recommended visualizations based on their \textit{expressiveness} (how well does it show the data and only the data) and their (perceptual) \textit{effectiveness}. 
Although the underlying rules may have been reviewed and refined since then, the basic principles of expressiveness and effectiveness are still vital in modern data-oriented visualization recommendation approaches (e.g. \cite{Wongsuphasawat2016a, Wongsuphasawat2017, Moritz2019}). 
%
However, basing recommendations mainly on expressiveness and perceptual effectiveness has its limits and does not necessarily help in solving specific tasks\cite{Lin2020}.
Our \visrec{} strategy implicitly covers expressiveness and effectiveness but also takes the task-based effectiveness of visualizations into account.


\changed{\emph{Task-oriented VisRec:}}{\textbf{Task-oriented VisRec:} }
Early work in task-oriented \visrec{} \cite{Wehrend1990, Casner1991, Roth1994, Zhou1998} focused on creating composited graphics for \textit{presentational} purposes and did not necessarily help overcome analytical gaps\cite{Amar2004}. 
Our work focuses on visualizations and strategies for \textit{exploratory} and \textit{confirmatory} data analysis. 
%
Modern insight-oriented\cite{Demiralp2017, LeeD2021} or task-oriented~\cite{ShenL2021} VisRec systems support a wide range of analytical tasks\cite{Amar2004}. 
Interestingly, their general VisRec approach is strikingly similar.
Each task is either associated with a fixed visualization or with visual features that are supposed to support this task.
However, the process of finding visualizations / visual features best supporting the respective tasks in the mentioned papers is not explicitly reported.
Draco\cite{Moritz2019} can learn weights to incorporate task-oriented efficiency in the recommendations.
Unfortunately, this requires an extensive corpus of labeled training data in the form of ranked examples, even for learning only a minimal number of very general tasks.
Furthermore, Draco only supports MDMV visualizations. 
Our VisRec approach bases its task-oriented recommendations on visual features which we extracted from the meta design study.
Additionally, our approach can not only handle MDMV visualizations but also complex object visualizations. 

\changed{\emph{Multi-view \visrec{}:}}{\textbf{\emph{Multi-view \visrec{}:}}} Solving high-level analysis tasks often requires the combination of multiple visualizations into a coherent whole.
VizDeck\cite{Key2012} allows for combining recommended visualizations from different sources into a dashboard, but it does not provide any guidance on how to combine them effectively.
DeepEye\cite{Luo2018} and MultiVision\cite{WuA2021} recommend combinations for multi-view visualizations, but their focus is on general-purpose dashboards and less on solving specific analysis tasks.

\emph{\textbf{Transparency and Explainability in \visrec{}:}} The importance of explanations in recommender systems is well established\cite{Tintarev2007} -- a fact often ignored in VisRec systems\cite{McNutt2020}.
Some systems (e.g., \cite{Schulz2013, Luo2018}) report the internal ranking scores to the users, but those most likely have no meaning to them. 
KG4Vis\cite{LiH2021} 
manages to explain why a visualization was recommended but is currently limited to visualizations encoding a maximum of two attributes. 
Our literature-based observations for task-based efficiency follow generally accepted visualization design rules and can be easily communicated to the users.


\textbf{\emph{\visrec{} Knowledge Encoding:}} 
Various methods exist for encoding VisRec knowledge, including machine\cite{HuK2019} or deep learning\cite{Luo2018} for rule acquisition, knowledge graphs\cite{HuK2019}, constraint-based rule definition with answer set programming \cite{Moritz2019}, and ontologies \cite{GilsonO2008, Polowinski2013, Lohfink2021}. 
Classical approaches \cite{Mackinlay1986, Casner1991} employ hard-coded rules, an approach also taken by CompassQL\cite{Wongsuphasawat2016a}, the recommendation strategy behind Voyager\cite{Wongsuphasawat2016, Wongsuphasawat2017}. 
For simplicity, \visrec{} rules encoded in RSVP are currently also hard-coded.\\\\
\textbf{Visualization Exploration: }
\label{sec:vis-exp}
\emph{Spreadsheet-like interfaces} for visualization exploration \cite{Chi1997} have been used to perform parameter space exploration and present the results to the user\cite{Jankun-Kelly2001}.
Voyager\cite{Wongsuphasawat2016,Wongsuphasawat2017} leverages visualization recommendations to present the user with a single optimal solution for a given data state.
One can argue, though, that it is important to present the same data through different perspectives, so the user can understand the data better, avoid possible misconceptions\cite{Roberts1998}, and foster the learning of unknown or less familiar visualizations\cite{van-den-Elzen2013}.
Van den Elzen\cite{van-den-Elzen2013} proposed an interface combining small multiples and large singles to show variations of the current state space via various visual mappings.
We build upon this idea to combine parameter space exploration with an unobtrusive visualization recommendation strategy for coordinated multiple views.\\\\
\textbf{Multi-view Visualization Construction:} 
\label{sec:vis-builder}
Improvise\cite{Weaver2004}, ComVis~\cite{Matkovic2008a}, and Visplore\cite{Piringer2009} are frameworks that allow creating applications suited for VPSA, but all of them require at least some level of programming, and they do not immediately encode visualization knowledge for domain experts.
Keshif~\cite{Yalcin2018} is a tool aimed at visualization novices to perform data exploration via multi-view applications. 
It favors aggregated views to explore big data tables.
The analysis of input-output models, however, often requires a focus on visualizations where the individual items are directly encoded and not abstracted away\cite{Cibulski2020}.
Tableau\cite{Tableau, Stolte2002} 
provides various visualizations suitable for VPSA but has limited support for complex objects and lacks sufficient guidance for choosing visualizations capable of solving high-level VPSA tasks.


\section{VPSA Meta Design Study}
\label{sec:meta-design-study}

This section outlines the procedure and outputs of our \textit{meta design study}. 

\subsection{Method ( \lbox{L1} )}
\label{sec:method}
\label{sec:methodology}
\label{sec:literature-survey}
We began by creating a \textit{coding table}, a common strategy often used in social and business research to quantitatively analyze large text corpora\cite{Bryman2016}.
%
Initially, we coded the twenty-one papers Sedlmair et al. used to derive the conceptual VPSA framework\cite{Sedlmair2014}.
After that, we extended this literature corpus of practical VPSA applications using a four-step strategy. 
First, we searched research databases using the query ``visual parameter space analysis.'' 
Second, we looked for papers that directly referenced the work by Sedlmair et al.\cite{Sedlmair2014}. 
Third, we examined the references of all papers found in the first two steps. 
Lastly, we searched for papers that referenced works in our current pool but did not cite the work by Sedlmair et al. 
In this manner, we identified twenty-four supplementary papers that fulfilled our selection criteria, resulting in a cumulative total of forty-five papers.

%

\colored[designspace]{
Our analysis revealed the need to categorize the applications described in those papers according to the following aspects:
the \textit{tasks} an application aimed to address, the \textit{sampling strategy} employed, the visualizations incorporated in an application, the number of \textit{dimensions} encoded by each visualization, the \textit{encoding channels} utilized in a specific visualization, and whether a visualization represented \textit{inputs}, \textit{outputs}, or \textit{both}.
At least two authors coded each paper, and the findings were discussed with the entire group.}
%
The supplemental material contains the list of forty-five papers and the \changed{filled-out}{populated} coding table, as well as a discussion of how our approach aligns with Furniss's\cite{Furniss2011} five stages of Grounded Theory.

\begin{Newcont}
  While the applications discussed in those papers may be domain-agnostic in several cases, the main application examples presented can be classified into the following categories: Engineering \& Material Sciences (16/45), Arts \& Design (10/45), and Biology \& Medical Domain (10/45). The remaining examples fall into the categories of Operations Research (2/45), Climate \& Geo-Science (3/45), and Machine Learning (2/45). Two analyzed applications dealt with generic datasets and could not be classified accordingly.  
\end{Newcont}

\subsection{Requirement Analysis ( \lbox{L2} )}
\label{sec:requirement-analysis}
\label{sec:requirements}
\changed{Based on our experience and the reviewed literature regarding domain experts (see \autoref{sec:domain-experts}), we identified two pivotal factors for the successful adoption of the proposed system by these professionals: \emph{Cost} and \emph{Trust}.}{
In \autoref{sec:domain-experts}), we identified \emph{Cost} and \emph{Trust} as two pivotal factors for the successful adoption of the proposed system by domain experts.}
\changed{These two aspects}{They} form the basis for \changed{the system's}{our} \textbf{Key Goals}, namely that the system must impose \textbf{\emph{Low Cost (\keyI{})}} on the user regarding time and cognitive effort\changed{.
The}{, and the} user must gain \textbf{\emph{High Trust (\keyII{})}} in the system and the general VPSA method.
In order to achieve those key goals, we derive a set of \textit{Design Goals} and \textit{Requirements}. 
%

%

\changed{We determined four \textbf{Design Goals} based on the design guidelines for domain experts by Wong et al. and our own experience.}{
\textbf{Design Goals: }Aligning with the design guidelines for domain experts by Wong et al.\cite{Wong2019}, we determined four design goals.}
The visualization design space must be \textbf{\emph{simple yet expressive (\dpI{})}} in order to solve high-level problems but not overwhelm visualization novices. 
A strategy to lower Cost and increase Trust is to \textbf{\emph{activate domain knowledge (\dpII{})}} which, for instance, can be achieved by including visualization options a domain expert most likely knows and showing them with familiar data. 
In order to improve Cost and Trust further, the system has to \textbf{\emph{promote learning (\dpIII{})}}, e.g., by including explanations and providing means so users can transfer existing visualization knowledge to more advanced visualization techniques. 
Last but not least, a crucial factor in gaining Trust is to \textbf{\emph{provide transparency (\dpIV{})}}, especially in terms of why a specific visualization was recommended to achieve a particular high-level task. 


\changed{We extracted general \textbf{System Requirements} complementing and supporting the Design Goals from the individual requirement analyses of the papers in the meta design study.}{ 
\textbf{System Requirements: } in order to support those Design Goals, we derived several requirements from the individual requirement analysis outlined in the papers of the meta design study.}
One requirement is that the system has to be \textbf{\emph{easy to set up (\reqI{})}} in terms of installation and initial data preparation. 
Another requirement we often found was that some kind of \textbf{\emph{overview (\reqII{})}} is necessary to understand the analyzed model better. 
Further, an application must provide means to \textbf{\emph{navigate a multi-dimensional parameter space  (\reqIII{})}} to explore the input-output relations of the model interactively.
The remaining generalizable requirements can be described as a variation or a detailed description to support one or several of the \textbf{\emph{analysis tasks (\reqIV{})}} outlined in \autoref{tab:vpsa-tasks}.
%

\label{sec:constraints}
\deleted{To put the scope of this work within reasonable boundaries, we
propose several }\textbf{Constraints}:
\deleted{First, }RSVP will provide \textbf{\emph{no sampling support (\conI{})}}. We assume that users are capable of sampling their model, at least by using random or Cartesian grid sampling. 
Most applications we researched focused on visualization solutions that provided multiple interconnected views to study a few hundred runs. 
For that reason, the tool should be capable of providing solutions for \textbf{\emph{no more than a couple of hundred runs (\conII{})}}. 
For a single run, some models can output a series of complex objects, e.g., time-varied images. 
We focus on visualizing single complex objects per run, i.e. \textbf{\emph{no series of complex objects (\conIII{}})} need to be supported. 
Further, integrating models for direct re-sampling (also known as integrated sampling\cite{Sedlmair2014}) is an often complicated and error-prone procedure.
Therefore, the tool will offer \textbf{\emph{no direct model interface (\conIV{})}}.

\begin{figure}[t]
    \centering
    \begin{subfigure}[b]{0.85\columnwidth}
    \includegraphics[width=\columnwidth,trim={0 20cm 0 4.5cm},clip]{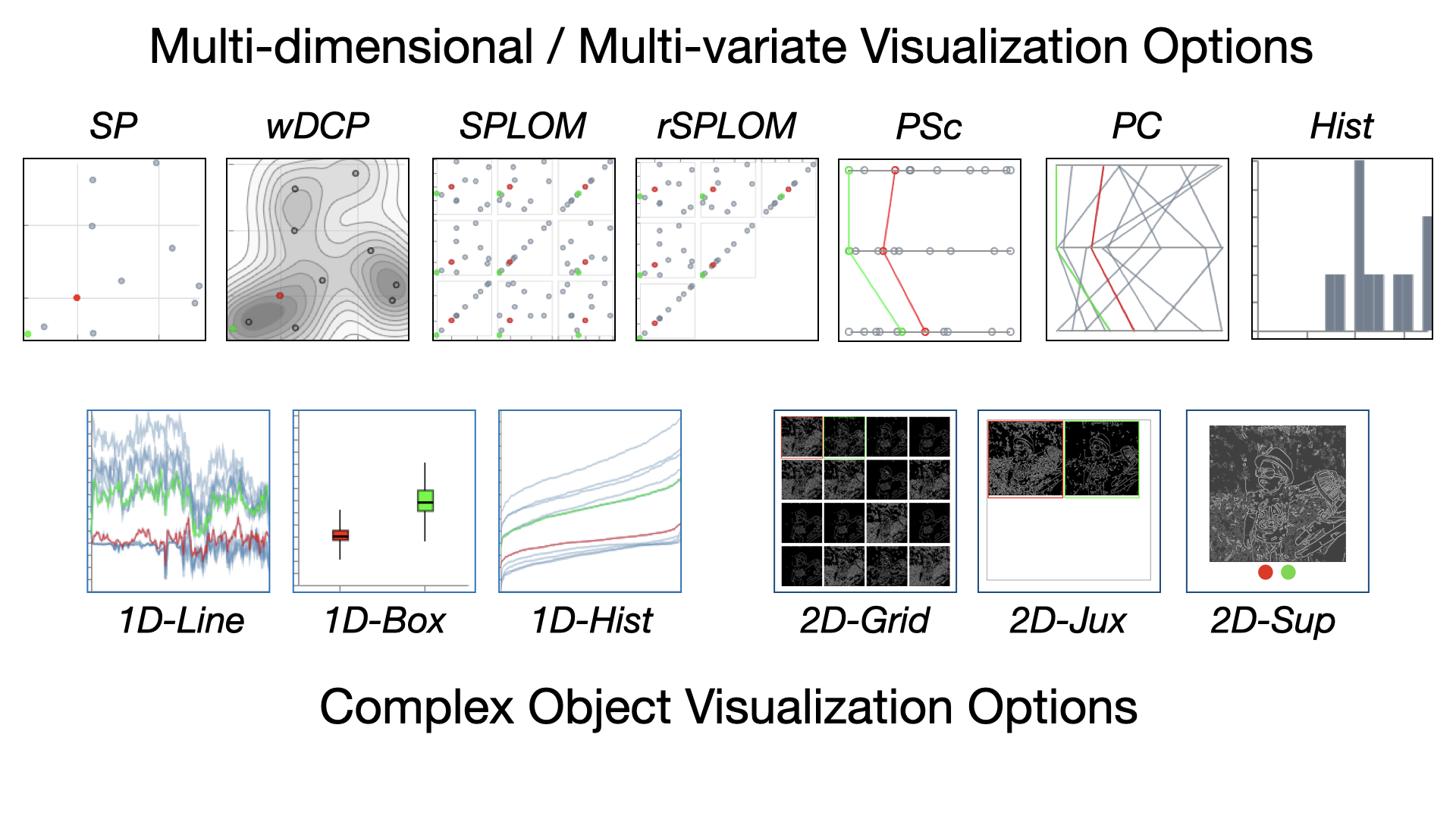}
    \end{subfigure}
    \begin{subfigure}[b]{0.98\columnwidth}
    \footnotesize
    \centering
\begin{tabular}{c l l} 
 \hline 
 \multirow{7}{*}{\rotatebox[origin=c]{90}{ MDMV}}
 &\spfull & \sp \Tstrut\\
  &\emph{(weighted)} \dcpfull & \dcp \\
 &\splomfull & \splom \\
 &\rsplomfull & \rsplom \\
 &\pscfull & \psc \\
 &\pcfull & \pc \\
 & \histfull & \hist \Bstrut\\ 
 \hline
 \textbf{Type} & \textbf{Visualization Options} & \textbf{Abbr}. \TBstrut\\ 
 \hline
 \multirow{6}{*}{\rotatebox[origin=c]{90}{Complex Object}}
 & \coIlinefull & \coIline \Tstrut\\
  & \coIboxfull & \coIbox \\
 & \coIhistfull & \coIhist \Bstrut\\

 \cline{2-3}
& \coIIgridfull &  \coIIgrid \Tstrut\\ 
  & \coIIjuxfull & \coIIjux \\
  & \coIIsupfull & \coIIsup \Bstrut\\
 \hline
 
 \end{tabular}
\end{subfigure}
    \begin{subfigure}[b]{0.85\columnwidth}
    \includegraphics[width=\columnwidth,trim={0 7.5cm 0 17.5cm},clip]{figures/Design-space/DesignSpace.001.png}
    \end{subfigure}
    \caption{Overview of the Visualization Design Space, split into multi-dimensional/multi-variate (MDMV) visualizations, and complex object visualization.}
    \label{fig:design-space}
\end{figure}
\subsection{Visualization Design Space ( \lbox{L3} )}
\label{sec:vis-design-space}

\renewcommand{\arraystretch}{1.5}
\begin{table*}[!b]
        \caption{Overview of the \visrec{}-strategy for the VPSA design space}

    \begin{subtable}[m]{0.60\textwidth}
    \centering

        \begin{tabular}{l l|l l|l}
        \multicolumn{1}{c}{Task} & Strategy & \multicolumn{2}{c}{MDMV} & Complex Object \\
        \hline
        \topt  & Overview  & \multirow{3}{*}{\rotatebox[origin=c]{90}{\makecell{Channel-\\based}}} & \ul{Spatial} expressivity   & \coIline{}, \coIIgrid{} \\
        \tfit  & Affiliation    &  & Overview + \ul{Color}               & \coIIjux{} \\
        \tunc  & Attenuation      &  & Overview + \ul{Brightness}            & \coIbox{} \\
        \tout  & Separation & \multirow{3}{*}{\rotatebox[origin=c]{90}{\makecell{Mark-\\based}}} & \ul{Point}-based (0D-mark) & (\coIline{})\\
        \tsens  & Con-/Divergence & & \ul{Line}-based (1D-mark) & \coIhist{}, \coIIsup{} \\
        \tpart & Summarization & & \ul{Area}-based (2D-mark) & \coIIgridminus{} \\
        \end{tabular}
        \bigskip
        \caption{high-level overview of the task-oriented VisRec strategy}
        \label{tab:high-level-vis-rec}
    \end{subtable}
    \begin{subtable}[m]{0.38\textwidth}
    \begin{tabular}{c|c|c|c}
        Dims\# & \incp{}\emph{-Reg} & \incp{}\emph{-Stoch}; \outdirder{}  \\
        \hline
        \emph{(1)} & (\psc{}+\hist{}) & (\psc{}) \\
        2 & \dcp+\hist & \sp \\
        3 & \splom+\dcp+\hist & \splom  \\
        4 & \pc+\dcp+\hist & \emph{(r)SPLOM}  \\
        5 & \pc+\dcp+\hist & \rsplom  \\
        6 & \pc+\hist & \rsplom  \\
        7-9 & \pc& \pc   \\
        10+ & (\pc) & \psc  \\
    \end{tabular}
    \medskip
    \caption{Spatial expressivity}
    \label{tab:spatial-information-clarity}
    \label{tab:spatial-expressivity}
    \end{subtable}

\end{table*}
\renewcommand{\arraystretch}{1.0}

\begin{Deleted}
\st{According to} \textit{\dpI{}}\st{, the design space must be expressive enough to perform VPSA tasks but simple enough not to overwhelm novice users. 
We derived a common design space by reviewing the applications from the meta design study and extracting visualizations that appeared in at least two different applications. 
However, not every visualization fitting this criterion ended up in the final design space. 
For instance, we found four occurrences of Heatmaps, three Density Plots, and two Contour Line Plots.
We combined those into \textit{(weighted) Density Contour Plots} (}\wdcp{}\st{) since Heatmaps would have required a dense grid sampling strategy which we wanted to avoid (}\conI{}). 
\st{As outlined in }\autoref{fig:vpsa-intro}\st{, data analyzed in VPSA typically consists of \textit{scalar parameter values} and often \textit{complex objects}, which can not be described with a single variable without losing information.
Therefore, the design space is conceptually split into \textit{multi-dimensional, multi-variate (MDMV)} visualizations and \textit{complex objects} visualization options.
The resulting design space is presented in }\autoref{fig:design-space}\st{
It consists of seven MDMV and six complex object visualization options, where the latter is split into three visualization options for 1D and three for 2D complex objects.
Several reviewed applications included a detailed view of a 3D object.
However, these views were typically associated with a unique research contribution and were, therefore, uncommon by definition.
Furthermore, in our experience, domain experts often have tools for visualizing details of their data that they would not like to give up.}
\end{Deleted}

\colored[designspace]{We derived a common design space meeting the requirements of \textit{\dpI{}} by reviewing the findings from the content coding stage and extracting visualizations that appeared in at least two different applications.}
An overview of the visualization options in the resulting design space is presented in \autoref{fig:design-space}.
\colored[designspace]{However, not every visualization fitting the previous criterion ended up in the final design space. 
For instance, we found four occurrences of Heatmaps, three Density Plots, and two Contour Line Plots.
We combined those into \textit{(weighted) Density Contour Plots} (\wdcp{}) since Heatmaps would have required a dense grid sampling strategy which we wanted to avoid (\conI{}). 
As previously outlined, data analyzed in VPSA typically consists of \textit{scalar parameter values} and often \textit{complex objects}, which cannot be described with a single variable without losing information.
Therefore, the design space is conceptually split into \textit{multi-dimensional, multi-variate (MDMV)} visualizations and \textit{complex objects} visualization options.
The final design space includes three 1D and three 2D complex object visualizations. 
It is noteworthy that several reviewed applications included a detailed view of a 3D object.
However, these views were typically associated with a unique research contribution, thus making them uncommon by definition.}

\subsection{VisRec Strategy ( \lbox{L4} )}
\label{sec:vis-rec}

\begin{figure}[t]
    \centering
    \includegraphics[width=\linewidth, trim={0 12cm 0 0},clip]{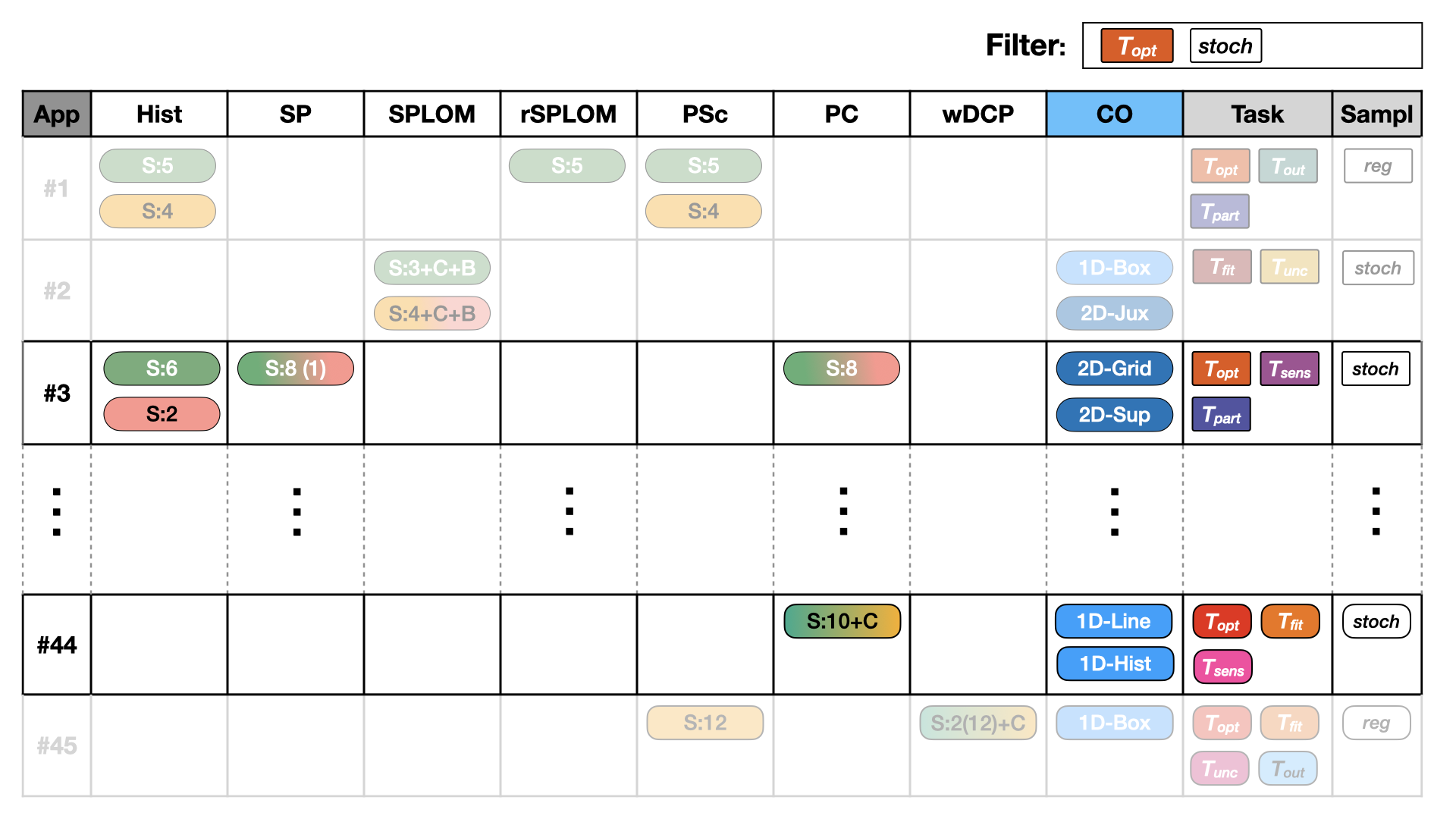}
    \caption{A visually enhanced representation of the coding table. 
    For the current filter criteria \topt{} and \vrtstoch{}, applications \#1 and \#2 are filtered out.
    Markers in MDMV visualization columns (white headers) contain information about how many dimensions were encoded (S)patially, if they encoded (C)olor or (B)rightness, and if they encoded 
    \Circled[fill color=inputs, outer color=white]{ inputs }, 
    \Circled[fill color=directout, outer color=white]{ direct outputs },
    or \Circled[fill color=derivedout, outer color=white]{ derived outputs }.
}
    \label{fig:visrec-table}
\end{figure}

The coding table described in \autoref{sec:method} provides the basis for our proposed VisRec strategy.
\autoref{fig:visrec-table} shows a visually enhanced example of that table.
Each row represents an application and columns represent visualization options (MDMV and complex objects), tasks, and the underlying sampling strategy used to generate the runs.  
The latter two form an \textit{evaluation space}\cite{MacLean1989} over the visualization design space.
This setup enables filtering for applications that address specific tasks and discovering the visualizations and settings they employ to accomplish their objectives.
\changed{A comprehensive analysis uncovered several compelling insights, enabling us to derive the rules for the VisRec strategy summarized in \autoref{tab:high-level-vis-rec}.}{A rigorous analysis of the papers in the meta design study and the subsequent content coding enabled us to derive the rules for the VisRec strategy summarized in \autoref{tab:high-level-vis-rec}.}
Our findings indicate that each task has a unique and fundamental problem-solving approach, regardless of the underlying data type. In the following paragraphs, we will present these approaches and provide further insights into our research.

\begin{figure*}[!b]
  \centering
  \includegraphics[width=\linewidth, trim={0 9.25cm 0 0},clip]{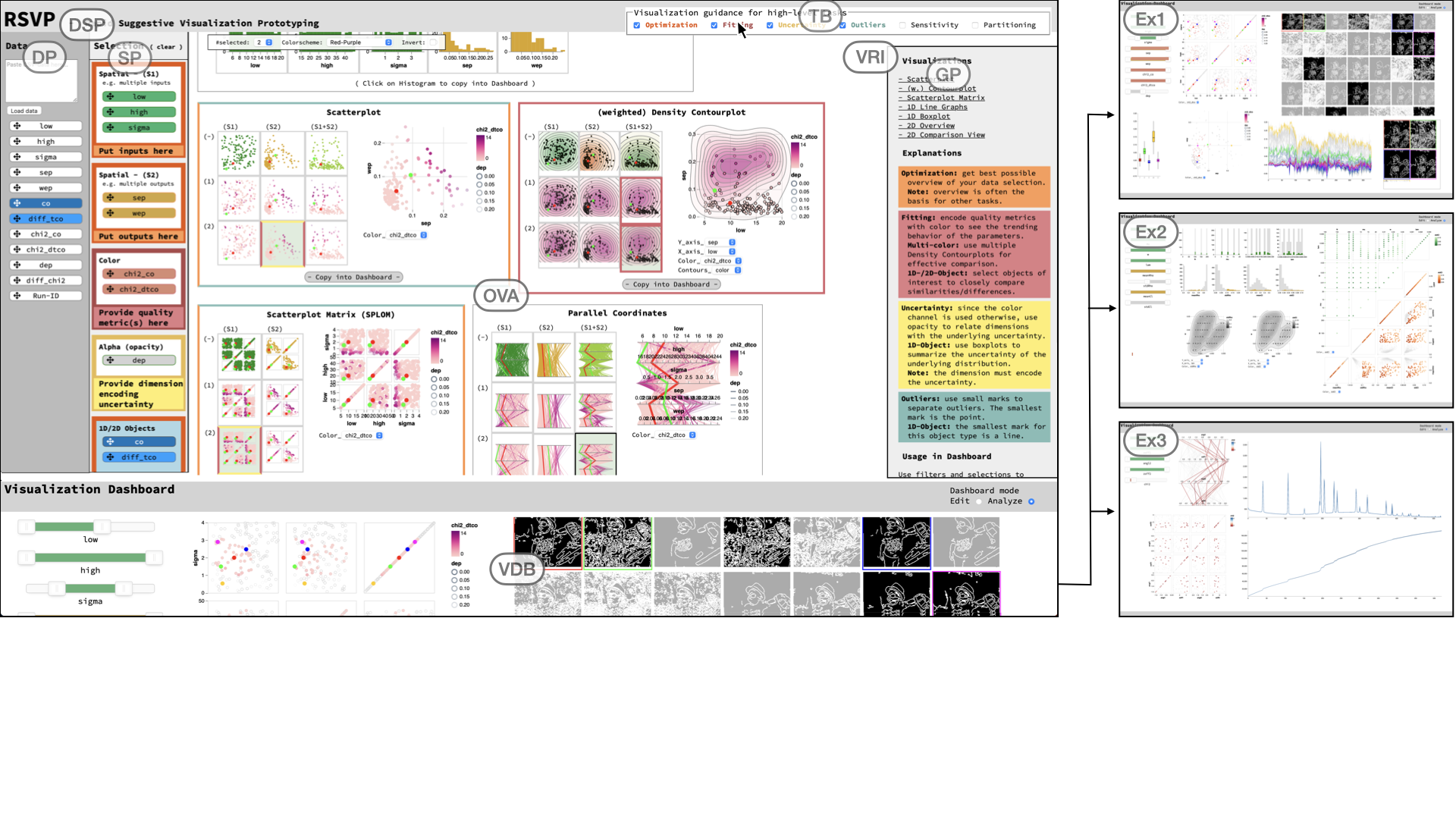 }
  \caption{\textbf{Left:} Overview of RSVP system for VPSA. \textit{System left}: Data-Selection Panel~\ldsp{}.
Dimensional markers in the data panel \ldp{} represent the user-provided CSV data.
Those markers can be dragged into the channel fields in the selection panel~\lsp{}, where white MDMV markers change their underlying colors. \textit{System center}: the Overview Area~\lova{} displays all available visualizations from the design space with user-selected data. The colors used in the visualizations match the marker colors in the selection panel.
\textit{System top-right}: the VisRec Interface \lvri{} is comprised of several components. 
Selecting tasks in the taskbar \ltb{} opens the guidance panel \lgp{} and injects recommendations directly into the overview area \lova{} and the selection panel \lsp{} in the form of colored frames. 
\textit{System bottom:} the Visualization Dashboard \lvdb{} enables users to perform interactive data analysis. Visualizations can be copied from the overview area into the dashboard.\\
\textbf{Right:} Dashboard instances \lexI{} - \lexIII{} showcase examples created from different data sets in RSVP within minutes. }
  \label{fig:teaser}
\end{figure*}

Achieving \textit{optimization} (\topt{}) requires providing the best possible overview of objects and parameters. Therefore, without an objective function, the requirement for overview (\reqII{}) is not just a general feature but a necessity for optimization.
In the context of MDMV, providing an overview means offering as much information about the data as possible while considering visual clutter, data occlusion, and cognitive overload. 
Based on the Mackinlay criteria\cite{Mackinlay1986}, we refer to the strategy of presenting spatial information clearly and concisely as ``spatial expressivity" (see \autoref{tab:spatial-information-clarity} for a summary of its high-level rules).
Providing an overview of complex objects is typically accomplished through overplotted function graphs (1D) or presenting options in a grid layout (2D).
Optimization often forms the basis for other, more specific tasks, which extend \topt{}'s overview with additional visual features or views.

Affiliating solutions is the primary approach for \textit{fitting} tasks (\tfit{}). 
The color channel is often used to represent the output of an objective function to establish connections between different MDMV visualizations. 
When dealing with 2D objects, we discerned a preference for juxtaposed comparison views.

When \textit{uncertainty} (\tunc{}) was quantified by a numerical parameter, the color channel was used to display it, preferably using black or white at opposite ends of the spectrum\cite{Torsney-Weir2011, Unger2012}.
If the color channel was already used for other purposes, the opacity channel was used instead\cite{Piringer2010, Weissenboeck2017}.
If the underlying distribution was available as a 1D object, selected items were displayed as Boxplots\cite{Berger2011, Booshehrian2012}.
Both \tfit{} and \tunc{} worked typically on top of visualizations suggested by \topt{}. 

To visually distinguish \textit{outliers} (\tout{}) from other data points, visualizations often used the smallest available marks for a given data type. This technique helped separate outliers from the rest of the data.
Six applications out of eight supporting this task used Scatterplots or Scatterplot Matrices (0D-mark) for MDMV visualizations (the remaining two\cite{Luboschik2014, Potter2009} using bespoke visualizations), and four applications out of five encoding 1D objects used overplotted Linegraphs (1D-mark).

In \textit{sensitivity} tasks (\tsens{}), the objective is to identify parameter settings with either low or high impact on model outputs.
Line-based visualizations help identify converging (dense) or diverging (sparse) sections, which makes Density Contour Plots and Parallel Coordinates the recommended choices for MDMV visualizations. 
Linegraphs are effective for identifying local sensitivities in 1D objects, while Cumulative Histograms can help in finding global thresholds.
Superpositioned views support determining sensitive regions for selected 2D objects.

The \textit{partitioning} task (\tpart{}) tries to identify different types of model behavior, especially when there is no objective function to guide the analysis.
The key is to find boundaries that separate groups that behave similarly.
Histograms are commonly used because they abstract away details and help to focus on the bigger picture.
When analyzing 2D objects, a Grid Layout showing elements within a selected range but hiding filtered ones can achieve a similar effect while still providing detailed information about the individual items.

From a high-level perspective, our findings suggest that tasks 
\topt{}, 
\tfit{}, and 
\tunc{} involve encoding specific \textit{channels}, whereas addressing
\tout{}, 
\tsens{}, and 
\tpart{} requires selecting appropriate \textit{marks}.
Within the context of our research, the specific marks and channels were even distinct for the various analysis tasks
(see MDMV-column in \autoref{tab:high-level-vis-rec}).

\section{RSVP System}
\label{sec:rsvp-system}
The \textit{Rapid Suggestive Visualization Prototyping} (RSVP) system is a practical implementation of the findings from the meta design study. 
In this section, we present a walk-through of the RSVP system, provide some technical details, and offer a qualitative result inspection about how the system aligns with previously outlined design goals and requirements.

\subsection{Walk-through}
\label{sec:walk-through}

\newcont{RSVP enables users without specific visualizations or programming skills to design dashboards tailored to their specific data and needs regarding VPSA.
The user interface tightly integrates an overview of available visualization options with an unobtrusive recommendation strategy to foster transparency and learnability.}
\changed{An overview of RSVP is presented in \autoref{fig:teaser} and consists of four main components:}{RSVP consists of four main components, as outlined in \autoref{fig:teaser}:}
the \textit{Data-Selection Panel} \ldsp{}, the \textit{Overview Area} \lova{}, the \textit{VisRec Interface} \lvri{}, and the \textit{Visualization Dashboard} \lvdb{}.
These components are shown in \autoref{fig:teaser} and will be described in detail in the following sections.\\

\begin{figure}[h]
    \centering
    \includegraphics[width=0.95\linewidth, trim={0 2.5cm 0 0}, clip]{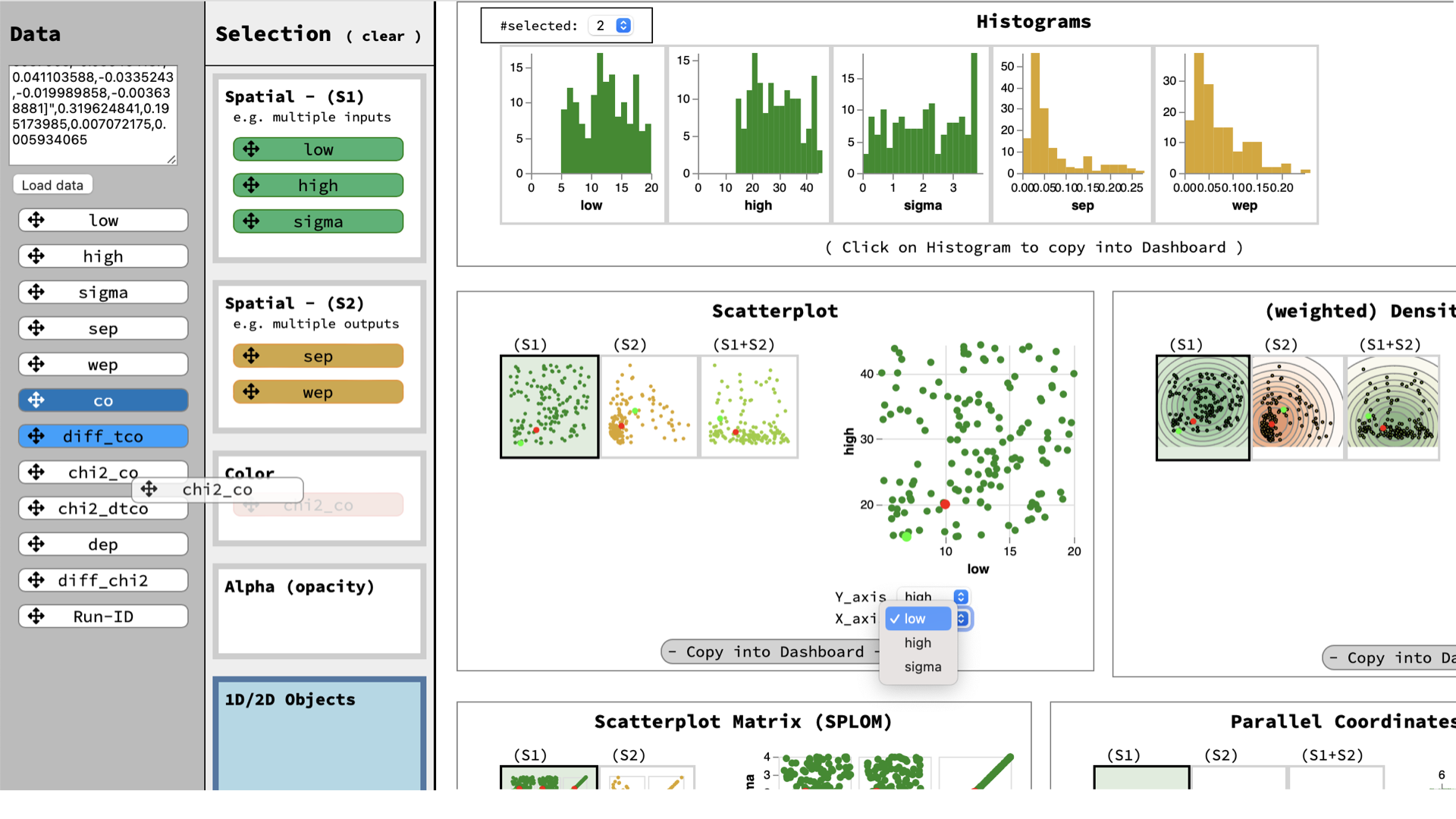}
    \caption{The Data-Selection Panel \ldsp{} and parts of the Overview Area \lova{}. Dimension  is actively moved into the Color field. The interactive switch for the Scatterplots x-axis shows the three available dimensions from Spatial field (S1). }
        \label{fig:walkthrough-dsp}
        \label{fig:walk-through-dsp}
\end{figure}

\textbf{DATA-SELECTION PANEL \ldsp{}} :
Data gets loaded in a simple CSV format directly into the textfield in the data panel \ldp{}.
The header must contain the names of the dimensions, and rows contain the details about the individual runs, namely the parameter settings and the results.
Once loaded, data dimensions are represented in the form of draggable markers (see \autoref{fig:walkthrough-dsp}).
The color of a marker represents the underlying data type.
RSVP detects three different data types: numerical parameters (scalar values, white), 1D objects (arrays, blue), and 2D objects (links to image files, dark blue).

The markers can be dragged and dropped into channel fields in the adjacent selection panel. 
Available channels for direct data encoding are spatial, color, and opacity (see \autoref{sec:vis-design-space}).
RSVP offers two spatial channel fields, S1 and S2, so MDMV visualizations split along 
inputs
and 
outputs
can be displayed simultaneously in the overview area using small multiples displays (SMDs). 
%
MDMV markers, when placed into channel fields, adjust their color to correspond with their appearance in the overview area, enhancing orientation. 
Complex object markers must be positioned within the object field to be visible in the overview area.

Please note that the colors utilized in Figures \ref{fig:vpsa-intro} and \ref{fig:visrec-table} are selected to align with the colors employed in the RSVP system for markers in channel fields 
Spatial S1, Spatial S2, and Color. 
These fields are frequently used to represent \Circled[fill color=inputs, outer color=white]{ inputs }, \Circled[fill color=directout, outer color=white]{ direct outputs }, and \Circled[fill color=derivedout, outer color=white]{ derived outputs }, respectively.\\

\begin{figure}[h]
    \centering
    \includegraphics[width=\linewidth, trim={0 12.6cm 0 0}, clip]{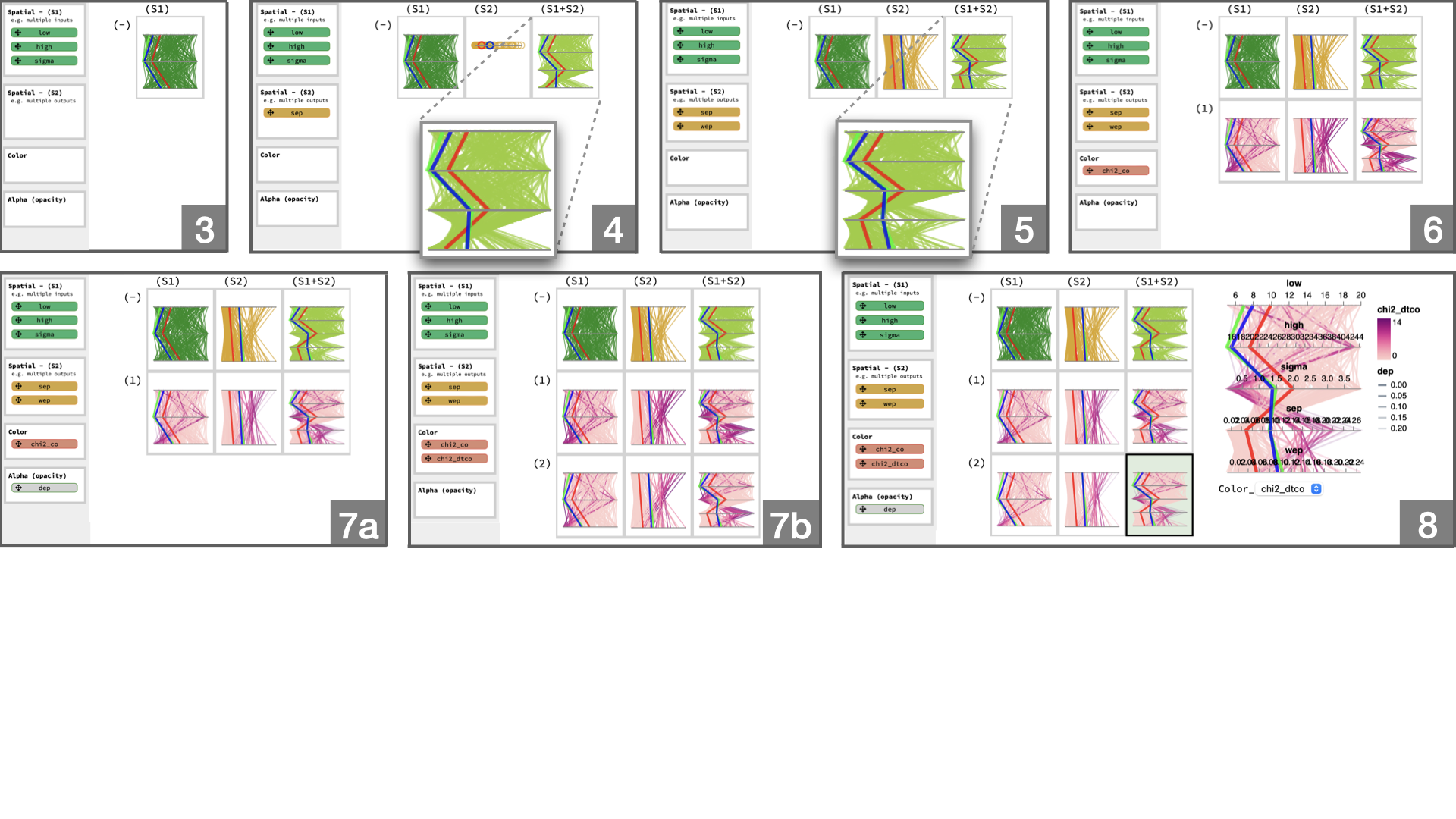}
    \caption{The SMD behavior using Parallel Coordinates as an example. The dark boxes indicate the numbers of encoded dimensions. 
Configurations \graybox{4} and \graybox{5} highlight the differences for the (S1+S2)-columns.
Configs \graybox{7a} and \graybox{7b} encode seven dimensions slightly differently, resulting in additional rows.
Configuration \graybox{\hspace{0.08cm}8\hspace{0.08cm}} is displayed with the according detail view for the currently selected small multiple.}
        \label{fig:walk-through-ova}
\end{figure}

\textbf{OVERVIEW AREA \lova{}} : displays all visualizations from the design space (see \autoref{fig:design-space}) in a 2-dimensional layout, directly encoding user-provided data from the selection panel \lsp{}.
The top area contains MDMV visualizations, and the bottom displays complex object visualizations.
With the exception of Histograms, MDMV visualization options consist of a Small Multiple Display (SMD) and a detail view~\cite{van-den-Elzen2013}.
Selecting a small multiple will load the according data settings in the detail view for this visualization. 
The SMDs show subsets and combinations of user-selected dimensions that could not be presented simultaneously otherwise. 
The layout of the SMDs tries to follow the same rules for all the visualization options incorporating it, depicted in \autoref{fig:walk-through-ova}.
Each view in a particular column encodes the same spatial dimensions. 
The first column encodes dimensions from the first spatial field (S1), and the second column from the second spatial field (S2). 
The third column encodes the combination of both spatial fields (S1+S2) (see the highlights in \autoref{fig:walk-through-ova} configurations \graybox{4} and \graybox{5} for an example).
Our literature analysis did not reveal an instance where inputs and outputs were combined into a single \textit{(r)SPLOM}. 
Therefore, \textit{SPLOM} and \textit{rSPLOM} do not provide a third column to avoid unnecessary rendering overhead. 
The first row in the SMD encodes spatial fields only, where the various small multiples are displayed with categorical colors that link them visually to the underlying spatial fields in the selection panel.
Additional rows encode the selections for the color and opacity channels. 
If either channel gets encoded more than once, multiple additional rows will be displayed (\autoref{fig:walk-through-ova}, configurations \graybox{6} through \graybox{\hspace{0.08cm}8\hspace{0.08cm}} show this particular behavior for different encodings). 

\begin{figure}[h]
    \centering
    \includegraphics[width=\linewidth, trim={0 15.5cm 0 0}, clip]{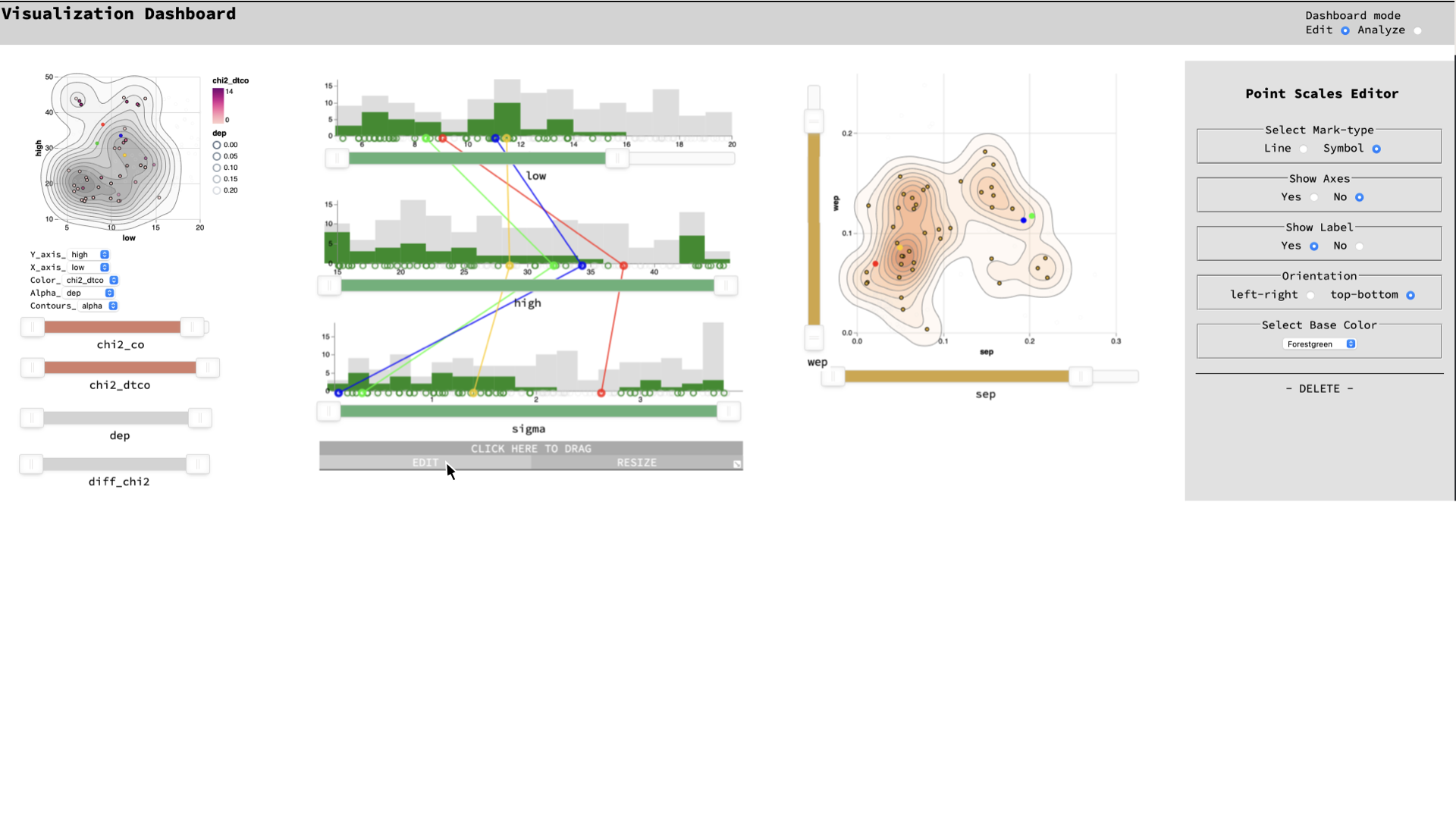}
    \caption{Visualization Dashboard \lvdb{} example in \emph{edit}-mode with the single-view editor (right) opened for the \textit{Point Scales} used in the parallel histograms (green). }
      \label{fig:walk-through-vdb}
      \label{fig:walkthrough-vdb}
\end{figure}

If a visualization can only display some of the dimensions from the selection panel, interactive switches will appear so the user can adapt them during analysis. 
Examples of such situations include cases where \textit{SP} or \textit{wDCP} have to encode three or more spatial dimensions (see \autoref{fig:walkthrough-dsp}), or any visualization has to encode more than one dimension with the color or opacity channel (see \autoref{fig:walk-through-vdb}). 
A panel at the top of \lova{} allows for global adjustments of the visualizations, like the number of pre-selected data points in the overview for comparison purposes or the color scheme for dimensions encoded in the color channel.\\

\textbf{VISREC INTERFACE \lvri{}} : consists of the \textit{taskbar} \ltb{} and the \textit{guidance panel} \lgp{}, and spans over the \textit{overview area} \lova{} and the \textit{selection panel}~\lsp{}.
RSVP can provide recommendations for up to four tasks simultaneously.
When seeking VisRec guidance, \topt{} is automatically selected, as it typically serves as the foundation for other tasks.
Each task is associated with a unique categorical color (see \autoref{tab:vpsa-tasks}), which makes it possible to track the recommendations for a specific task across the different sections. 

The taskbar \ltb{} in the top right shows the tasks for which users can seek recommendations. 
Hovering over a specific task provides a brief explanation about it. 
Selecting one of the tasks triggers the recommendation process.
Recommendations are provided as colored frames around the visualization types and single instances in the small multiples in the overview area \lova{}.
For tasks optimization~\topt{}, fitting \tfit{}, and uncertainty \tunc{}, dimensions have to be encoded according to the channels supporting the respective tasks (see \autoref{tab:high-level-vis-rec}).
RSVP frames the channel fields in the selection~panel~\lsp{} with the respective task color and provides context information on which kind of dimensions are supposed to be encoded (\textit{inputs}, \textit{outputs}, \textit{derived} values and values \textit{quantifying uncertainty}).

\begin{figure}[t]
    \centering
    \includegraphics[width=\linewidth, trim={0 11.5cm 0 0}, clip]{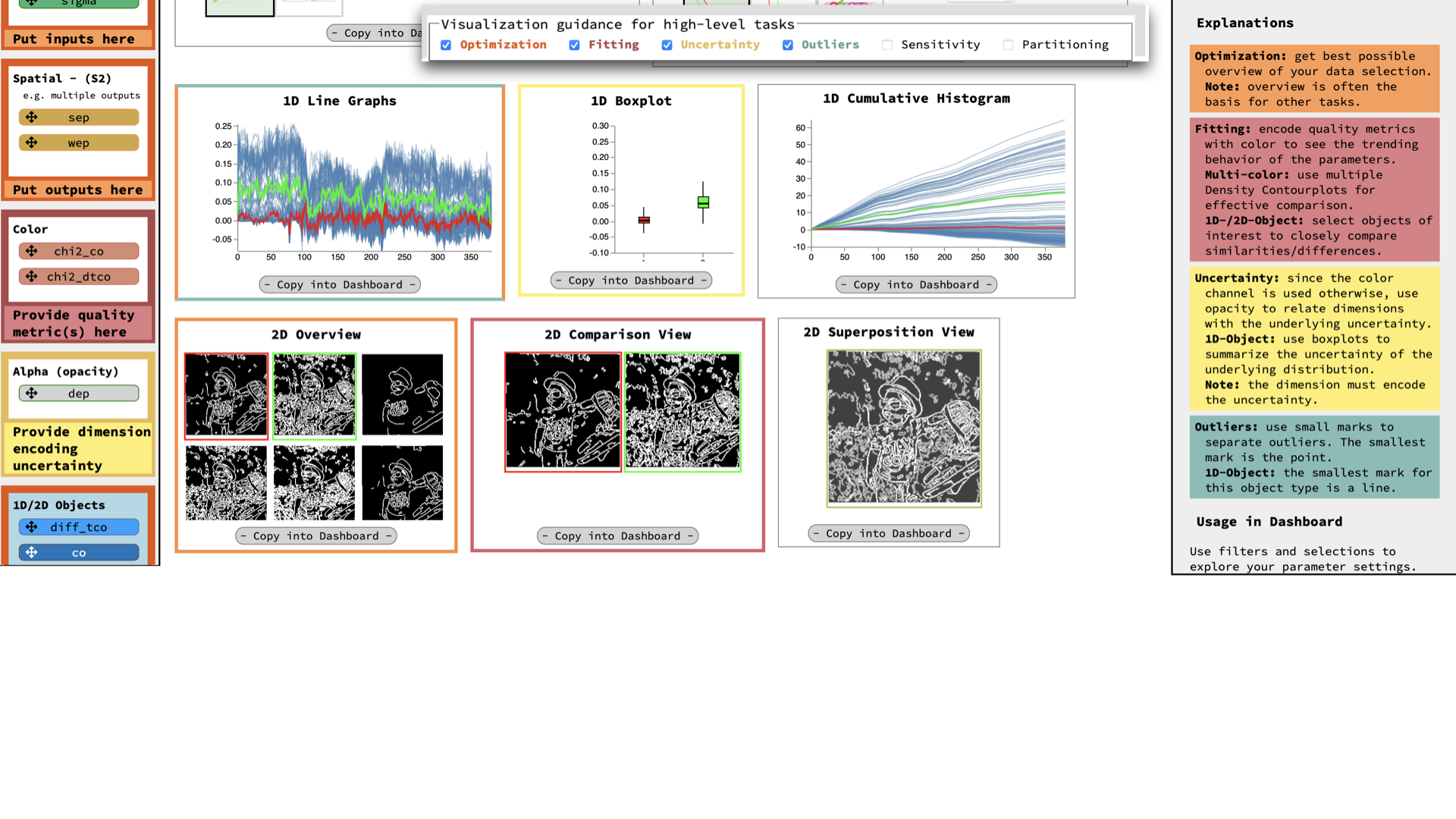}
    \caption{Complementary view showing the complex objects for the recommendations in \autoref{fig:teaser}.}
      \label{fig:rsvp-guidance}
      \label{fig:walk-through-vri}
      \label{fig:walkthrough-vri}
\end{figure}


The Guidance Panel \lgp{} provides helpful additional information regarding the recommendations. 
It consists of three parts. 
On top is a list of the recommended visualization types. 
This list links to the respective options in the Overview Area \lova{}. 
The middle section contains explanations and background information on how recommended visualizations are supposed to support respective tasks.
At the bottom are hints on how to interact with the visualization dashboard to solve these tasks.

Figures \ref{fig:teaser} and \ref{fig:walk-through-vri} show an example recommendation for tasks \topt{}, \tfit{}, \tunc{}, and \tout. 
Further examples for recommendations are provided in \autoref{sec:case-study-I}, the supplemental material, and the accompanying video.\\


\textbf{VISUALIZATION DASHBOARD \lvdb{} :} visualizations can be copied from the Overview~Area~\lova{} into the Visualization Dashboard \lvdb{}.
The dashboard allows the combination of an arbitrary number of visualizations to perform multi-view analysis.
Individual visualizations can even be combined into more complex ones. 
\autoref{fig:walk-through-vdb} shows an example where Point Scales, Histograms, and interactive sliders are combined into parallel histograms. 
A slider appears for each dimension loaded into the dashboard, providing a consistent way of filtering data points across all visualizations.
The color of the slider matches the color of the dimension in the selection panel. 
Selecting a single data point within a visualization will also highlight the same element in all the other visualizations in the dashboard. 
The dashboard offers two modes, \emph{edit} and \emph{analyze}.
In \textit{edit}-mode, hovering over a visualization or slider reveals a context menu that allows for manipulating this particular element, like moving, resizing, or changing its attributes.
The attributes can be edited via the single-view editor.
When in \textit{analyze mode}, dashboard design and attributes can not be changed, and the user can focus on analyzing the dataset. 

\begin{Newcont}
    \subsection{Technical Aspects}
RSVP is a single-page web application built with standard web technologies (HTML, CSS, JavaScript). 
It operates without a backend; all computations occur within the browser. 
The application leverages several helper libraries for various tasks, including parsing CSV data\footnote{\url{https://www.papaparse.com}}, rendering HTML templates\footnote{\url{https://underscorejs.org}}, enabling drag-and-drop functionality\footnote{\url{https://bevacqua.github.io/dragula/}} in the data selection panel \ldsp{}, and implementing dragging and resizing\footnote{\url{https://interactjs.io}} as well as the data sliders\footnote{\url{https://refreshless.com/nouislider/}} in the visualization dashboard \lvdb{}. 
MDMV visualizations and 1D complex object visualizations are generated using the Vega visualization toolkit \cite{Satyanarayan2014}, while 2D complex objects are implemented using plain HTML and CSS.
The recommendation algorithm is designed as a series of cascading rules, following a similar approach outlined in BOZ \cite{Casner1991}. 
Further technical details about the implementation of the VisRec strategy can be found in the supplemental material.
\end{Newcont}

\subsection{Qualitative Result Inspection (QRI)}
\label{sec:qri}

In this section, we provide the rationales of how our proposed solution fulfills the requirements and design goals outlined in \autoref{sec:requirement-analysis}.
The two Key Goals \textit{Low Cost (Key-I)} and \textit{High Trust (Key-II)} will be discussed in more detail in \autoref{sec:findings}.

\newcont{\textbf{ Easy setup (\reqI{}):}} RSVP is a web application \changed{that does not require an account.}{that can be used without an account.}
It requires a simple CSV file as input, which can be created with any modern spreadsheet software.
Further, the system does not require a specific sampling strategy, but the user can choose the most feasible one. 
In the dashboard \lvdb{}, visualizations and sliders for filtering can be arranged and resized via drag-and-drop. 
All these aspects ensure an \changed{}{easy setup} for the user. 

\newcont{\textbf{Overview (\reqII{}): }} RSVP provides an overview of basically all elements associated with user interaction.
It shows \deleted{all} the user-provided data dimensions in the data panel \ldp{} and all encodable channels in the selection panel \lsp{}.
The taskbar \ltb{} presents the available analysis tasks, and the overview area \lova{} shows all the available visualizations \newcont{directly encoding data}.
\deleted{This area also encodes all visualizations with data the user selects, offering}{This enables} a quick overview of the parameters and complex objects from many different perspectives.
Visualizations offered by RSVP are extracted from applications mainly featuring \textit{global-to-local} navigation strategies.
\changed{Hence, dashboards composed of these visualizations for interactive analysis focus on an initial overview of the data by design.}{Hence, visualization dashboards \lvdb{} composed of these visualization options feature an initial overview of the data by design. }

\newcont{\textbf{Navigating Parameter Space (\reqIII{}): }} Available operations for navigation during an analytical session \newcont{in the visualization dashboard \lvdb{}} are filtering, zoom, and selection of individual runs, therefore following Shneiderman's information-seeking mantra (``Overview first, zoom and filter, then details-on-demand")\cite{Shneiderman1996}.
To navigate the design space, the system presents MDMV visualizations in small multiples and large singles, akin to the design of van den Elzen and van Wijk \cite{van-den-Elzen2013} for visualization exploration. 
RSVP supports the user in navigating the vast design space by suggesting appropriate visualizations to solve user-defined high-level tasks. 
This support also includes suggestions for navigating the parameter space by indicating to the user how to encode the various dimensions in the parameter space for visual exploration. 

\newcont{\textbf{Analytical Task Support (\reqIV{}):}} RSVP supports domain scientists in solving analytical tasks by indicating which visualizations to choose and how to encode them. 
It further informs the users how to interact within the dashboard to achieve those tasks. 

\newcont{\textbf{Simple yet expressive (\dgI{}): }}
Our approach focused on creating a design space that balances simplicity for novice users with the expressiveness needed for complex analytical tasks.
Instead of favoring complicated special purpose views, our approach focuses on finding common and likely easier-to-understand visualization options, which can further be composed into complex multi-view visualizations, supporting specific, high-level tasks.

\textbf{Activate Domain Knowledge (\dgII{}):}
Practical recommendations for activating domain knowledge are using terminology and visualizations familiar to users\cite{Wong2019}.
Since RSVP is domain-agnostic, it avoids using domain-specific terminology in labels or explanations.
However, users are likely familiar with the terminology used in the data they provide.
Moreover, the dimensions and data points are likely meaningful to them.
The key idea is that the data will initially help improve understanding of the various visualizations. 
RSVP presents all available visualizations immediately encoded with the user-provided data. 
This might not only improve understanding but seeing the data through different lenses of multiple visualizations might provide immediate insights or raise exciting questions\cite{Roberts1998, Roberts1999}.
Since several provided visualizations (like scatterplots, histograms, and line charts) are the subject of teaching within the K-12 curriculum\cite{Lee2016}, users are likely familiar with at least some of them. 
 
\begin{figure}
    \centering
    \includegraphics[width=0.98\columnwidth, trim={0 0 0 3cm}, clip]{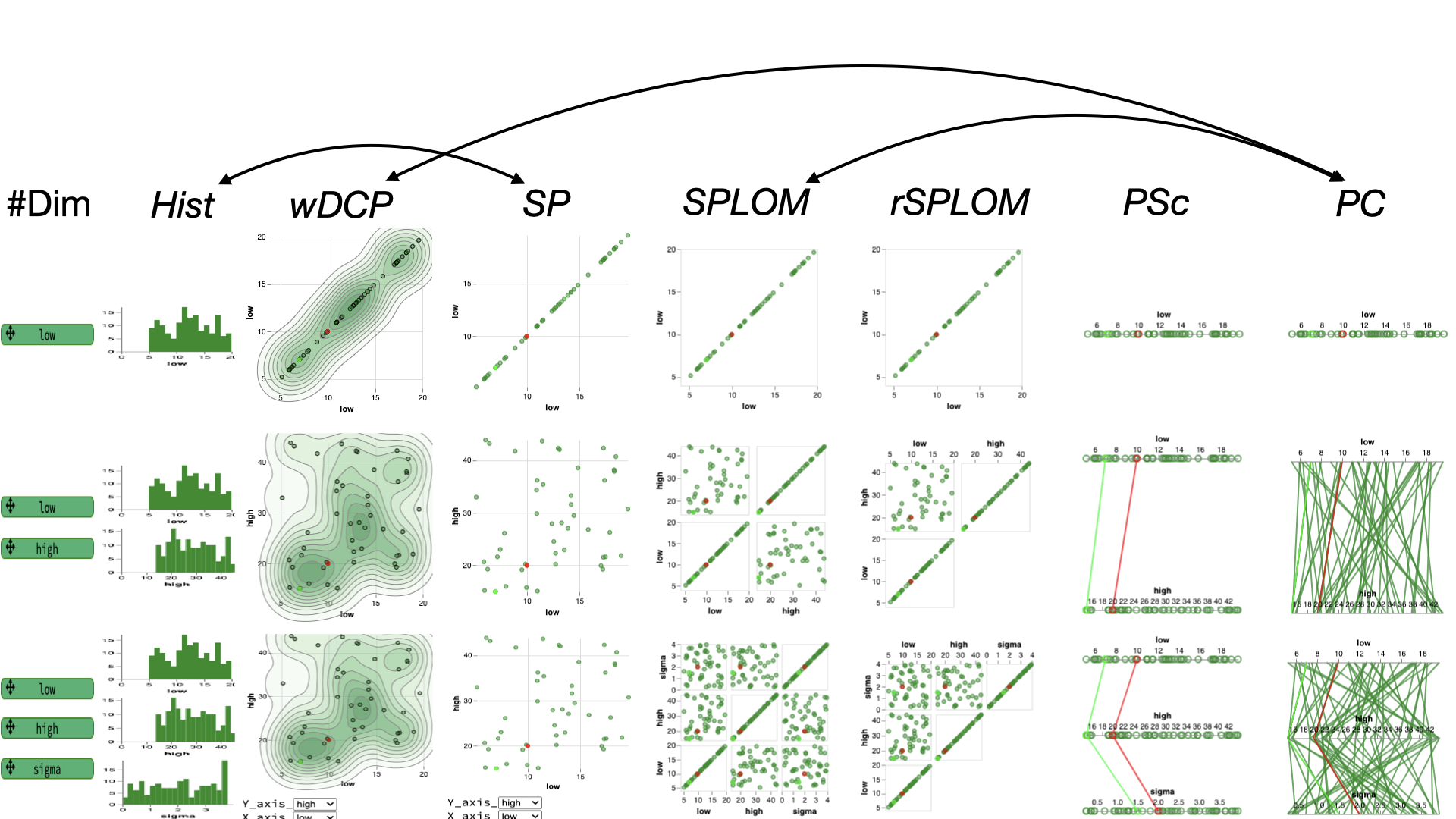}
    \caption{
    RSVP promotes learning by adjacency and progressive reveal. 
    Columns show which visualizations are adjacent in the Visualization Overview \lova{} (the arrows on top show further adjacencies in the 2-dimensional layout) and how they are supposed to show similarities and differences between adjacent visualizations.
    Rows show how the complexity of visualizations is progressively revealed when adding more dimensions. 
    }
    \label{fig:PromoteLearning}
\end{figure}

\textbf{Promote Learning (\dgIII{})}:
RSVP promotes learning about new visualizations by positioning similar visualization options closer together in the overview area and encoding them directly with user-provided data.
The idea is to make it easier for users to spot and compare similarities and differences between known and previously unknown visualizations. 
It should also improve the understanding of familiar visualizations in the context of the current problem. 
Furthermore, to make learning about visualizations even more accessible, the complexity of each visualization is initially reduced and then revealed progressively\cite{Wong2019, Ruchikachorn2015} when additional dimensions get added. 
As presented in \autoref{fig:PromoteLearning}, when only a single dimension is encoded (visualizations requiring at least two spatial dimensions encode the same dimension on both axes), several MDMV-visualization options look very similar. 
Adding a second dimension shows how previously similar visualizations begin to differ from each other. 
Adding a third dimension further distinguishes visualizations capable of spatially encoding more than two dimensions (like SPLOM or PC).
After each step, the users can familiarize themselves with the more complicated visualizations.
To avoid misunderstandings caused by non-self-descriptive tasks, our system explains each task when hovering over them in the taskbar. Additionally, RSVP explains to users why recommended visualizations are supposed to support specific tasks.

\textbf{Provide Transparency~(\dgIV{}):} These explanations are not only helpful for learning but additionally provide transparency and should ultimately support building trust in the system \cite{Dasgupta2017}.
Recommendations are presented in a non-obtrusive way. 
RSVP presents all available visualizations for a given datatype and highlights recommended ones (a guidance strategy known as ``orientation")\cite{Ceneda2016}. 
This approach allows users to draw conclusions about recommended visualizations and make alternative decisions if necessary.

\section{Evaluation}
\label{sec:evaluation}

The three most prevalent evaluation strategies we found in the meta design study, reflecting general evaluation trends in the visualization community\cite{Khayat2019}, are \changed{case studies}{\textit{case studies}}, \changed{usability studies}{\textit{usability studies}}, and \textit{qualitative result inspections (QRI})\cite{IsenbergT2013}.
The latter we already discussed in \autoref{sec:qri}. 
In this section, we will present the results of a usability study and two real-world case studies from different problem domains utilizing RSVP.
Additionally, the supplemental material presents and discusses another case study that was conducted using an earlier prototype of RSVP.

%

\subsection{Usability Study: Edge Detection Algorithm}
\label{sec:usability-study}
For this study, we recruited four undergraduate students majoring in media informatics and two graduate students specializing in computer graphics. 
Our choice for the number of study participants is based on recommendations for determining the sample size for usability tests\cite{Turner2006, Lewis1994}.
The recruits, four males and two females, were between 21 and 36 years old ($\mu=25.83, \sigma=5.01$). 
The rationale for recruiting these users stems from the potential benefits they could gain from using a tool like RSVP. 
They frequently interact with diverse computational models, where understanding the connections between inputs and outputs is crucial for fostering learning and comprehension. 
However, such problems do not warrant the collaboration with a visualization expert in order to perform a design study.
Participants had to analyze data sets generated from the edge detection algorithm used in the running example, utilizing the same generic image in order to avoid domain complexity.
Nonetheless, the problem of creating a contour outline of an image finds application in many different areas (see \autoref{sec:background}).

\colored[details]{Sessions were conducted online via Zoom using a hosted version of RSVP.
Each session took 60 minutes.
The participants received a five-minute introduction to VPSA and a five-minute tutorial on how to use RSVP.
Afterward, they had ten minutes to explore RSVP's various functions and ask questions.
The primary analysis session took thirty minutes, during which participants had to create dashboards to analyze three increasingly complex data sets. 
Each dataset was introduced briefly at the start of each iteration and came with a challenge that participants needed to address within ten minutes.}
The challenges were (1) identifying optimal parameterizations for the data set, (2) determining sensitive areas and which dimensions influenced them, and (3) comparing and evaluating the quality of different objective functions.
\autoref{fig:teaser} - \lexI{} shows an example dashboard created during an analysis session.
The nature of the questions was open-ended, which means there were no right or wrong answers. 
Instead, at the end of each iteration, the users had to rate their confidence in their findings on a five-point Likert scale (see \autoref{tab:confidence}), which is a common evaluation approach in such open-ended settings \cite{Dasgupta2017b}.

\renewcommand{\arraystretch}{1.5}

\begin{table}[ht]
\centering
\vspace{0.5cm}
\caption{Confidence ratings of the study participants, ranging from \emph{not confident at all}
(1) to \emph{very confident} (5)}
\label{tab:confidence}

    \begin{tabular}{l c c c c c c c}
        & \textbf{P1} & \textbf{P2} & \textbf{P3} & \textbf{P4} & \textbf{P5} & \textbf{P6} & \(\varnothing\)  \\
        \hline
     \textbf{Challenge 1} & 4 & 4 & 5 & 5 & 2 & 4 & 4\\ 
     \textbf{Challenge 2} & 5 & 4 & 3 & 4 & 3 & 5 & 4\\ 
     \textbf{Challenge 3} & 5 & 4 & 5 & 5 & 5 & 5 & 4.833 \\
     \hline
     & \textbf{4.67} & \textbf{4} & \textbf{4.33}& \textbf{4.67}& \textbf{3.67} & \textbf{4.67} & \textbf{4.277}
\end{tabular}
\vspace{-.5cm}
\end{table}
\renewcommand{\arraystretch}{1.0}


During the final interview, we asked the participants to fill out a \emph{System Usability Scale} (SUS) survey~\cite{Brooke2013}, a well-known and widely adopted approach to measuring the usability of a system.
\deleted{The questionnaire has alternating positive and negative questions, with the intent that users would have to think about the answers instead of just rushing through them.}
An abbreviated version of the questions, along with the answers given by the participants, are presented in \autoref{fig:sus-results}. 
The complete results are available in the supplemental material.
Converting the users' answers into the final scores yields results between 70 and 92.5 points, with a mean of 82.5 points. 
According to Brooke, average scores above 80 are considered \emph{very good}~\cite{Brooke2013}.
\begin{figure}[h]
    \centering
    \includegraphics[width=0.98\linewidth, trim={6cm 11.5cm 3cm 0}, clip]{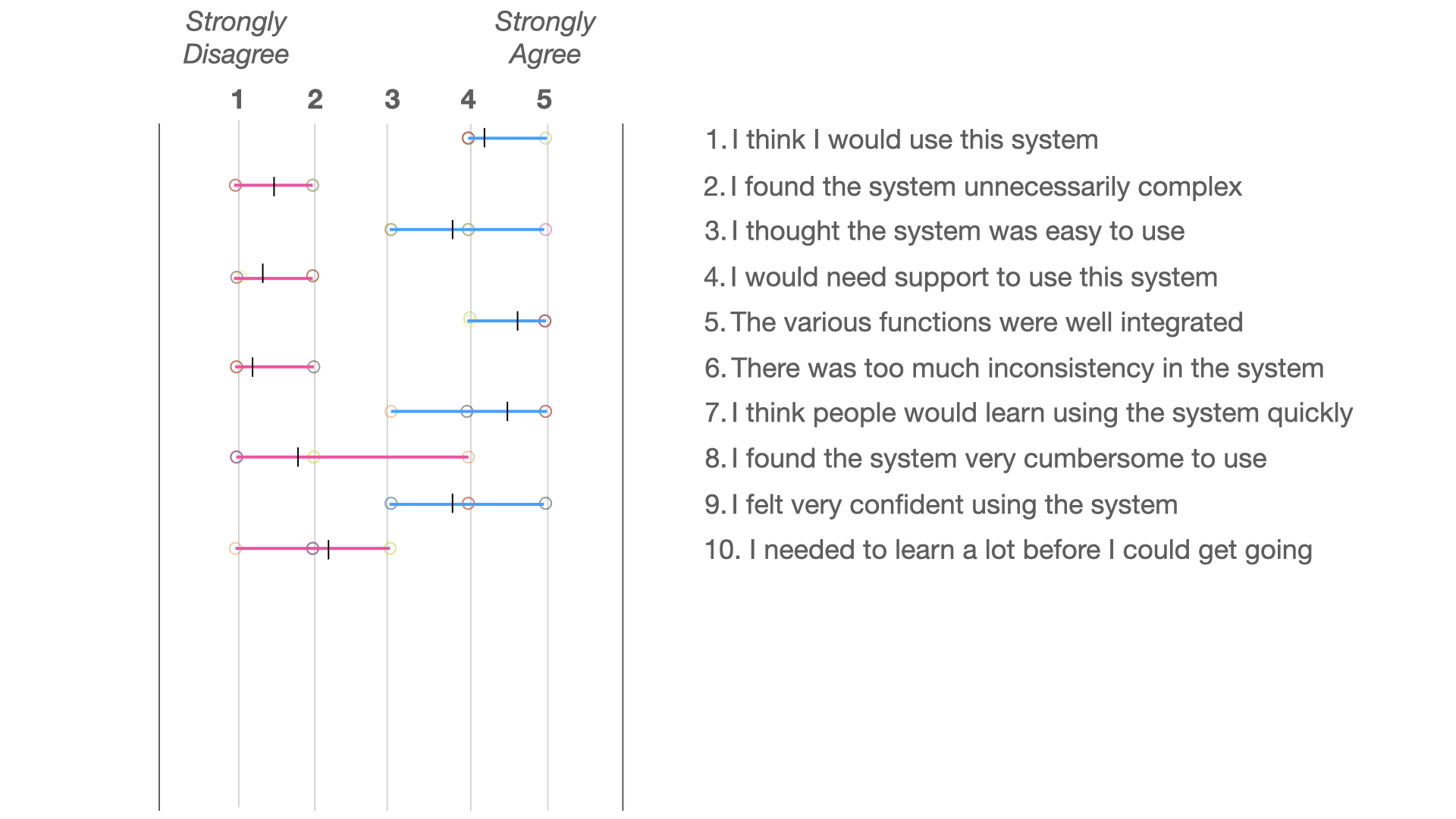}
    \caption{Results for the SUS survey, showing range (colored lines) and means (vertical black bars) of given answers. 
    Positive questions (odd-numbered, blue lines) aim for a high score, and negative questions (even-numbered, red lines) aim for a low score. }
        \label{fig:sus-results}
\end{figure}

\subsection{Case Study I: Crystal Powder Diffraction}
\label{sec:case-study}
\label{sec:case-study-I}
Our first case-study user is a physics professor who specializes in solid-state materials.
He studies crystalline structures using a procedure known as \emph{powder diffraction} and compares the results to simulated model data using a $\chi^2$-metric to learn about molecular behavior inside the crystal.
The knowledge gained from such experiments helps the physicist to grow new crystals with desired features.  

\begin{figure}[!b]
    \centering
    \includegraphics[width=\linewidth, trim={0 7cm 0 0 },clip]{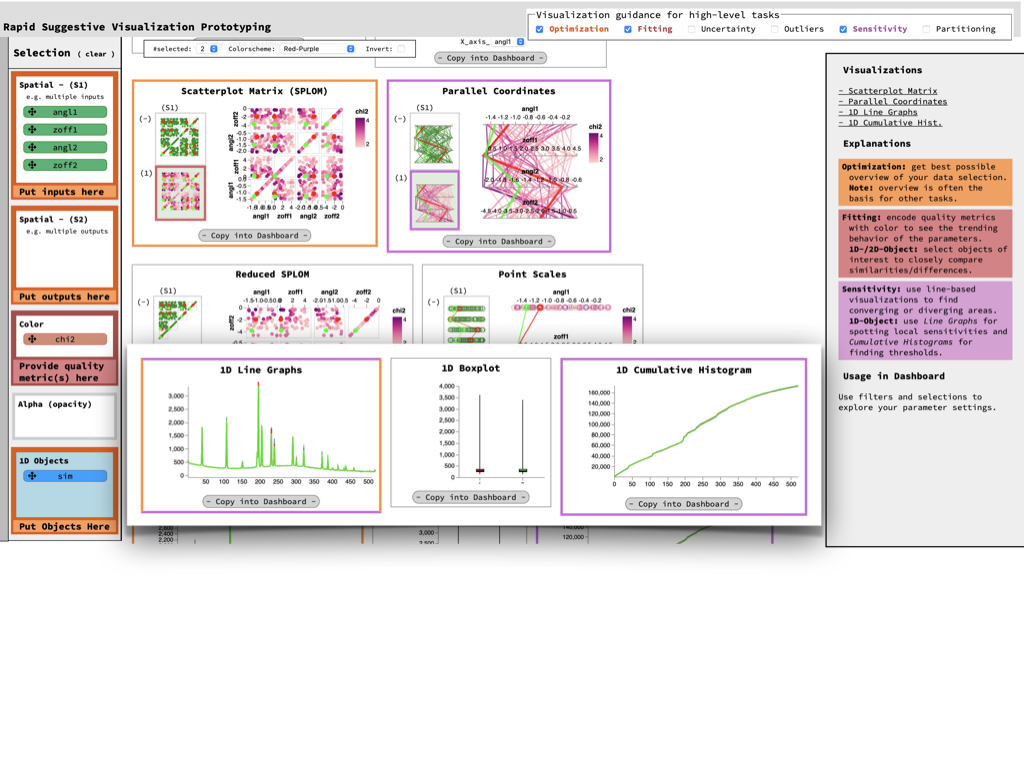}
    \caption{Recommendations for tasks optimization \topt{}, fitting \tfit{}, and sensitivity \tsens{} for the Powder-dataset.}
    \label{fig:case-study-I-visrec}
\end{figure}

The physicist provided fifty random samples for four parameters of interest and 1D outputs and derived $\chi^2$-metrics. 
\colored[details]{Due to timing constraints, the analysis was divided into two sessions: one in-person and one online.
At the beginning of the in-person session, he received a brief introduction to RSVP, followed by time to experiment with the system and ask questions.}
The recommendations that underpin the dashboard depicted in \autoref{fig:teaser}-\lexIII{}, which showcases the most noteworthy insights, are shown in \autoref{fig:case-study-I-visrec}.

\begin{figure}[!t]
    \centering
    \includegraphics[width=0.7\linewidth, trim={0 8cm 0 0 },clip]{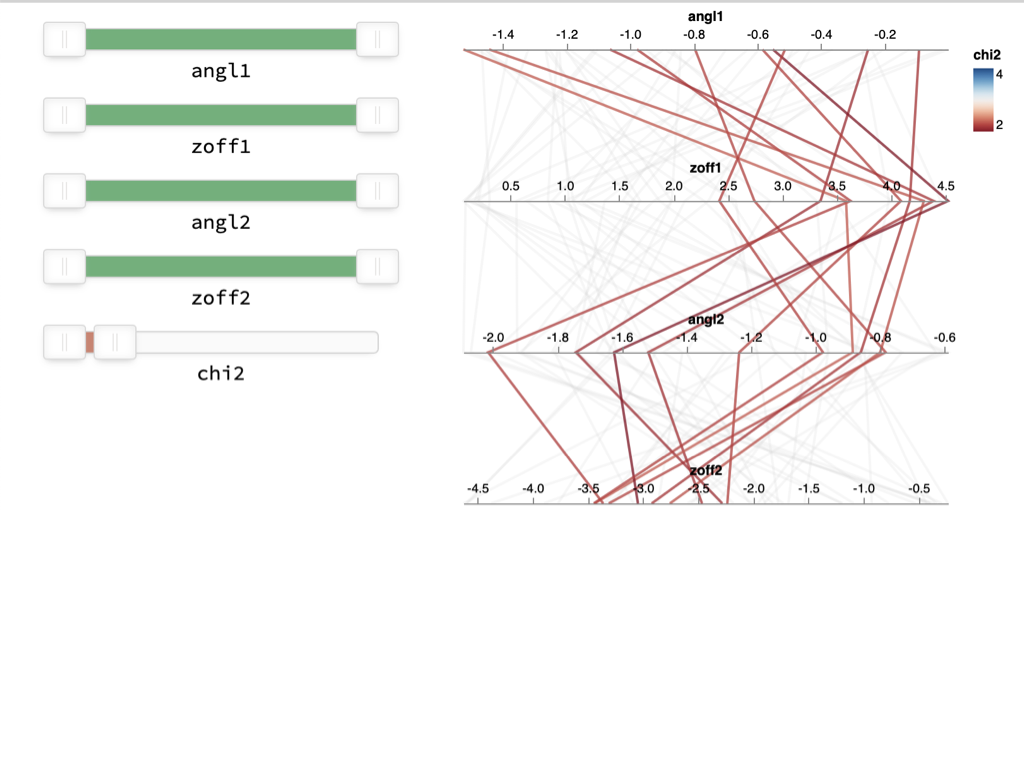}
    \caption{Detail view of the filters and Parallel Coordinates from \lexIII{}. }
    \label{fig:case-study-I-pc}
\end{figure}

The Parallel Coordinates (see \autoref{fig:case-study-I-pc}) show that parameter settings to obtain a low ${\chi}^2$ result scatter over the entire range for parameters \emph{angl1} and \emph{angl2}, but they become narrower for \emph{zoff1} and \emph{zoff2}. 
This behavior indicates that \emph{angl1} and \emph{angl2} have little to no influence to get a good ${\chi}^2$ result, whereas \emph{zoff1} and \emph{zoff2} do. 
The physicist was aware of this behavior but was surprised that such a low-effort approach using a coarse sampling strategy combined with a relatively simple visualization could reveal what he had learned from long computational cycles and multiple sequential experiments. 

Studying the complex object visualizations revealed insights previously unknown to the scientist. 
Generally speaking, the physicist tries to figure out how the molecules fit together inside the crystal. 
The curves in the 1D-Linegraph represent crystal structures in reciprocal space, a Fourier transform of the real space\cite{Kittel2018}.
The peak locations and spacings represent the periodicity of the crystal in three space dimensions, and peak heights indicate details about the molecular structure.
Usually, the domain scientist studies these curves one at a time instead of overplotting the results of several experiments in a single graph.

\begin{figure}[b]
    \centering
    \includegraphics[width=0.8\linewidth, trim={12cm 8cm 0 0 },clip]{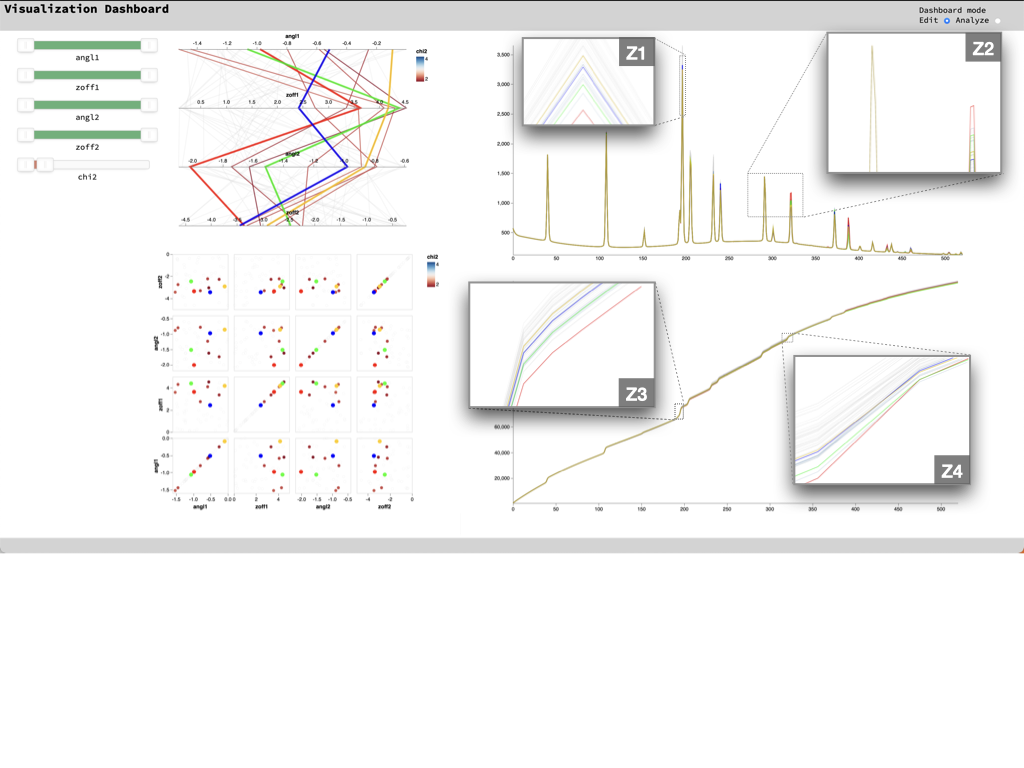}
    \caption{Showing the 1D-Linegraph and the 1D-Cumulative Histogram from dashboard \lexIII{}, with several good candidates selected and presenting zooms for multiple areas of interest.}
    \label{fig:case-study-I-vdb}
\end{figure}

As presented in \autoref{fig:case-study-I-vdb}, the four input parameters highly influence the peak at position 196, a well-known circumstance to the scientist. 
However, he was surprised by the high variance of suitable candidates for good ${\chi}^2$ results (see \autoref{fig:case-study-I-vdb}-\graybox{Z1}). 
He expected the margin of error at this position to be way lower for suitable candidates.
Another finding was that the peak at position 291 is invariant under these four parameters, meaning it remains static for any parameterization (see \graybox{Z2}, left peak).
It will require further research on why the molecules are invariant to depth (zoff) and rotation (angl) at this particular frequency.
Combining these findings and studying them within the \textit{1D Cumulative Histogram} view revealed further insights.
Highlighting runs with good ${\chi}^2$ results show that they do not change their positions relative to each other between the peaks at locations 196 and 291 (see \graybox{Z3} and \graybox{Z4}, respectively).
In other words, the area between these two peaks is stable regarding the objective function. 
From a global perspective, the peak with the highest variability is followed by a relatively stable area, culminating in a perfectly stable state at position 291 before massive changes set in again immediately after this location (the right side of \graybox{Z4} shows how the highlighted runs begin crossing each other). 
The scientist was completely unaware of this global pattern that actually challenged his understanding of this crystal entirely. 

\deleted{Based on these findings, the physics professor is now considering the possibility of re-examining previously analyzed models using RSVP.
He wants to look for crystal behavior that could help them understand open questions they might have missed using classical analysis methods.}


\begin{Newcont}
\subsection{Case Study II: Lighting Design Optimization}  

Our second domain scientist is a researcher who aims to automate the placement of light sources in official buildings by adapting the lighting configuration through a gradient-based optimization approach. 
Office environments and official buildings must meet specific lighting standards for different areas, like desk surfaces. 
Current commercial lighting design tools necessitate manual adjustments of luminaires to reach the desired outcome. 
This new approach would allow designers to directly set target radiances for surfaces, eliminating the need to manually adjust the luminaires to achieve the intended results.


The researcher was intrigued by the general VPSA approach and its potential benefits to his work. He conducted 685 runs of a two-level office room with four lamps. The positioning of the lamps was unconstrained, allowing them to be placed anywhere in the room. Each lamp's position in 3D space constituted twelve parameters, along with the objective function results for each configuration. Additionally, he provided gradients for the optimization step, including the gradient norm, adding thirteen more parameters, for a total of twenty-six to analyze. Generating the samples and converting them into the required format took approximately forty-five minutes.

\colored[details]{The analysis session was conducted in person on a 2021 16-inch MacBook Pro equipped with an Apple M1 Max chip and 64GB of memory, running a locally hosted version of RSVP. 
After receiving a brief introduction to RSVP, the researcher began with the analysis, during which he could ask questions about the system at any time.}
He experimented with combining the dimensions in different ways and studied the recommendations and what they meant. During the session, he created numerous dashboards and had several interesting insights. 
The dashboard presented in \autoref{fig:lighting-design-optimization} summarizes some of the most interesting ones. 
This is the outcome of RSVP's recommendation for task optimization when encoding all parameters (PSc for both inputs and outputs) and subsequently adding the recommendation when only including the y-dimensions of the lights (SPLOM).
\begin{figure}[!t]
    \centering
    \includegraphics[width=\linewidth]{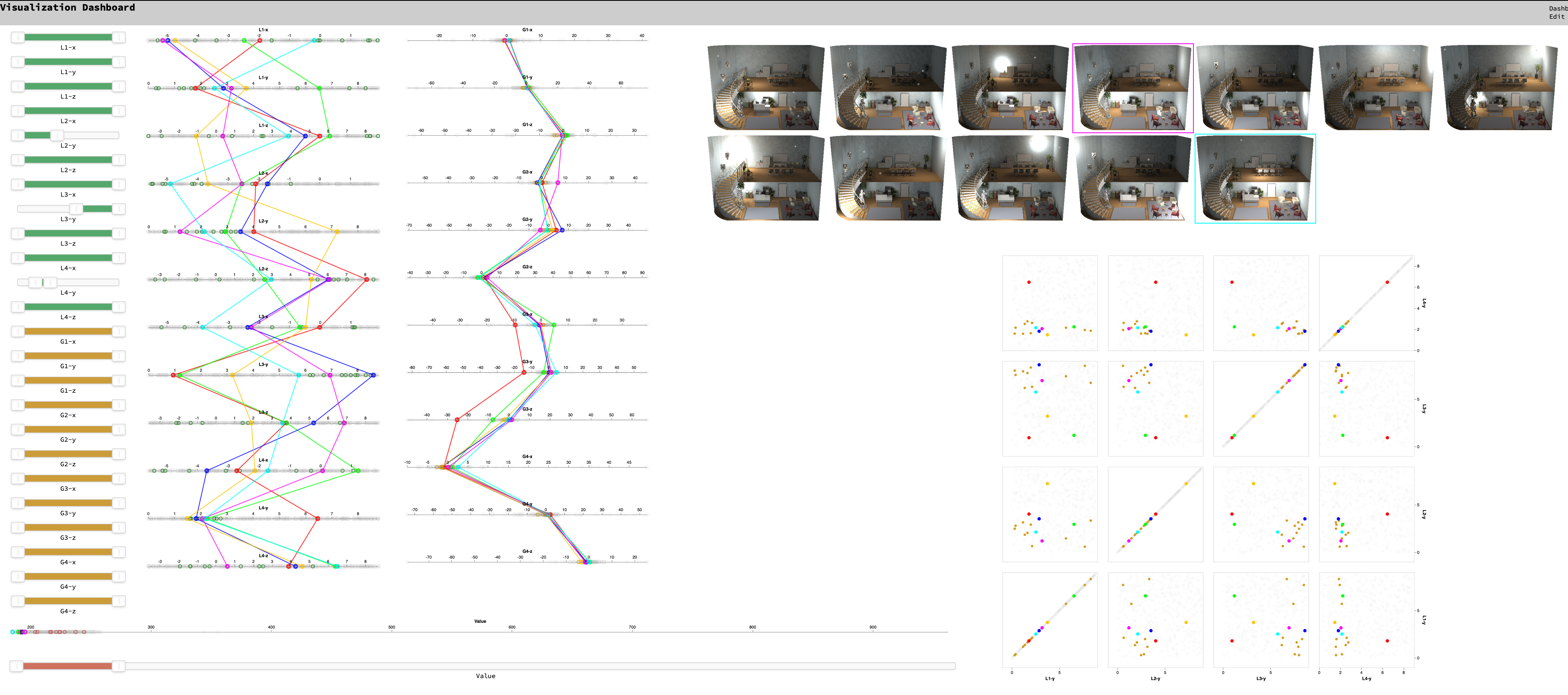}
    \caption{\newcont{Dashboard created during the Case Study on Automated Lighting Optimization}}
    \label{fig:lighting-design-optimization}
\end{figure}

In this dashboard, the objective function value is filtered to a relatively low level, highlighting options with favorable values. 
An interesting observation emerged regarding the y-positions of the lamps, indicating a necessity for at least two lamps on the bottom level and one on the top level of the office, while the position of the remaining lamp could be more flexible. 
The researcher could deduct this from the images and confirm it with the recommended plots.
Although not surprising, since effective lighting is mainly required for three surfaces in this particular room, this insight prompted the realization that a simple sampling strategy could address a challenge in their algorithm: determining the minimum number of lamps to place in a room initially
Apparently, a straightforward random sampling strategy could be used with varying numbers of lights, and those yielding feasible objective values could then serve as the starting points for their optimization algorithm.

The other insight pertained more to the system itself. 
While all values are below a required threshold, examining the rendered images reveals that not all of them are visually pleasing. 
Several results display bright spots on the wall (see \autoref{fig:office-ex1}), which should be avoided in efficient lighting design. 
On the other, elements that should receive at least some light can appear rather dark, like the staircase in \autoref{fig:office-ex2}.
Both issues can be addressed by adjusting the illumination constraints on these objects. While relatively easy to achieve in theory, it is often overlooked in practice to add all constraints that are not part of the requirement specification. RSVP could be utilized for debugging these configurations. Rendering a second image from a different perspective would essentially provide a 360° view of the office, allowing one to scrutinize good results for possible mistakes stemming from missing negative constraints.
\begin{figure}[h]
\centering
\begin{subfigure}[m]{0.48\linewidth}
    \centering
    \includegraphics{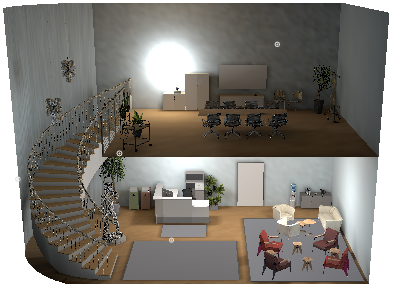}
    \caption{Bright Spot}
    \label{fig:office-ex1} 
\end{subfigure}%
\begin{subfigure}[m]{0.48\linewidth}
    \centering
    \includegraphics{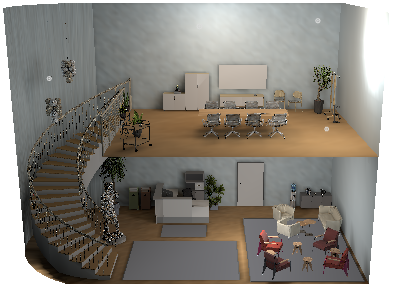}
    \caption{Poorly Lit Staircase}
    \label{fig:office-ex2} 
\end{subfigure}
\caption{Office Examples}
\label{fig:office-examples}
\end{figure}


\end{Newcont}


\section{Discussion}
\label{sec:discussion}
In this section, we present our findings regarding the evaluations and our two key design goals, followed by additional analysis concerning the design aspects of RSVP. We conclude this section by outlining the limitations and potential directions for future research.

\subsection{Findings}
\label{sec:findings}

Our basic assumption for the meta design study proved to be correct. 
It was possible to extract an expressive visualization design space and a task-oriented visualization recommendation strategy from existing literature, utilizing insights from the conceptual VPSA framework. 
In the following, we will analyze how our design achieves our two key goals.

\textbf{\textit{Low Cost (Key-I)} :}
The physicist reported that sampling the model and formatting the data took approximately sixty minutes. 
This estimate includes the time required to figure out how to perform the sampling. 
He believes this process could be reduced to a few minutes with some exercise.

Five out of six participants of the usablity study were unaware of at least one visualization offered by RSVP.
However, no participant had trouble interpreting and using any of them. 
All the users mentioned that it was beneficial to see the visualization options side by side encoded with actual data, which made learning by comparison easy.
Three participants noted that the step-by-step introduction was helpful for them. 
One of them noted: \emph{"Seeing all the visualizations [encoded] with all the dimensions immediately would probably have been overwhelming. But seeing how similar they are in the beginning was very interesting and made it easy to follow as you added new dimensions."}

The domain scientist was pleased to see his data visualized right away.
He had prior experience with a design study that took months and several iterations before he saw actual visualizations of his data.
\emph{"Back then, I entered the design study with the prospect of learning something about visualizing my data and finding out if it could help me with my research. RSVP did all of that, just really fast."}

Regarding the SUS questionnaire, we argue that \textit{high usability} can be interpreted as \textit{low cost} of using a system.
As previously reported, RSVP achieved an average score of 82.5 points, which is considered to be \emph{very good}\cite{Brooke2013}. 

We take those findings as positive indicators that RSVP achieves \keyI{}, imposing \textit{low costs} on the user in terms of time and cognitive effort required for using it.

\textbf{\textit{High Trust (Key-II)} :} 
The domain scientist admitted that he was initially skeptical about the VPSA approach.
He could not imagine that such a coarse sampling could reveal anything about the data that he did not already know.
However, fast revelations and confirming existing knowledge about the data helped to build trust and further spurred his curiosity. 
Consecutive findings and insights led to the decision to adopt RSVP for future research. 
He issued the following statement: 
\emph{"Although the instruments for taking physical measurements have improved over the years, the way scientists analyze their data has not changed much. 
VPSA, in general, could change how scientists study their models, and RSVP, in specific, could probably help them to achieve this."}

Trustworthiness in VisRec may not be captured by simple measures, but Dasgupta et al. suggest that trust can be defined as self-calibrated confidence in the analysis outcome \cite{Dasgupta2017}.
Therefore, the confidence ratings obtained from the usability study indirectly reflect the reliability of both the VPSA approach and the VisRec strategy.
The overall confidence score was 4.17 points, which means that usability testers were \emph{rather confident} in their findings.
Interestingly, the challenges with the highest complexity had the highest confidence ratings (4.83 points on average). 
This finding is similar to observations by Dasgupta et al. \cite{Dasgupta2017} where they reported that domain scientists' trust in a visual analysis system increased with the complexity of the analysis tasks.
The users explained this behavior by the experience gained during the previous challenges and that they would not have known how to evaluate the objective functions otherwise.
We take these findings as proof that RSVP also fulfills \keyII{}, making users feel \emph{high trust} in both the RSVP system and the VPSA method in general.

    \subsection{Further Design Analysis}

As previously discussed, a meta design study diverges from a classical design study in its method of deriving domain knowledge. While the latter emphasizes user-centric design, the former derives domain knowledge not from direct interaction with individuals but rather from written reports. This is also how assumptions about domain scientists are formed. However, to enhance the overall usability of such a tool or to potentially address the needs of individual scientists, iterative and user-centric design remain essential. 
Therefore, we will delve into these aspects in this section.
\colored[tableau]{
Furthermore, existing frameworks like ComVis~\cite{Matkovic2008a} and Visplore~\cite{Piringer2009} generally allow for the creation of more powerful dashboards compared to those possible with RSVP.
However, they are not readily accessible to domain scientists, requiring a visualization expert to design and develop a customized application or dashboard before analysis can begin. 
In contrast, RSVP allows domain scientists to independently develop their own VPSA dashboards without the assistance of a visualization expert.}
\colored[tableau]{
Tableau, however, does not require programming and can effectively create all the visualizations that RSVP can.}
This raises the question: what factors would motivate choosing RSVP over Tableau? 
To address this, we will compare RSVP to Tableau and attempt to answer this question.

\textbf{\textit{Iterative Design:}}
The design study discussed in the supplementary material represents an early application of RSVP, demonstrating its initial functionality. 
At this stage, RSVP only featured a single spatial encoding field, necessitating the sequential creation of visualizations for inputs and outputs. Recommendations were provided in the form of a ranked list of visualizations deemed effective for specific tasks. While this study provided interesting insights for the user, it showed the necessity of collaboration with visualization experts familiar with RSVP and VPSA for the development of comprehensive dashboards.

This experience prompted a redesign of RSVP, introducing two spatial fields to enable simultaneous input and output dimension encoding. Additionally, it prompted a reassessment of the coding table concerning the Vis Rec strategy, culminating in the strategy detailed in Section 3.3.

Furthermore, prior to conducting the usability study, two pilot studies involving an undergraduate student and a Ph.D. student were conducted, resulting in further usability enhancements. 
These enhancements included clarifications on task meanings, explanations within the dashboard, and adjustments to color schemes. While the initial color differentiations between tasks and encoding channels were distinct, efforts were made after the pilot studies to align them more closely, facilitating task-channel associations. Moreover, the color scheme for the third column in small multiples was revised to reflect a blend of the colors used in the first two columns, enhancing participants' understanding of the data representation.

\textbf{\textit{User-oriented Design:}} 
In situations where specific users have requirements that can not be addressed by RSVP, a traditional design study may be warranted.
Nonetheless, RSVP continues to play a crucial role by serving as a bridge between domain scientists and visualization researchers, effectively acting as a liaison \cite{Simon2015}. Domain scientists can utilize RSVP to construct dashboards with their own data, simplifying the explanation of data, tasks, and system limitations to visualization researchers. Consequently, these researchers can better grasp the actual needs of domain scientists and expedite the design of potential solutions.

RSVP was designed with extensibility in mind for such cases. It provides a mechanism whereby visualization designers proficient in creating visualizations with Vega can develop new visualizations. As long as these visualizations adhere to certain requirements regarding data setup, RSVP can seamlessly integrate them into the visualization dashboard \lvdb{}. This integration has the potential to substantially reduce the time required for design studies, which in some instances can extend over several years \cite{Booshehrian2012}. Although this mechanism within RSVP is still in its early stages, it will be refined in future iterations of the system.

\textbf{\textit{Comparison to Tableau:}}
The draggable markers in RSVP's data-selection panel \ldsp{} were inspired by Tableau. However, Tableau does not allow for editing two different visualizations simultaneously, unlike RSVP, which facilitates this through two distinct spatial fields. 
This is a critical aspect when creating VPSA applications, as input and output dimensions are often encoded in separate views.

Similar to VisRec in RSVP, Tableau's "Show Me" feature also utilizes small multiples of different visualizations, employs colored frames to highlight recommended visualizations, and provides hints to the user on how to encode a specific visualization. 
However, the small multiples in Tableau are static icons and do not display actual user data. To view them encoded, the user must instantiate a visualization. If a user wishes to compare multiple visualizations, they must instantiate them individually and compare them sequentially. 
In RSVP, all available visualizations are instantiated simultaneously, allowing for easy comparison. 
This approach aims to address potential issues with visualization literacy and save time for users, which is particularly important for domain scientists.
Furthermore, recommendations in Tableau are generic suggestions based on certain data features, while recommendations in RSVP are task-oriented, tailored to problems commonly encountered in VPSA.

Tableau can handle 1D objects in the form of additionally loaded and linked data sheets. 
2D objects can be added as links to images, similar to how it is done in RSVP. 
However, converting links into images is not straightforward in Tableau and requires considerable expertise to accomplish. 
This becomes even more challenging when juxtaposing or superimposing selected images. 
Furthermore, Tableau does not provide guidance on combining these complex objects with other visualizations in the dashboard to facilitate analysis and achieve specific high-level tasks.

In summary, Tableau offers greater overall power compared to RSVP. 
However, RSVP is highly efficient and effective for handling VPSA-related problems. This comparison is akin to comparing a domain-specific language (DSL) with a general-purpose programming language: while a general-purpose language can theoretically achieve anything, it may be complex to achieve. 
In contrast, a DSL simplifies achieving specific tasks by making operations irrelevant to the domain impossible\cite{McNutt2022}.

\subsection{Limitations and Future Work}
\label{sec:limitations}
\label{sec:future-work}

\colored[smalln]{\textbf{\textit{Number of Participants in Usability Study:}}
Our decision for $n=6$ follows recommendations aimed at optimizing sample size for usability tests to maximize return on investment (ROI) \cite{Turner2006, Lewis1994}.
}
\colored[smalln]{We set our problem discovery goal at a value of 0.9, which we considered sufficient for a prototype at this stage. 
Based on a preliminary study involving two users and a previous case study, we estimated a problem discovery rate $p=0.33$ per user. 
In order to meet our goal with this rate, six participants were necessary \cite{Turner2006}. 
If we wanted to enhance the robustness of the results, $n=8$ could accommodate variability in the discovery rate up to 0.08 points. 
In other words, if our estimated discovery rate was overly optimistic at 0.33, but the actual rate was 0.25, $n=8$ would achieve our goal of 0.9.
Conversely, if our estimation of 0.33 was accurate, $n=8$ would increase the problem discovery probability to 0.95. 
These considerations pertain to the evaluation of a single prototype.
If the SUS was used to assess user preferences between two competing prototypes, eight to twelve participants would be necessary to ensure robust results~\cite{Tullis2004}. 
However, since our study currently does not involve a comparison between prototypes, this recommendation does not immediately apply to our situation.}

\textbf{\textit{Predictions:}}
\label{sec:predictions}
The dataflow model presented in \autoref{fig:vpsa-dataflow} is an abbreviated version of the model proposed by Sedlmair et al.
The complete version includes an alternative prediction step, which allows analyzing data points that have not been sampled and simulated yet but were approximated with a surrogate model.
Using prediction allows the continuous analysis of the parameter space. 
However, it requires more complicated visualizations and advanced navigation techniques, and keeping selections consistent across discrete and continuous views would be rather complicated, if possible at all.
Showing complex objects for non-simulated data points would be another challenge.
Training deep learning based surrogate models capable of approximating complex objects for not simulated points can take up to several hours on a supercomputer\cite{HeW2019}, which is not feasible for most users. 
Despite these challenges, including predictive analysis capabilities in a generic visualization tool would undoubtedly appeal to many users performing VPSA.

\newcont{\textbf{\textit{Transferability:}}}
We have shown how a meta design study can help extract a visualization design space and a suitable visualization recommendation strategy for VPSA. 
\changed{\st{Nevertheless, we believe this general approach is transferable to other problems.}}{ However, we believe this approach can be applied to other problem domains}
\changed{\st{For instance, many of the analyzed papers in our study feature complex objects and spend a fair amount of time describing techniques and approaches to transform them into lower-dimensional representations for analytical purposes.}}{For example, many of the papers analyzed in our study deal with complex objects and devote considerable effort to describing techniques for transforming them into lower-dimensional representations for analysis.}
Selecting appropriate transformations can be challenging, and users may not be aware of all possible transformations for their analyzed objects.
To the best of our knowledge, assisting users in selecting suitable transformations for complex objects into alternative representations offering enhanced analytical capabilities is currently a research gap that could potentially be addressed by a meta design study.

\begin{Newcont}
\textbf{\textit{Scalablity:}}
As previously stated, VPSA is an approach that facilitates the analysis of multi-dimensional input-output-related problems, where the dimensions carry semantic meaning. 
For high-dimensional problems encountered in fields like machine learning and deep learning, alternative strategies exist Hohman et al., 2019.
There is no definitive threshold distinguishing between multi-dimensional and high-dimensional problems~\cite{Sedlmair2014}, but the analysis of the applications in the meta design study revealed that most of them featured examples with less than ten dimensions, and none of them exceeded twenty.

RSVP provides visualization options and recommendations for up to fifteen input and output dimensions, respectively, with support for up to one thousand runs (\conII{}). If these limits are significantly exceeded, alternative visualization options and a potentially faster rendering solutions (such as a WebGL/WebGPU implementation of the renderer in Vega) may be necessary.


\textbf{\textit{Extensibility:}}
As previously explained, the visualization dashboard is already extensible with new Vega visualizations (though currently undocumented), but the visualization options in the overview area are not currently extensible. Additionally, the VisRec rules are presently derived manually, which complicates the introduction of new rules or updates to existing ones when new papers or applications are added to the knowledge base. Converting our hard-coded rules into constraints for answer-set-programming or utilizing a framework like Knowledge Rocks for handling visualization knowledge might mitigate this problem, but further research is necessary in this regard. Moreover, automating the extraction of new VisRec rules from design study papers is a task for future research.
\end{Newcont}

\newcont{\textbf{\textit{Sampling Support:}}}
\changed{\st{In our experience, users can often}}{Users often can} sample their model using random or Cartesian grid sampling but are incapable of using more sophisticated strategies like Latin Hypercube sampling~\cite{Mckay2000} or uniform sampling on a space-filling curve~\cite{He2016}.
However, advanced visualization techniques like the previously discussed ones also necessitate improved sampling strategies. 
Furthermore, Latin Hypercube Sampling would probably improve uncertainty and sensitivity analysis\cite{Helton2003}. 
We envision a tool like Hanpuku~\cite{Bigelow2017}, which would support domain scientists in sampling their models with more sophisticated strategies and create outputs that could be directly used by tools like RSVP. 

\newcont{\textbf{\textit{Automated Dashboard Design:}}}
RSVP provides multi-view recommendations and hints to users on how to interact with them, but it does not support arranging those views in the dashboard. 
The size of individual views related to the amount of data they encode could be determined using Pixnostics\cite{Schneidewind2006}, Scagnostics\cite{Wilkinson2005}, and Pargnostics\cite{Dasgupta2010}. 
Layout proposals could be based on simple heuristics\cite{Bach2018}, or they could be learned from online dashboards\cite{LinY2022}. 
A domain-oriented recommendation approach could also be considered. 
Domain scientists tag and share dashboards they deem helpful, and the system could recommend those to users facing a similar problem. 
The similarity could be established by analyzing those tags and data\cite{Oppermann2020b} or via an ontology-based approach\cite{Lohfink2021,Sobral2020}.

\section{Conclusion}
\label{sec:conclusion}
\begin{Deleted}
\st{We introduced RSVP, a VPSA knowledge system for designing visualization dashboards that empowers domain scientists to quickly prototype custom solutions tailored to their specific data and analytical needs. 
Intending to facilitate the prototyping process, RSVP implements a task-oriented, multi-view visualization recommendation (VisRec) strategy to guide users in meeting their analytical demands. 
To meet the diverse needs of domain scientists, RSVP provides an overview of all available visualization options and seamlessly integrates the VisRec strategy to ensure ease of use. 
Additionally, the system promotes learning by explaining its recommendations and supporting users in transferring knowledge from existing visualizations to new ones. 

We gathered the formalized VPSA knowledge implemented in the system by conducting an extensive meta design study over the body of work on VPSA. 
This comprehensive study covered data, tasks, and requirements and thoroughly examined the visualizations and encodings used to address them effectively. 
Due to the structured results of this process, we were able to derive a VisRec strategy, utilizing the insights gained from the data-task-abstraction process as a basis for evaluating and selecting options from the extracted visualization design space. 
This work demonstrates the effectiveness of using a meta design study to derive a visualization design space and corresponding VisRec strategy from existing work in VPSA. 
However, we discussed the possibility of extending this approach to other problem domains, allowing the creation of general and versatile tools for visual data analysis that employ a task-oriented VisRec strategy.

We conducted a usability study showing that the system is user-friendly and can be easily adopted by students and users with average data visualization skills. 
In addition, case studies have shown that the VisRec strategy implemented in RSVP can provide task-oriented recommendations that can lead domain scientists to previously unknown insights.
These results indicate that RSVP has the potential to make VPSA accessible to a wide range of users in different fields and industries. 
In upcoming releases, we plan to enhance the system by offering sampling support and expanding the visualization options for different types of complex objects, further expanding the capabilities of RSVP and enabling users to tackle their domain-specific problems.}
\end{Deleted}

\begin{Changed}

We presented RSVP, a system based on an extensive meta design study on the subject of VPSA, designed to enable domain scientists wihtout programming skills to create custom visualization dashboards quickly.
It features a task-oriented visualization recommendation strategy and a drag-and-drop interaface for creating said dashbaords. 
RSVP not only guides users through their data analysis needs but also promotes learning and provides explanations for its recommendations. 
A usability study and case studies confirm RSVP's user-friendliness and its ability to provide valuable insights, suggesting its applicability across various fields. Future updates will focus on enhancing RSVP's capabilities regarding the expansion of visualization options, recommendations, and support for more complex data types.
\end{Changed}

\section*{Acknowledgements}
We would like to thank Silvia Miksch for the valuable discussions and her helpful insights. 
This research was partially supported by NSF award 2007436 and WWTF award ICT19-041.


%





\ifCLASSOPTIONcaptionsoff
  \newpage
\fi



\bibliographystyle{IEEEtran}
\bibliography{rsvp}
\begin{IEEEbiography}[{\includegraphics[width=1in,height=1.25in,trim={0.1in 0.3in 0.1in 0}, clip,keepaspectratio]{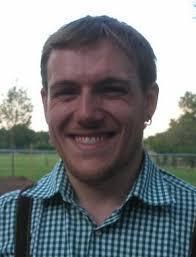}}]{Manfred Klaffenboeck}
is a doctoral candidate at TU Wien's Institute for Visual Computing and Human-Centered Technology in Vienna, Austria.
He earned a Magister degree in Movie and Media Science in 2013, followed by a BSc in Computer Science in 2016 and an MSc in Media Informatics in 2018, all from the University of Vienna. 
His research integrates computer graphics, computer vision, data visualization, and media analysis, with a current focus on semi-automated multiverse analysis leveraging (visual) parameter space exploration, visual analytics, and knowledge assistance. 
\end{IEEEbiography}

\begin{IEEEbiography}[{\includegraphics[width=1in,height=1.25in,trim={0.45in 0 0.45in 0}, clip,keepaspectratio]{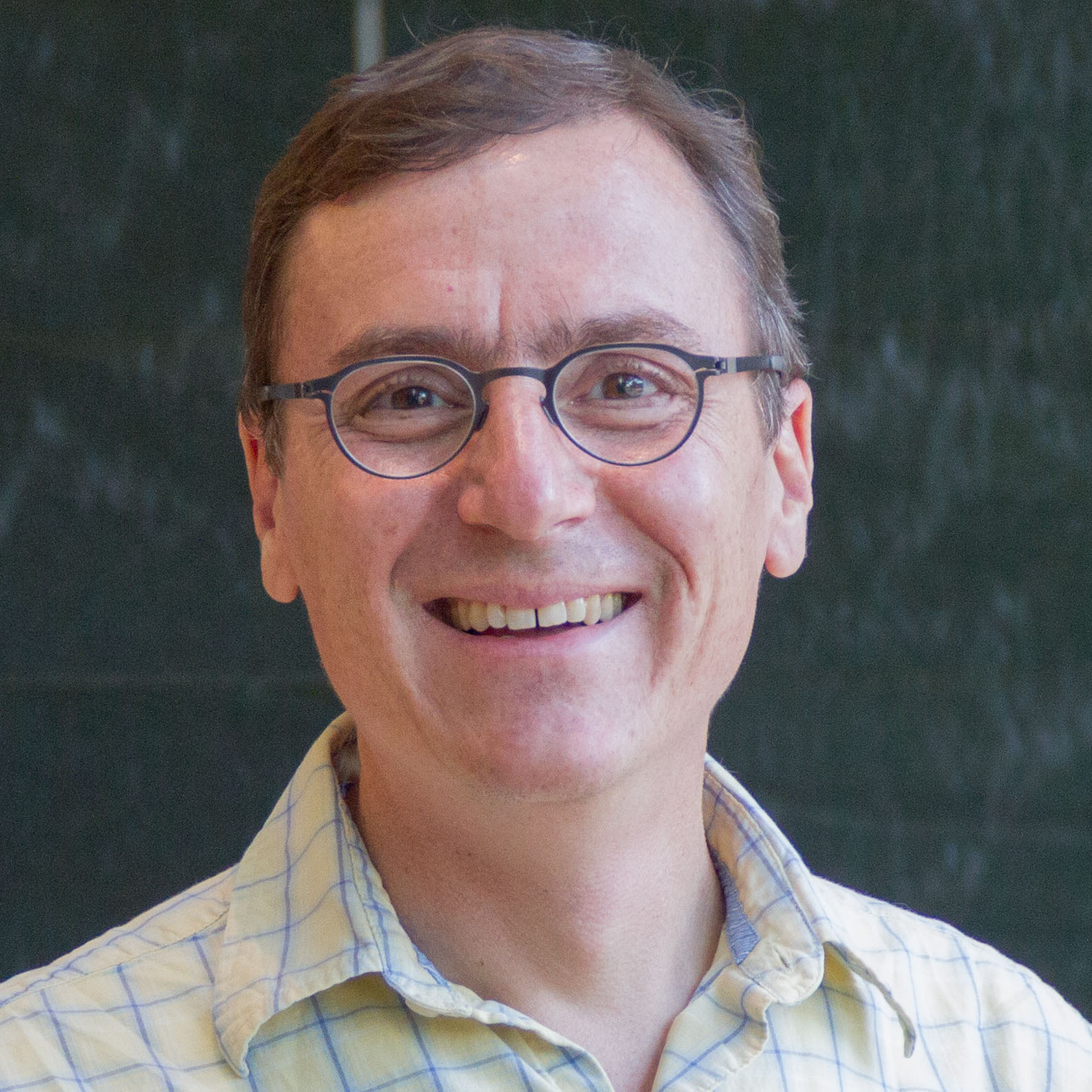}}]{Michael Gleicher}
is a Professor in the Department of Computer Sciences at the University of Wisconsin, Madison. Prof. Gleicher is founder of the Department's Visual Computing Group and  co-directs both the Visual Computing Laboratory and the Collaborative Robotics Laboratory at UW-Madison. His research interests span the range of visual computing, including data visualization, robotics, and virtual/extended reality. His recent work includes exploring perceptual issues in visualization, the use of visual simulation for robotics, and geometric approaches to enhance robot perception and interaction. 
He earned his Ph. D. in Computer Science (1994) from Carnegie Mellon University, and earned a B.S.E. in Electrical Engineering from Duke University (1988). 
In 2023-2024, Prof. Gleicher holds a concurrent appointment as a Design Scholar at Amazon Robotics. This work is not associated with Amazon.
\end{IEEEbiography}

\begin{IEEEbiography}[{\includegraphics[width=1in,height=1.25in,trim={0.55in 0 0.55in 0}, clip,keepaspectratio]{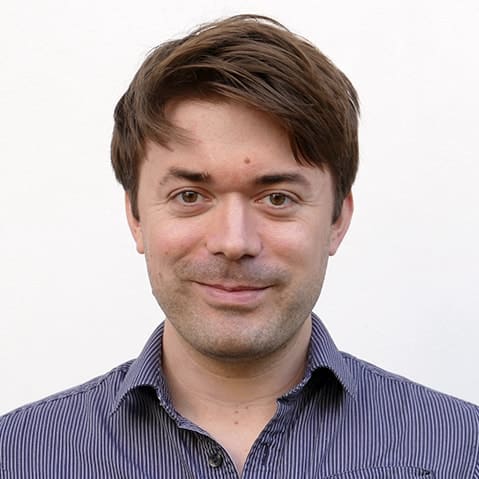}}]{Johannes Sorger}
 received his computer science degree in visual computing from the TU Wien at the Institute of Computer graphics and Algorithms, where he also finished his PhD in collaboration with the VRVis research center. Johannes’ main research interests are centred around the application of visualization as an enabling technology. For his work on the visualization of neuronal networks Johannes received the 2014 OCG Incentive Award, as well as the Best Paper Award at the 2013 IEEE Symposium on Biological Data Visualization.
During his postdoc, he was heading the visualization team at CSH Vienna where he researched novel approaches in Immersive network analytics. 
\end{IEEEbiography}

\vfill

\begin{IEEEbiography}[{\includegraphics[width=1in,height=1.25in,trim={0.25in 0 0.25in 0}, clip,keepaspectratio]{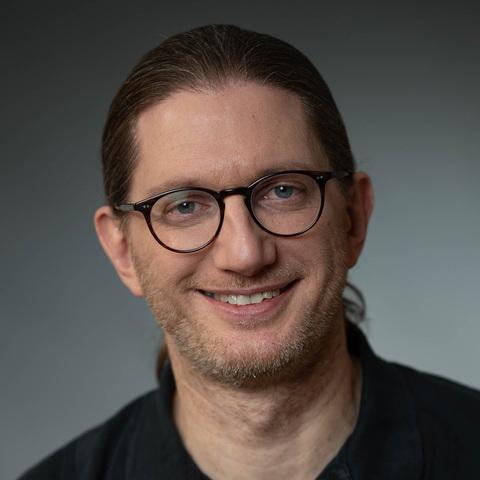}}]{Michael Wimmer}
is currently a full professor at the Institute of Visual Computing and Human-Centered Technology at TU Wien, where he heads the Rendering and Modeling Group. He is also the director of the Center for Geometry and Computational Design (GCD). His academic career started with his M.Sc. in 1997 at TU Wien, where he also obtained his Ph.D. in 2001.
His research interests are real-time rendering, computer games, real-time visualization of urban environments, point-based rendering, reconstruction of urban models, procedural modeling, shape modeling, and computational design.
\end{IEEEbiography}

\begin{IEEEbiography}[{\includegraphics[width=1in,height=1.25in,trim={0in 0.2in 0in 0},clip,keepaspectratio]{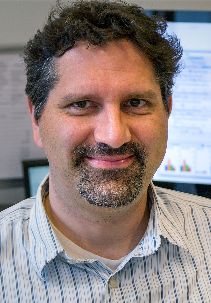}}]{Torsten Möller}
(Senior Member, IEEE) received the Vordiplom (BSc) degree in mathematical computer science from the Humboldt University of Berlin, Germany, and the PhD in computer and information science from Ohio State University in 1999. He has been a professor of computer science at the University of Vienna, Austria, since 2013. Between 1999 and 2012, he served as a computing science faculty member at Simon Fraser University, Canada. He is a senior member of ACM, IEEE, and a member of Eurographics. His research interests include algorithms and tools for analyzing and displaying data with principles rooted in computer graphics, human-computer interaction, signal processing, data science, and visualization.
\end{IEEEbiography}





\vfill


\end{document}